\documentclass{emulateapj}
\usepackage{natbib}
\newcommand{\hi}{\mbox{\rm \ion{H}{1}}}
\newcommand{\hii}{\mbox{\rm \ion{H}{2}}}
\newcommand{\htwo}{\mbox{\rm H$_2$}}

\newcommand{\halpha}{\mbox{\rm H$\alpha$}}

\newcommand{\xco}{\mbox{$X_{\rm CO}$}}
\newcommand{\xcounits}{\mbox{cm$^{-2}$ (K km s$^{-1}$)$^{-1}$}}
\newcommand{\acounits}{\mbox{M$_\odot$~pc$^{-2}$~$({\rm K\,km\,s}^{-1})^{-1}$}}
\newcommand{\sigsfr}{\mbox{$\Sigma_{\rm SFR}$}}

\newcommand{\sightwo}{\mbox{$\Sigma_{\rm mol}$}}
\newcommand{\siggas}{\mbox{$\Sigma_{\rm gas}$}}

\shorttitle{Molecular Gas and Star Formation in Nearby Disk Galaxies}
\shortauthors{Leroy et al.}

\begin{document}

\slugcomment{Accepted to the Astronomical Journal}
\title{Molecular Gas and Star Formation in Nearby Disk Galaxies}

\author{Adam K. Leroy\altaffilmark{1},
  Fabian Walter\altaffilmark{2},
  Karin Sandstrom\altaffilmark{2},
  Andreas Schruba\altaffilmark{3},
  Juan-Carlos Munoz-Mateos\altaffilmark{1},
  Frank Bigiel\altaffilmark{4},
  Alberto Bolatto \altaffilmark{6},
  Elias Brinks\altaffilmark{7},
  W.J.G. de Blok\altaffilmark{8,9},
  Sharon Meidt\altaffilmark{2},
  Hans-Walter Rix\altaffilmark{2},
  Erik Rosolowsky\altaffilmark{10},
  Eva Schinnerer\altaffilmark{2},
  Karl-Friedrich Schuster\altaffilmark{11},
  Antonio Usero\altaffilmark{5},
  }
\altaffiltext{1}{National Radio Astronomy Observtory, 520 Edgemont Road, Charlottesville, VA 22903, 
USA}
\altaffiltext{2}{Max Planck Institute f\"ur Astronomie, K\"onigstuhl 17, 69117, Heidelberg, Germany}
\altaffiltext{3}{California Institute for Technology, 1200 E California Blvd, Pasadena, CA 91125}
\altaffiltext{4}{Theoretische Astrophysik, Albert-Ueberle-Str. 2, 69120, Heidelberg, Germany}
\altaffiltext{5}{Observatorio Astron\'{o}mico Nacional, C/ Alfonso XII, 3, 28014, Madrid, Spain}
\altaffiltext{6}{Department of Astronomy, University of Maryland, College Park, MD, USA}
\altaffiltext{7}{Centre for Astrophysics Research, University of Hertfordshire, Hatfield AL10 9AB, 
United Kingdom}
\altaffiltext{8}{ASTRON, Netherlands Foundation for Radio Astronomy, Postbus 2, 7990 AA Dwingeloo, The Netherlands}
\altaffiltext{9}{Astrophysics, Cosmology and Gravity Centre, Department of Astronomy, University of 
Cape Town, Private Bag X3, 
Rondebosch 7701, South 
Africa}
\altaffiltext{10}{University of British Columbia, Okanagan Campus, Kelowna, BC Canada}
\altaffiltext{11}{IRAM, 300 rue de la Piscine, 38406 St. Martin d\textquoteright H\`{e}res, France}

\begin{abstract}
We compare molecular gas traced by $^{12}$CO(2-1) maps from the HERACLES survey,
with tracers of the recent star formation rate (SFR) across 30 nearby
disk galaxies. We demonstrate a first-order linear correspondence
between $\Sigma_{\rm mol}$ and $\Sigma_{\rm SFR}$ but also find
important second-order systematic variations in the apparent molecular
gas depletion time, $\tau_{\rm dep}^{\rm mol} = \Sigma_{\rm mol} /
\Sigma_{\rm SFR}$. At the 1~kpc common resolution of HERACLES, CO
emission correlates closely with many tracers of the recent
SFR. Weighting each line of sight equally, using a fixed $alpha_{\rm CO}$ equivalent 
to the Milky Way value, our data yield a molecular
gas depletion time, $\tau_{\rm dep}^{\rm mol}=\Sigma_{\rm
mol}/\Sigma_{\rm SFR} \approx 2.2$~Gyr with 0.3 dex $1\sigma$ scatter, in very
good agreement with recent literature data. We apply a
forward-modeling approach to constrain the power-law index, $N$, that
relates the SFR surface density and the molecular gas surface density,
$\Sigma_{\rm SFR} \propto \Sigma_{\rm mol}^N$. We find $N = 1 \pm0.15$
for our full data set with some scatter from galaxy to galaxy. This also agrees with recent work, but we 
caution that a power law treatment oversimplifies the topic given that 
we observe correlations between $\tau_{\rm dep}^{\rm mol}$ and other local and 
global quantities. The strongest of these are a decreased $\tau_{\rm dep}^{\rm mol}$ in low-mass,
low-metallicity galaxies and a correlation of the kpc-scale $\tau_{\rm
dep}^{\rm mol}$ with dust-to-gas ratio, D/G. These correlations can be
explained by a CO-to-H$_2$ conversion factor ($\alpha_{\rm CO}$) that
depends on dust shielding, and thus D/G, in the theoretically expected
way. This is not a unique interpretation, but external evidence of
conversion factor variations makes this the most conservative
explanation of the strongest observed $\tau_{\rm dep}^{\rm mol}$
trends. After applying a D/G-dependent $\alpha_{\rm CO}$, some 
weak correlations between $\tau_{\rm dep}^{\rm mol}$ and local conditions persist.  In particular, we 
observe lower $\tau_{\rm dep}^{\rm mol}$ and enhanced CO excitation associated with nuclear gas
concentrations in a subset of our targets. These appear to reflect
real enhancements in the rate of star formation per unit gas and
although the distribution of $\tau_{\rm dep}$ does not appear bimodal
in galaxy centers, $\tau_{\rm dep}$ does appear multivalued at fixed
$\Sigma_{\rm mol}$, supporting the the idea of ``disk'' and
``starburst'' modes driven by other environmental parameters.
\end{abstract}

\keywords{galaxies: evolution --- galaxies: ISM --- radio lines: galaxies --- stars: formation}

\section{Introduction}
\label{sec:intro}

The relationship between gas and star formation in galaxies plays a
key role in many areas of astrophysics. Its (non)evolution over cosmic
time informs our understanding of galaxy evolution at high redshift
\citep{DADDI10,TACCONI10,GENZEL10}. The small-scale efficiency of star
formation is a key input to galaxy simulations and scaling relations
measured for whole galaxies provide important benchmarks for the
output of these simulations. Measurements of gas and star formation at
large scales give context for studies focusing on parts of the Milky
Way \citep[e.g.,][]{LADA10,HEIDERMAN10} and the nearest galaxies
\citep[e.g.,][]{SCHRUBA10,CHEN10}. Ultimately, a quantitative understanding of
the gas-stars cycle is needed to understand galaxy evolution, with
implications for the galaxy luminosity function, the galaxy
color-magnitude diagram, the structure of stellar disks, and chemical
enrichment among other key topics.

Recent multiwavelength surveys make it possible to estimate the
surface densities of gas and recent star formation in dozens of nearby
galaxies. This has lead to several studies of the relationship between
gas and stars. Many of these focus on a single galaxy
\citep[e.g.,][]{HEYER04,KENNICUTT07,BLANC09,VERLEY10,RAHMAN11} or a
small sample \citep[e.g.,][]{WILSON09,WARREN10}. Restricted by the
availability of complete molecular gas maps, studies of large sets
of galaxies
\citep[e.g.,][]{YOUNG96,KENNICUTT98A,ROWND99,MURGIA02,LEROY05,SAINTONGE11B} mostly
use integrated measurements or a few low-resolution pointings per
galaxy.

From 2007-2010, the HERA CO-Line Extragalactic Survey \citep[HERACLES,
  first maps in][]{LEROY09} used the IRAM 30-m
telescope\footnote{IRAM is supported by CNRS/INSU (France), the MPG
  (Germany) and the IGN (Spain).} to construct maps of CO emission
from 48 nearby galaxies. The common spatial resolution of the 
survey is $\sim 1$~kpc, sufficient to place many resolution elements across a
typical disk galaxy. Because the targets overlap surveys by {\em Spitzer}
\citep[mostly SINGS and LVL,][]{KENNICUTT03,DALE09} and GALEX
\cite[mostly the NGS,][]{GILDEPAZ07}, a wide variety of
multiwavelength data are available for most targets. In this paper, we
take advantage of these data to compare tracers of molecular gas and
recent star formation at $1$~kpc resolution across a large sample of
30 galaxies.

This paper builds on work by \citet[][hereafter L08]{LEROY08} and
\citet[][hereafter B08]{BIGIEL08}. They combined the first HERACLES
maps with data from The \hi\ Nearby Galaxies Survey \citep[THINGS][]{WALTER08},
SINGS, and the GALEX NGS data to compare \hi , CO, and tracers of recent
star formation in a sample of nearby galaxies. In the disks of large
spiral galaxies, they found little or no dependence of the star
formation rate per unit molecular gas on environment. The fraction of
interstellar gas in the molecular phase, on the other hand, varies
strongly within and among galaxies, exhibiting correlations with
interstellar pressure, stellar surface density, and total gas surface
density among other quantities \citep[][L08]{WONG02,BLITZ06}. They
advocated a scenario for star formation in disk galaxies in which star
formation in isolated giant molecular clouds is a fairly universal
process while the formation of these clouds out of the atomic gas reservoir depends sensitively on
environment \citep[see also,][]{WONG09,OSTRIKER10}.

The full HERACLES sample spans a wider range of masses,
morphologies, metallicities, and star formation rates (SFRs) than the spirals studied
in L08 and B08. \citet{SCHRUBA11} and \citet{BIGIEL11} used these
to extend the findings of L08 and B08. Using stacking techniques, \citet{SCHRUBA11} demonstrated
that correlations between star formation tracers and CO emission
extend into the regime where atomic gas dominates the ISM,
$\Sigma_{\rm HI} > \Sigma_{\rm mol}$. This provides the strongest
evidence yet that star formation in disk galaxies can be separated
into star formation from molecular gas and the balance between atomic
and molecular gas, a hypothesis that has a long history
\citep[e.g.,][and references therein]{YOUNG91}. \citet{BIGIEL11}
demonstrated that a fixed ratio of CO to recent star formation rate remains
a reasonable description of the ensemble of 30 galaxies. 

In this paper we expand on L08, B08, \citet{BIGIEL11}, and \citet{SCHRUBA11} and examine the
general relationship between molecular gas and SFR in nearby disk galaxies. We divide the analysis into two main
parts. In Section \ref{sec:scaling}, we consider the scaling relation linking $\Sigma_{\rm SFR}$
to $\Sigma_{\rm mol}$, the molecular analog to the ``Kennicutt-Schmidt law'' or ``star formation law.''
We show the distribution of data in $\Sigma_{\rm SFR}$-$\Sigma_{\rm mol}$ parameter space
using different weightings (Section \ref{sec:combined}) and examine how this distribution changes with
different approaches to physical parameter --- varying the choice of SFR tracer, the 
CO transition studied, the processing of SFR maps, or the adopted conversion factor (Section \ref{sec:tracers}). 
Using an expanded version of the Monte Carlo modeling approach of \citet{BLANC09}, we carry
out power law fits to our data, avoiding some of the systematic biases present in previous work (Section \ref{sec:index}). 
Finally, we compare our results to a wide collection of literature data, demonstrating an emerging consensus
with regard to the region of parameter space occupied by $\Sigma_{\rm SFR}$-$\Sigma_{\rm mol}$ data,
if not the interpretation (Section \ref{sec:lit}). We conclude that in the disks of normal, massive star-forming galaxies,
to first order the relationship between $\Sigma_{\rm SFR}$ and $\Sigma_{\rm mol}$ can be described by
a single depletion time with a factor of two scatter. This expands and reinforces the results of L08, B08, and \citet{SCHRUBA11}.

In Section \ref{sec:tdep_vary}, we show important second-order deviations from this simple picture, which can
easily be missed by comparing only $\Sigma_{\rm SFR}$ and $\Sigma_{\rm mol}$. We find systematic variations 
in the molecular gas depletion time, $\tau_{\rm dep}^{\rm mol}$ as a function of global galaxy properties (Section \ref{sec:global}) and 
local conditions (Section \ref{sec:local}). This analysis includes the first resolved comparison of $\tau_{\rm dep}^{\rm mol}$ 
to the local dust-to-gas ratio, which is expected to play a key role in setting the CO-to-H$_2$ conversion factor, $\alpha_{\rm CO}$ and we discuss
the possibility that $\alpha_{\rm CO}$ drives some of the observed $\tau_{\rm dep}^{\rm H2}$ variations. We explicitly consider the central
regions of our targets (Section \ref{sec:centers}) and show strong evidence for lower $\tau_{\rm dep}^{\rm mol}$, i.e., more efficient star
formation, in galaxy centers compared to galaxy disks - a phenomenon that we discuss in context of recent proposals
for ``disk'' and ``starburst'' modes of star formation \citep{DADDI10,GENZEL10}. In our best-resolved targets
we examine how the scatter in $\tau_{\rm dep}^{\rm mol}$ depends on spatial scale (Section \ref{sec:scatter}). The relationship appears
much shallower than one would expect for uncorrelated averaging in a disk. This suggests
either a high degree of large-scale synchronization in the star formation process or, more likely, 
widespread systematic, but subtle, variations in $\tau_{\rm dep}^{\rm mol}$ due to still-undiagnosed drivers.

Thus our conclusions may be abstracted to: {\em molecular gas and star formation exhibit a first order one-to-one scaling but
we observe important second order variations about this scaling}. These include likely conversion factor effects, efficient nuclear
starbursts, and weak systematic variations in $\tau_{\rm dep}^{\rm mol}$ that emerge considering scale-dependent scatter or global 
galaxy properties. The remainder of this section presents a brief background, Section \ref{sec:data} describes our data and physical parameter estimation, 
Sections \ref{sec:scaling} and \ref{sec:tdep_vary} motivate these conclusions, and Section \ref{sec:summary} synthesizes these 
results and identifies several key future directions.

\subsection{Background}
\label{sec:background}

Following \citet{SCHMIDT59,SCHMIDT63}, astronomers have studied the
relationship between gas and the SFR for more than 50 years.  Most
recent work follows \citet{KENNICUTT89,KENNICUTT98A} and compares the
surface densities of SFR, $\Sigma_{\rm SFR}$, and neutral (\hi +H$_2$)
gas mass, $\Sigma_{\rm gas}$. Recent work focuses heavily on the power
law relationship between these surface densities (the
``Schmidt-Kennicutt law'' or the ``star formation law''):
\begin{equation}
\sigsfr=A\times\siggas^N~.
\end{equation}

An alternative approach treats the ratio of gas and SFR as the
quantity of interest \citep[][L08]{YOUNG86,YOUNG96}. This ratio can be
phrased as a gas depletion time ($\Sigma_{\rm mol}/\Sigma_{\rm SFR}$) or its inverse,
a star formation efficiency ($\Sigma_{\rm SFR}/\Sigma_{\rm mol}$). These share the same physical
meaning, which is the SFR per unit gas. Both convolve a timescale with
a true efficiency, for example the lifetime of a molecular cloud with
the fraction of gas converted to stars over this time. We focus exclusively on {\em molecular} gas in 
this paper and phrase this ratio as the molecular gas depletion time,

\begin{equation}
\tau_{\rm dep}^{\rm mol} = \frac{\Sigma_{\rm mol}}{\Sigma_{\rm SFR}}~,
\end{equation}

\noindent which is the time for star formation to consume the current
molecular gas supply.

The state of the field is roughly the following. \citet{KENNICUTT98A}
demonstrated a tight, non-linear ($N = 1.4 \pm 0.15$) scaling between
galaxy-averaged $\Sigma_{\rm SFR}$ and $\Sigma_{\rm gas}$ spanning
from normal disk galaxies to starbursts. Including \hi\ improved the
agreement between disks and starbursts, but most of the dynamic range
and the nonlinear slope was driven by the contrast between the disks and merger-induced starbursts, 
especially the local ultraluminous infrared galaxy (ULIRG) population. The adoption 
of a single conversion factor for all systems also had 
significant impact; adopting the ``ULIRG" conversion factor suggested by \citet{DOWNES98} for the 
\citet{KENNICUTT98A} starburst data drives the implied 
slope to $N \sim 1.7$. Subsequent
studies resolved galaxy disks --- often as radial profiles --- and usually
revealed distinct relationships between $\Sigma_{\rm SFR}$, $\Sigma_{\rm HI}$, and
$\Sigma_{\rm mol}$ with shallower indices $N$ for H$_2$ than \hi\ \citep[][B08,
  L08]{WONG02,HEYER04,KENNICUTT07,SCHRUBA11}. This suggests that the immediate
link is between SFR and H$_2$. A more aggressive conclusion, motivated
by the steep, relatively weak relation between $\Sigma_{\rm SFR}$
and $\Sigma_{\rm HI}$ is that star formation in galaxy disks may be
broken into two parts: (1) the formation of stars in molecular clouds and (2)
the balance between H$_2$ and \hi\ \citep[][L08, B08]{WONG02,BLITZ06}.

The fraction of dense molecular gas also appears to be a key
parameter. \citet{GAO04} found a roughly fixed ratio of SFR to HCN
emission, a dense gas tracer, extending from spiral galaxies to
starbursts. Over the same range, the ratio of SFR to total H$_2$,
traced by CO emission, varies significantly (though the two relate roughly linearly in their normal 
galaxy sample). Galactic studies 
also
highlight the impact of density on the SFR on cloud scales
\citep{WU05,HEIDERMAN10,LADA10}. It remains unclear how the dense gas
fraction varies inside galaxy disks, but merger-induced starbursts do
show high HCN-to-CO ratios \citep{GAO04,GARCIABURILLO12}.

SFR tracers and CO can both be observed at high
redshift. \citet{GENZEL10} demonstrated broad
consistency between local H$_2$-SFR relations and those at $z \sim
1-3$. One key difference at high-$z$ is the existence of disk galaxies
with $\tau_{\rm dep}^{\rm mol}$ only slightly lower than those in
local disk galaxies but H$_2$ surface density, $\Sigma_{\rm mol}$, as high as that found in
starbursts in the local universe
\citep{DADDI10,TACCONI10}. Merger-driven starbursts with similar
$\Sigma_{\rm mol}$ can have much lower $\tau_{\rm dep}^{\rm mol}$, suggesting the
relevance of another parameter to set $\tau_{\rm dep}^{\rm
  mol}$. Density and the dynamical timescale are both good candidates \citep[][]
{DADDI10,GENZEL10}.

Meanwhile, investigations of the Milky Way and the nearest galaxies
have attempted to connect observed scaling relations to the properties
of individual star-forming regions. These are able to recover the
galaxy-scale relations at large scale but find enormous scatter in the
ratios of SFR tracers to molecular gas on small scales
\citep{SCHRUBA10,CHEN10,ONODERA10}. Detailed studies of Milky Way and
LMC clouds suggest the time-evolution of individual star-forming
regions as a likely source of this scatter
\citep{MURRAY10,KAWAMURA09}, with the volume density of individual clouds a
key parameter \citep{HEIDERMAN10,LADA10}.

For additional background we refer the reader to the recent review by
\citet{KENNICUTT12}.

\section{Data}
\label{sec:data}

\begin{deluxetable}{lccccccc}
\tablecaption{Sample} 

\tablehead{ 
\colhead{Galaxy} & 
\colhead{$D$} & 
\colhead{res.} &
\colhead{$i$} & 
\colhead{$PA$} & 
\colhead{$r_{25}$} &
\colhead{$r_{25}$} & 
\colhead{Multi-} 
\\
\colhead{} & 
\colhead{[Mpc]} & 
\colhead{[kpc]} & 
\colhead{[$\arcdeg$]} & 
\colhead{[$\arcdeg$]} & 
\colhead{[$\arcmin$]} & 
\colhead{[kpc]} &
\colhead{scale}
\\
\colhead{(1)} & 
\colhead{(2)} & 
\colhead{(3)} & 
\colhead{(4)} & 
\colhead{(5)} & 
\colhead{(6)} & 
\colhead{(7)} &
\colhead{(8)}
}

\startdata
  NGC\,0337 & 19.3\tablenotemark{K} & 1.24\tablenotemark{a} & 51 & 90 & 1.5 & 10.6 & \nodata \\
  NGC\,0628 & 7.2\tablenotemark{K} & 0.46 & 7 & 20 & 4.9 & 10.4 & $\checkmark$ \\
  NGC\,0925 & 9.1\tablenotemark{K} & 0.59 & 66 & 287 & 5.4 & 14.3 & \nodata \\
  NGC\,2403 & 3.2\tablenotemark{W} & 0.21 & 63 & 124 & 7.9 & 7.4 & $\checkmark$ \\
  NGC\,2841 & 14.1\tablenotemark{K} & 0.91 & 74 & 153 & 3.5 & 14.2 & \nodata \\
  NGC\,2903 & 8.9\tablenotemark{W} & 0.57 & 65 & 204 & 5.9 & 15.2 & $\checkmark$ \\
  NGC\,2976 & 3.6\tablenotemark{K} & 0.23 & 65 & 335 & 3.6 & 3.8 & \nodata \\
  NGC\,3049 & 19.2\tablenotemark{K} & 1.24\tablenotemark{a} & 58 & 28 & 1.0 & 2.7 & \nodata \\
  NGC\,3184 & 11.8\tablenotemark{K} & 0.76 & 16 & 179 & 3.7 & \nodata \\
  NGC\,3198 & 14.1\tablenotemark{K} & 0.91 & 72 & 215 & 3.2 & 13.0 & \nodata \\
  NGC\,3351 & 9.3\tablenotemark{K} & 0.60 & 41 & 192 & 3.6 & 10.6 & \nodata \\
  NGC\,3521 & 11.2\tablenotemark{K} & 0.72 & 73 & 340 & 4.2 & 12.9 & \nodata \\
  NGC\,3627 & 9.4\tablenotemark{K} & 0.61 & 62 & 173 & 5.1 & 13.8 & $\checkmark$ \\
  NGC\,3938 & 17.9\tablenotemark{K} & 1.15\tablenotemark{a} & 14 & 15 & 1.8 & 6.3 & \nodata \\
  NGC\,4214 & 2.9\tablenotemark{W} & 0.19 & 44 & 65 & 3.4 & 2.9 & \nodata \\
  NGC\,4254 & 14.4\tablenotemark{K} & 0.93 & 32 & 55 & 2.5 & 14.6 & \nodata \\
  NGC\,4321 & 14.3\tablenotemark{K} & 0.92 & 30 & 153 & 3.0 & 12.5 & \nodata \\
  NGC\,4536 & 14.5\tablenotemark{K} & 0.94 & 59 & 299 & 3.5 & 14.9 & \nodata \\
  NGC\,4559 & 7.0\tablenotemark{K} & 0.45 & 65 & 328 & 5.2 & 10.7 & \nodata \\
  NGC\,4569 & 9.86\tablenotemark{K} & 0.64 & 66 & 23 & 4.6 & 26.5 & \nodata \\
  NGC\,4579 & 16.4\tablenotemark{K} & 1.06\tablenotemark{a} & 39 & 100 &  2.5 & 15.0 & \nodata \\
  NGC\,4625 & 9.3\tablenotemark{K} & 0.60 & 47 & 330 & 0.7 & 1.9 & \nodata \\
  NGC\,4725 & 11.9\tablenotemark{K} & 0.77 & 54 & 36 & 4.9 & 13.2 & $\checkmark$ \\
  NGC\,4736 & 4.7\tablenotemark{K} & 0.30 & 41 & 296 & 3.9 & 5.3 & $\checkmark$ \\
  NGC\,5055 & 7.9\tablenotemark{K} & 0.51 & 59 & 102 & 5.9 & 17.3 & \nodata \\
  NGC\,5194 & 7.9\tablenotemark{W} & 0.52 & 20 & 172 & 3.9 & 9.0 & $\checkmark$ \\
  NGC\,5457 & 6.7\tablenotemark{K} & 0.43 & 18 & 39 & 12.0 & 25.8 & $\checkmark$ \\
  NGC\,5713 & 21.4\tablenotemark{K} & 1.38\tablenotemark{a} & 48 & 11 & 1.2 & 9.5 & \nodata \\
  NGC\,6946 & 6.8\tablenotemark{K} & 0.44 & 33 & 243 & 5.7 & 9.8 & $\checkmark$ \\
  NGC\,7331 & 14.5\tablenotemark{K} & 0.94 & 76 & 168 & 4.6 & 19.5 & \nodata 
\enddata
\label{tab:sample}
\tablecomments{Sample used in this paper. Columns give (1) galaxy name; (2) adopted distance in Mpc; (3) FWHM spatial resolution of HERACLES data at that distance, in kiloparsecs; (4) adopted inclination and (5) position angle in degrees; adopted radius of the the B-band 25$^{\rm th}$ magnitude isophote, used to normalize the radius in (6) arcminutes and (7) kiloparsecs. Most analysis in this paper considers data inside $0.75~r_{25}$. Column (8) indicates if the galaxy is close and large enough for the multiscale analysis in Section \ref{sec:scatter}.}
\tablenotetext{a}{Too distant to convolve to 1~kpc resolution. Included in analysis at native 
resolution.}
\tablenotetext{K,W}{Distance adopted from K: \citet{KENNICUTT11} or W: \citet{WALTER08}.}
\end{deluxetable}

\subsection{Data Sets}

We use HERACLES CO(2-1) maps to infer the distribution of
H$_2$ and GALEX far-ultraviolet (FUV), {\em Spitzer} infrared (IR),
and literature H$\alpha$ data to trace recent star formation. We supplement these with \hi\ data used 
to mask the CO, derive 
kinematics, and measure the dust-
to-gas ratio and with near-IR data used to estimate the stellar surface density, $\Sigma_*$.

{\em HERACLES CO:} The HERA CO Line Extragalactic Survey (HERACLES)
used the Heterodyne Receiver Array \citep[HERA,][]{SCHUSTER04} on the
IRAM 30m telescope to map CO(2-1) emission from $48$
nearby galaxies, of which we use 30 in this paper (see Section \ref{sec:sample}). HERACLES 
combines an IRAM Large Program 
and several
single-semester projects that spanned from 2007 to
2010. \cite{LEROY09} presented the first maps \citep[see
  also][]{SCHUSTER07}. The additional data here were observed and
reduced in a similar manner. The largest change is a revised estimate of the main beam
efficiency, lowering observed intensities by $\approx 10\%$. This
propagates to a revised CO (2-1)/(1-0) line
ratio estimate, so our estimates of $\Sigma_{\rm mol}$ are largely
unaffected compared to B08 and L08. The HERACLES cubes cover out to radii of
$r_{25}$ with angular resolution $13\arcsec$ and typical $1\sigma$
sensitivity $20$~mK per 5~km~s$^{-1}$ channel.

We integrate each cube along the velocity axis to produce maps of the
integrated intensity along each line of sight. We wish to avoid
including unnecessary noise in this integral and so restrict the
velocity range over which we integrate to be as small as possible
while still containing the CO line, i.e., we ``mask'' the cubes. To be
included in the mask a pixel must meet one of two conditions: 1) lie
within $\pm 25$~km~s$^{-1}$ of the local mean \hi\ velocity
\citep[derived from THINGS,][supplemented by new and archival \hi
]{WALTER08} or 2) lie in part of the spectrum near either two
consecutive channels with SNR above $4$ or three consecutive channels
with SNR above $3$. Condition (2) corresponds to traditional radio
masking \citep[e.g.,][]{HELFER03,WALTER08}. Condition (1) is less
conventional, but important to our analysis. Integrating over the
\hi\ line, which is detected throughout our targets, guarantees that
we have an integrated intensity measurement along each line of sight,
even lines of sight that lack bright CO emission \citep[see][for detailed
  discussion of this approach]{SCHRUBA11}. This avoids a traditional
weakness of masking, that nondetections are difficult to deal with
quantitatively. We calculate maps of the statistical uncertainty in
the integrated CO intensity from the combination of the mask and
estimates of the noise derived from signal-free regions. The result is
an integrated intensity and associated uncertainty for each line of
sight in the HERACLES mask.

{\em SINGS and LVL IR:} We use maps of IR emission from 3.6--160$\mu$m
from the {\em Spitzer} Infrared Nearby Galaxies Survey
\citep[SINGS,][]{KENNICUTT03} and the Local Volume Legacy survey
\citep[LVL,][]{DALE09}. We describe the processing of these maps in \citet[][hereafter L12]
{LEROY12}.

{\em SINGS, LVL, and Literature H$\alpha$:} Both SINGS and LVL
published continuum-subtracted H$\alpha$ images for most of our
sample. We supplement these with literature maps, particularly from the GoldMine and Palomar-Las 
Campanas surveys. L12 
describe our approach to these 
maps (masking, \ion{N}{2} correction, flux scaling, background subtraction) and list the source of the 
H$\alpha$ data for each 
galaxy.

{\em GALEX UV:} For 24 galaxies, we use NUV and FUV maps from the
Nearby Galaxy Survey \citep[NGS,][]{GILDEPAZ07}. For one galaxy, we use
a map from the Medium Imaging Survey (MIS) and we take maps for five
targets from the All-sky Imaging Survey (AIS). L12 describe our processing.

{\em BIMA+12-m and NRO 45-m CO (1-0) Maps:} A subset of
our targets have also been observed by the BIMA SONG \citep{HELFER03}
or the Nobeyama CO Atlas of Nearby Spiral Galaxies
\citep{KUNO07}. Where these data are available, we apply our HERACLES
masks to these maps and measure CO (1-0) intensity. We
only use BIMA SONG maps that include short-spacing data from the Kitt
Peak 12-m.

{\em THINGS and Supplemental {\sc Hi}:} We assemble \hi\ maps for all
targets, which we use to mask the CO, estimate the dust-to-gas ratio,
explore 24$\mu$m cirrus corrections, and derive approximate rotation curves. These come from 
THINGS
\citep{WALTER08} and a collection of new and archival VLA data
(programs AL731 and AL735). These supplemental \hi\ are C+D
configuration maps with resolutions $13\arcsec$--$25\arcsec$. We
reduced and imaged these in a standard way using the CASA package.

\subsection{Physical Parameter Estimates}
\label{sec:param}

Following standard practice in this field, we estimate physical parameters from observables. Despite the intrinsic
uncertainty involved in this process, these estimates play a fundamental role in enhancing our 
understanding of the physics of galaxy and star formation, as
demonstrated from the earliest works in this subfield \citep{YOUNG89,KENNICUTT89}. We adopt an approach largely oriented to physical quantities, but 
discuss the impact of our assumptions throughout.

{\em CO Intensity to H$_2$:} We convert CO (2-1) intensity to \htwo\ mass via

\begin{equation}
\label{eq:xco}
\Sigma_{\rm mol} [{\rm
    M_{\sun}\,pc^{-2}}]=6.3~\left(\frac{0.7}{R_{21}}\right)~\left(\frac{\alpha_{\rm
    CO}^{1-0}}{4.35}\right)~I_{CO}~[{\rm K\,km\,s^{-1}}],
\end{equation}

\noindent where $R_{21}$ is the CO(2-1)-to-CO(1-0) line ratio and $\alpha_{\rm CO}$ is the
CO(1-0)-to-\htwo\ conversion factor. By default, we adopt a Galactic conversion factor,
$\alpha_{\rm CO}^{1-0} = 4.35$~M$_\odot$~pc$^{-2}$~$({\rm K\,km\,s}^{-1})^{-1}$ equivalent to $X_{\rm 
CO} =2\times10^{20}\,{\rm cm^{-2}\,(K\,km\,s^{-1})^{-1}}$ \citep[][]{STRONG96,DAME01} and a line ratio of $R_{21}=0.7$. 
This line ratio is slightly lower than the $R_{21} = 0.8$ derived by \citet{LEROY09}, reflecting the revised efficiency used in the reduction. 
The appendix motivates this value using integrated flux ratios and follow-up spectroscopy of HERACLES targets. 
Equation \ref{eq:xco} and all ``$\Sigma_{\rm mol}$'' in this paper include a factor of 1.36 to
account for helium. Because we consider only molecular gas, any results that we derive using a fixed 
$\alpha_{\rm CO}$ can be straightforwardly restated in terms of CO intensity.

We adopt this ``Galactic'' $\alpha_{\rm CO}^{1-0}$ to facilitate clean comparison to  previous work, but improved
estimates exist for HERACLES. \citet{SANDSTROM12} solved directly for the CO-to-H$_2$ conversion factor
across the HERACLES sample using dust as an independent tracer of the gas mass. They find
a somewhat lower average $\alpha_{\rm CO}^{1-0} \approx 3.1$. We quote this as a CO (1-0) conversion factor,
though \citet{SANDSTROM12} directly solve for the CO (2-1) conversion factor. They find a CO(2-1)
conversion factor of $\alpha_{\rm CO}^{2-1} \approx 4.4$~\acounits , compared to our ``Galactic'' CO(2-1) conversion factor 
$\alpha_{\rm CO}^{2-1} = 6.3$~\acounits . \citet{SANDSTROM12} find $\approx 0.4$~dex point-to-point scatter, of 
which $\approx 0.3$ may be intrinsic with the remainder solution uncertainties. Because
\citet{SANDSTROM12} solve directly for a CO (2-1) conversion factor using the same HERACLES data
employed in this paper, these values should be borne in mind when reading our results. Our results
remain pinned to a Galactic CO(1-0) conversion factor of $\alpha_{\rm CO}^{1-0} = 4.35$~\acounits\ that 
may be $\approx 30\%$ too high, on average. As a result, a systematic bias of $\approx 30\%$ in $\Sigma_{\rm mol}$ appears plausible with
factor of two variations in the conversion factor point-to-point.

In addition to a fixed conversion $\alpha_{\rm CO}$, we consider the effects of variations
in $\alpha_{\rm CO}$ due to decreased dust shielding at low metallicity and variations in
the linewidth, optical depth, and temperature of CO in galaxy centers. Our ``variable'' $\alpha_{\rm CO}$
builds on the work of \citet{SANDSTROM12}, who compare \hi , CO(2-1), and $\Sigma_{\rm 
Dust}$ in 22 HERACLES galaxies and \citet{WOLFIRE10}, who consider the effects of dust shielding on the
``CO-dark'' layer of molecular clouds, where most H is \htwo . The $\alpha_{\rm CO}$ prescription 
combines three terms

\begin{equation}
\label{eq:alphaco}
\alpha_{\rm CO} = \alpha_{\rm CO}^{0}~c_{\rm CO-dark} \left( D/G \right)~c_{\rm center} \left( r_{\rm 
gal} \right)~.
\end{equation}

\noindent Here $\alpha_{\rm CO}^{0} = 6.3$~\acounits\ is our fiducial CO (2-1) conversion factor in the disk
of a galaxy at solar metallicity (Equation \ref{eq:xco}). The term $c_{\rm CO-dark}$ represents a correction to the H$_2$ mass to reflect 
the H$_2$ in a CO-dark layer not directly traced by CO emission. We calculate this factor following \citet
{WOLFIRE10}, assuming that all GMCs share a fixed surface density, $\left< \Sigma_{\rm GMC} \right>$
and adopting a linear scaling between the dust-to-gas ratio and metallicity \citep[see also][]{GLOVER11}. In this case

\begin{equation}
\label{eq:ccodark}
c_{\rm CO-dark} \left( D/G^\prime \right) \approx 0.65 \exp \left( \frac{0.4}{D/G^\prime~\Sigma_{\rm 100}} \right)~.
\end{equation}

\noindent Here $D/G^\prime$ is normalized to our adopted ``Galactic'' value of 
0.01, with the normalization constructed to yield $\alpha_{\rm CO} = \alpha_{\rm CO}^{0}$ for $D/G^\prime = 1$.
$\Sigma_{\rm 100} = \left< \Sigma_{\rm GMC} \right> / 100$~M$_\odot$~pc$^{-2}$. The appendix
presents this calculation in detail. 

We consider two cases:  $\left< \Sigma_{\rm GMC}\right>=100$~M$_\odot$~pc$^{-2}$ (``$\Sigma=100$'') and 
$\left< \Sigma_{\rm GMC}\right>=50$~M$_\odot$~pc$^{-2}$ (``$\Sigma=50$''). ``$\Sigma=100$'' reflects a typical surface 
density that is often assumed and observed for extragalactic GMCs \citep[e.g.,][]{BOLATTO08,NARAYANAN12}. Over the 
range of $D/G^\prime$ that we consider, this prescription reasonably resembles the shallow power law dependences of 
$\alpha_{\rm CO}$ on metallicity calculated from simulations by \citet[][]{FELDMANN12} and \citet{NARAYANAN12}, 
who both suggest $\alpha_{\rm CO} \sim Z^{-0.7}$.  ``$\left< \Sigma_{\rm GMC}\right>=50$~M$_\odot$~pc$^{-2}$'' yields a 
steeper dependence of $\alpha_{\rm CO}$ on metallicity over our range of interest but we will see in
Sections \ref{sec:global} and \ref{sec:local} that it offers a simple way to account for most of the dependence of
$\tau_{\rm dep}^{\rm H2}$ on dust-to-gas ratios in our observations.
Low surface density GMCs or significant contribution of ``translucent'' ($A_V \sim 1$--$2$~mag) gas to the
overall CO emission are supported by observations of the Milky Way \citet{HEYER09} and \citet{LISZT10}, LMC
observations by \citet{HUGHES10,WONG11}, and M31 (Schruba et al. in prep.) but may not be appropriate for
more actively star-forming systems (Hughes et al., submitted). We return to this issue in Section \ref{sec:tdep_vary}.

We calculate $\alpha_{\rm CO} (D/G^\prime)$ using $D/G^\prime$ derived for fixed $\alpha_{\rm CO}$. Given observations
of $\Sigma_{\rm HI}$, $I_{\rm CO}$, $\Sigma_{\rm dust}$, and a prescription for $\alpha_{\rm CO} (D/G^\prime)$,  one can simultaneously 
solve for $\alpha_{\rm CO}$ and $D/G^\prime$. The solution is often multivalued and unstable, though not intractable. However, after 
experimentation and comparison with the self-consistent {\em Herschel}-based results of \citet{SANDSTROM12}, we found
that the process does not clearly improve our estimates. In the interests of clarity and simplicity, we work with {\em only} $D/G$ 
calculated using fixed $\alpha_{\rm CO}$ throughout the paper. This simplification biases our $\alpha_{\rm CO}$ estimate 
high by $\approx 8\%$ (``$\Sigma=100$'') and $\approx 15\%$ (``$\Sigma=50$'').
With improved $\Sigma_{\rm dust}$ estimates, we expect that the self-consistent treatment will become necessary.

The third term, $c_{\rm center}$, accounts for depressed values of $\alpha_{\rm CO}$ in the centers of galaxies. \citet{SANDSTROM12} find 
such depressions in the centers of many systems  \citep[see also][]{ISRAEL09B,ISRAEL09A}. 
These likely reflect the same line-broadening and temperature effects that drive the commonly invoked ``ULIRG conversion factor'' \citep{DOWNES98}, 
though the depression observed by \citet{SANDSTROM12} have lower magnitude than the factor of 5 depression found
by \citet{DOWNES98}. \citet{SANDSTROM12} could not identify a unique observational driver for these depressions, though they correlate well 
with stellar surface density. Instead, they appear to be present with varying magnitudes in the centers of most 
systems with bright central CO emission. Following their recommendation, we apply this correction 
where $r_{\rm gal} < 0.1 r_{\rm 25}$ in systems that have such central CO concentrations. Whenever available, 
we adopt $c_{\rm center}$ directly from \citet{SANDSTROM12}, taking the factor by which the central 
$\alpha_{\rm CO}$ falls below the mean for the disk of that galaxy. For systems with central CO concentrations but not
in the sample of \citet{SANDSTROM12}, we apply a factor of 
two depression, again following their recommendations. The appendix presents additional details.

{\em SFR from H$\alpha$, UV, and 24$\mu$m Emission:} L12 combined UV, H$\alpha$, and IR 
emission to estimate the recent star formation rate surface density, $\Sigma_{\rm SFR}$, at
1~kpc resolution (the limiting common physical resolution of the HERACLES survey) for our sample. 
We adopt their estimates and refer the reader to that work for detailed discussion. Briefly, our 
baseline estimate of $\Sigma_{\rm SFR}$ combines H$\alpha$ and infrared emission at 24$\mu$m via

\begin{eqnarray}
\label{eq:sfr}
\Sigma_{\rm SFR}~\left[ {\rm M}_\odot~{\rm yr}^{-1}~{\rm kpc}^{-2} \right] &=& 634~I_{\rm H\alpha}~\left[ {\rm erg~s~sr}^{-1} \right] + \\
&&0.00325~I_{24~\mu{\rm m}}~\left[ {\rm MJy~sr}^{-1} \right]
\end{eqnarray}

\noindent where $I_{\rm H\alpha}$ and $I_{\rm 24\mu m}$ refer to the line-integrated H$\alpha$ intensity and intensity at $24\mu$m.

The H$\alpha$ emission  captures direct emission from \hii\ regions powered by massive young stars while the 24$\mu$m 
emission accounts for recent star formation obscured by dust. Before estimating $\Sigma_{\rm SFR}$, we 
correct our 24$\mu$m maps for the effects of heating of dust by a weak, pervasive 
radiation field (i.e., a ``cirrus'') with magnitude derived from modeling the infrared spectral energy distribution. The cirrus removed 
corresponds to the expected emission from the local dust mass illuminated by a quiescent radiation field, 
typically $\sim 0.6$ times the Solar neighborhood interstellar radiation field (see L12 for details). We derive the 
appropriate weighting for the combination of H$\alpha$ and 24$\mu$m emission based on 
comparing our processed H$\alpha$ and 24$\mu$m maps to literature estimates of H$\alpha$ extinction. The 
resulting linear combination resembles that of \citet{KENNICUTT07} but places slightly more weight on the 24$
\mu$m term. For comparison, we also estimate $\Sigma_{\rm SFR}$ from combining FUV and 24$
\mu$m emission and taking H$\alpha$ alone while assuming a typical 1~magnitude of extinction.

L12 estimate a substantial uncertainty in the absolute calibration of ``hybrid'' UV+IR or H$\alpha$+IR 
tracers, with magnitude $\approx 50\%$. In addition to this overall uncertainty in the calibration, they derive 
a point-to-point uncertainty in $\Sigma_{\rm SFR}$ of $\approx 0.15$~dex based on intercomparison of different estimates.

{\em Dust Properties:} In order to measure dust properties, we convolve the {\em Spitzer} 24, 70, and
160$\mu$m data and the CO and \hi\ maps to the resolution of the {\em Spitzer} 160$\mu$m data. 
At this resolution, we build radial profiles of each band and then fit the dust models of \citet{DRAINE07A} to these profiles. 
These fits, presented in L12, provide us with radial estimates of the dust-to-gas
ratio, $D/G$, and are used to help account for ``cirrus'' contamination when estimating $\Sigma_{\rm SFR}$. Note that the $\sim 40\arcsec$ resolution of the 
160$\mu$m data used to measure these dust properties is significantly coarser than the 1~kpc resolution used for the rest of our data. Where possible, we have 
compared our {\em Spitzer}-based dust masses to masses estimated using the improved SED coverage offered by {\em Herschel} \citep[e.g.,][]{ANIANO12}; 
above $\Sigma_{\rm dust} \approx 0.05$~M$_\odot$~pc$^{-2}$ the median offset between the {\em Herschel} and {\em Spitzer} based dust masses is only $
\approx10\%$; however the dust masses derived for individual rings using only {\em Spitzer} do scatter by $\approx 0.3$~dex (a factor of two) compared to {\em 
Herschel}-based dust masses and show weak systematic trends with the sense that {\em Spitzer} underestimates the mass of cooler ($\sim 15$~K) dust in the 
outskirts of galaxies \citep[both consistent with the analysis of][]{DRAINE07B}. We expect that once {\em Herschel} images become available, they will significantly 
improve the accuracy of dust-based portion of this analysis.

{\em Stellar Mass:} To estimate the stellar mass for whole galaxies, we draw 3.6$\mu$m fluxes from 
\citet{DALE07,DALE09}, convert to a luminosity using our adopted distance, and apply a fixed $3.6\mu$m 
mass-to-light ratio. Based on comparison to \citet{ZIBETTI09}, we use 

\begin{equation}
\label{eq:sigstar}
\Sigma_* \left[ {\rm M}_\odot~{\rm pc}^{-2} \right] = 200~I_{\rm 3.6}~\left[  {\rm MJy~sr}^{-1} \right]~,
\end{equation}

\noindent which is $\sim 30\%$ lower than L08. This value
is uncertain by $\sim 50\%$. 

We estimate the stellar surface density, $\Sigma_{\rm *}$, for each kpc-sized element from the contaminant-corrected 3.6$\mu$m 
maps of \citet{MEIDT12}. Starting from a reprocessing of the SINGS data \citep[as part of the S4G survey][]{SHETH10}, they 
used independent component analysis to remove contamination by young stars and hot dust from the overall maps. These contaminants make a minor 
contribution to the overall 3.6$\mu$m flux but may be important locally. We convert the contaminant-corrected 3.6$\mu$m maps to $\Sigma_*$ estimates 
using Equation \ref{eq:sigstar}.

{\em Rotation Velocities:} Following L08 and \citet{BOISSIER03} we work with a simple two-parameter fit to the rotation curve of 
each galaxy

\begin{equation}
v_{\rm rot} \left(r_{\rm gal} \right) = v_{\rm flat} \left[ 1 - \exp \left( \frac{-r_{\rm gal}}{l_{\rm flat}}\right) 
\right]
\end{equation}

\noindent with $v_{\rm flat}$ and $l_{\rm flat}$ free parameters. We derive these from fits to the 
rotation curves of \citet{DEBLOK08} wherever they are available. Where these are not available, we carry out our own tilted ring fits to the combined \hi\ and CO first 
moment maps. We use these fits to calculate the orbital time $\tau_{\rm orb} =  2 \pi r_{\rm gal} / v_{\rm rot}$ for each line of sight.

\subsection{Sample and Galaxy Properties}
\label{sec:sample}

We present measurements for galaxies meeting the following criteria:
1) a HERACLES CO map containing a clear CO detection (S/N$>5$ over a significant area and 
multiple channels), 2) {\em Spitzer} data at 24$\mu$m, and 3) inclination $\lesssim 75\degr$. The first
condition excludes low mass galaxies without CO detections \citep[these are discussed in][]
{SCHRUBA12}. The second removes a few targets with saturated or incomplete {\em Spitzer}
coverage. The third removes a handful of edge-on galaxies. We are left with the $30$ disk galaxies listed in Table \ref{tab:sample}.

For each target, Table \ref{tab:sample} gives the distance, physical resolution of the HERACLES maps 
at that distance, inclination, position angle, and optical radius. The Table notes note the subset of galaxies that 
that are close and large enough for us to carry out the multi-resolution analysis in Section \ref{sec:scatter}. We adopt 
distances from \citet{KENNICUTT11}  where possible and from \citet{WALTER08} elsewhere. We take orientations 
from \citet{WALTER08} and from LEDA \citep{PRUGNIEL98} and NED elsewhere.

Table \ref{tab:integrated} reports integrated and disk-average properties for our sample. We report our integrated stellar mass 
estimate, galaxy morphology, metallicity and dust-to-gas ratio at $\approx 0.4~r_{25}$, average gas mass and star formation rate
surface density inside $0.75~r_{25}$, our parameterized rotation curve fit, and the orbital time at $0.4~r_{25}$. We take 
metallicities from \citet{MOUSTAKAS10}, averaging their PT05 and KK04 strong-line calibrations. They 
argue that these two calibrations bracket the true metallicity and that the relative ordering of
metallicities is robust \citep[see also][]{KEWLEY08}, but the uncertainty in the absolute value is considerable. For cases where
\citet{MOUSTAKAS10} do not present a metallicity, we draw one from the recent compilations by
\citet{MARBLE10} and \citet{CALZETTI10}.

\begin{deluxetable*}{lccccccccc}
\tablecaption{Sample Properties} 
\tablehead{ 
\colhead{Galaxy} & 
\colhead{log($M_{*}$)} &
\colhead{Morphology} & 
\colhead{$z$} & 
\colhead{$D/G$} & 
\colhead{$\left< \Sigma_{\rm HI+H2} \right>$} &
\colhead{$\left< \Sigma_{\rm SFR} \right>$} &
\colhead{$v_{\rm flat}$} & 
\colhead{$l_{\rm flat}$} &
\colhead{$\left<\tau_{\rm orb}\right>$}
\\
\colhead{} & 
\colhead{log$_{10}$ [M$_{\sun}$]} & 
\colhead{T-Type} & 
\colhead{[12+log[O/H]]} & 
\colhead{} &
\colhead{[M$_{\sun}$~pc$^{-2}$]} & 
\colhead{[$10^{-3}$ $\frac{\mbox{ M$_{\sun}$~yr$^{-1}$}}{\mbox{kpc$^{2}$}}$]} &
\colhead{[km s$^{-1}$]} & 
\colhead{[kpc]} &
\colhead{[$10^8$ yr]} 
\\
\colhead{(1)} & 
\colhead{(2)} & 
\colhead{(3)} & 
\colhead{(4)} & 
\colhead{(5)} &
\colhead{(6)} & 
\colhead{(7)} &
\colhead{(8)} & 
\colhead{(9)} &
\colhead{(10)} 
}

\startdata
  NGC\,0337 & 9.9 & 6.7 & 8.51 & 0.004 & 21 & 14 & 130 & 2.9 & 2.4 \\
  NGC\,0628 & 10.0 & 5.2 & 8.69 & 0.012 & 15 & 4.0 & 200 & 0.8 & 1.3 \\
  NGC\,0925 & 9.7 & 7.0 & 8.52 & 0.004 & 7.5 & 1.3 & 140 & 6.9 & 4.4 \\
  NGC\,2403 & 9.6 & 6.0 & 8.57 & 0.009 & 10 & 3.3 & 120 & 0.95 & 1.6 \\
  NGC\,2841 & 10.7 & 3.0 & 8.88 & 0.037 & 4.6 & 1.4 & 310 & 2.3 & 1.2 \\
  NGC\,2903 & 10.4 & 4.0 & 8.90\tablenotemark{c} & 0.012 & 12 & 5.7 & 210 & 2.4 & 2.0 \\
  NGC\,2976 & 9.0 & 5.2 &  8.67 & 0.008 & 7.6 & 4.4 & 88 & 1.1 & 1.4 \\
  NGC\,3049 & 9.5 & 2.5 &  8.82 & 0.005 & 8.0 & 10 & 180 & 3.0 & 1.5 \\  
  NGC\,3184 & 10.2 & 6.0 & 8.83  & 0.018 & 14 & 2.8 & 200 & 2.5 & 1.8 \\
  NGC\,3198 & 10.0 & 5.2 & 8.62 & 0.012 & 8.4 & 2.3 & 150 & 3.0 &  2.6 \\
  NGC\,3351 & 10.1 & 3.1 & 8.90 & 0.018 & 8.5 & 5.2 & 200 & 1.1 & 1.2 \\
  NGC\,3521 & 10.7 & 4.0 & 8.70 & 0.012 & 22 & 7.8 & 229 & 1.5 & 1.5 \\
  NGC\,3627 & 10.5 & 3.1 & 8.67 & 0.016 & 13 & 7.7 & 190 & 1.1 & 1.8 \\
  NGC\,3938 &10.3 & 5.1 & 8.71\tablenotemark{c} & 0.018 & 22 & 7.9 & 140 & 0.73 & 1.6 \\
  NGC\,4214 & 8.7 & 9.8 & 8.36\tablenotemark{c} & 0.0038 & 9.2 & 8.4 & 350 & 11 & 2.0 \\
  NGC\,4254 & 10.5 & 5.2 &  8.79  & 0.01 & 47 & 18 & 170 & 1.4 & 1.6 \\
  NGC\,4321 & 10.6 & 4.1 & 8.84  & 0.012 & 30 & 9.0 & 229 & 1.8 & 1.4 \\
  NGC\,4536 & 10.2 & 4.3 & 8.61 & 0.005 & 13 & 6.8 & 180 & 0.7 & 2.0 \\
  NGC\,4559 & 9.5 & 6.0 & 8.55 & 0.005 & 10 & 1.8 & 100 & 2.1 & 2.9 \\
  NGC\,4569 & 10.2 & 2.4 & 8.88\tablenotemark{c} & 0.017 & 8.5 & 1.9 & 220 & 3.2 & 1.8 \\
  NGC\,4579 &10.7 & 2.8 & 8.93\tablenotemark{c} & 0.021 & 13 & 3.8 & 270 & 1.7 & 1.2 \\
  NGC\,4625 & 8.9 & 8.8 & 8.70 & 0.011 & 8.5 & 6.6 & 27 & 0.53 & 2.3 \\
  NGC\,4725 & 10.5 & 2.2 & 8.73  & 0.03 & 4.5 & 0.75 & 220 & 1.1 & 1.9 \\
  NGC\,4736 & 10.2 & 2.4 & 8.66 & 0.008 & 17 & 10 & 170 & 0.25 & 0.77 \\
  NGC\,5055 & 10.5 & 4.0 & 8.77 & 0.02 & 18 & 4.1 & 200 & 0.71 & 1.7 \\
  NGC\,5194 & 10.5 & 4.0 & 8.87 & 0.02 & 53 & 20 & 210 & 0.58 & 1.0 \\
  NGC\,5457 & 10.4 & 6.0 & 8.46\tablenotemark{c} & 0.013 & 10 & 2.4 & 210 & 1.2 & 2.7 \\
  NGC\,5713 & 10.3 & 4.0 & 8.64 & 0.006 & 54 & 37 & \nodata & \nodata & \nodata \\
  NGC\,6946 & 10.5 & 5.9 & 8.73 & 0.007 & 37 & 21 & 190 & 1.2 & 1.5 \\
  NGC\,7331 & 10.8 & 3.9 &  8.68 & 0.01 & 16 & 4.4 & 260 & 1.9 & 1.9 \\
\enddata
\label{tab:integrated}
\tablecomments{Properties of sample galaxies. Columns give (1) galaxy name; (2) integrated stellar mass of whole galaxies based on 3.6$\mu$m flux of \citet{DALE07,DALE09}; (3) morphology; (4) ``characteristic'' metallicity at 0.4~$r_{25}$ from \citet{MOUSTAKAS10}, averaging their PT05 and KK04 calibrations; (5) dust-to-gas ratio at 0.4~$r_{25}$ based on our modeling of {\em Spitzer} data; (6) average \hi+H$_2$ surface density inside $0.75~r_{25}$; (7) average star formation rate surface density inside $0.75~r_{25}$; parameters for simple rotation curve fit, (8) $v_{\rm flat}$ and (9) $l_{\rm flat}$; and (10) orbital time at $0.4~r_{25}$ based on the rotation curve.}
\tablenotetext{c}{Metallicity from compilation of \citet{CALZETTI10} and \citet{MARBLE10} or \citet
{KENNICUTT03B} (NGC~5457).}
\end{deluxetable*}

\subsection{Methodology}

We sample our targets at 1~kpc resolution. This is fine enough to 
isolate many key physical conditions in the interstellar medium (ISM): 
metallicity, coarse kinematics, gas and stellar surface density. At the same time, we
expect to average several star forming regions in each element \citep[e.g.,][]{SCHRUBA10}, with
$M_{\rm mol} \gtrsim 10^7$~M$_\odot$ and $M_* \gtrsim 10^4$~M$_\odot$
formed over the last $\sim 5$~Myr in each element. This minimizes
concerns about evolution of individual regions, sampling the IMF, and
drift of stars or leakage of ionizing photons from their parent
region.

We convolve each map to have a symmetric gaussian beam with FWHM
1~kpc. For the {\em Spitzer} 24$\mu$m maps we first convert from the
MIPS PSF to a $13\arcsec$ gaussian beam using a kernel kindly provided
by K. Gordon, then we convolve to $1$~kpc. This exercise effectively
places our targets at a common distance but does not account for
foreshortening along the minor axis. Five galaxies are too distant to
convolve to 1~kpc. We mark these in Table \ref{tab:sample} and include
them in our analysis at their native resolution.

We sample each map to generate a set of intensity measurements. The
sampling points are distributed on a hexagonal grid with points spaced
by 0.5\,kpc, one half-resolution element. At each sampling point we
measure CO(2-1) intensity, {\sc Hi} intensity, a suite of star formation rate tracers (described in L12), 
dust properties, and $\Sigma_*$. We use these to estimate physical conditions as described 
above and in L12, taking into account the inclination of the galaxy.

We also identify a sample of galaxies to study the effects of physical
resolution. Nine galaxies, marked in Table \ref{tab:sample}, have both
the proximity and extent to allow us to test the effect of physical resolution
on our results. We convolve these to a succession of physical resolutions
from 0.6 to 2.4~kpc for further analysis (Section \ref{sec:scatter}).

We treat regions with $\Sigma_{\rm SFR} < 10^{-3}$~M$_{\odot}$~yr$^{-1}$~kpc$^{-2}$ or $I_{\rm CO} <
2.5 \times \sigma_{\rm CO}$ as upper limits and consider only points with
$r_{\rm gal} < 0.75~r_{25}$ --- the HERACLES maps contain signal
outside this radius \citep{SCHRUBA11} but mostly not significant
emission over individual lines-of-sight. In total we have $\sim
14,500$ lines of sight with at least one significant measurement, of
which $1,900$ have CO upper limits and $1,650$ have SF upper
limits. Points for which neither measurement is significant are not
considered in the analysis. Nyquist  sampling the maps in a hexagonal 
pattern leads to oversampling by a factor of $\sim 5$,
so that this corresponds to $> 2,000$ independent measurements. The
maximum ($2.5\sigma$) upper limit on $\Sigma_{\rm mol}$ is $\approx
6$~M$_\odot$~pc$^{-2}$, the median upper limit is $\approx
2.6$~M$_\odot$~pc$^{-2}$.

\subsection{Literature Data}
\label{sec:litdata}

We compare our results to recent measurements of SFR and molecular
gas. These employ a variety of sampling schemes and SFR tracers. We adjust each to match our adopted
CO-to-H$_2$ conversion factor and IMF. Contrasting our approach with
these data illuminates the impact of methodology and allows us to
explore whether diverse observations yield consistent results under matched assumptions.

\citet{KENNICUTT98A} presented disk-averaged measurements for 57
normal spiral galaxies and 15 starburst galaxies. He used literature
CO with a fixed $\alpha_{\rm CO}$ to estimate $\Sigma_{\rm mol}$. To
estimate $\Sigma_{\rm SFR}$, he used H$\alpha$ in disk galaxies and IR
emission in starbursts.

\citet{CALZETTI10} estimated disk-averaged $\Sigma_{\rm SFR}$ for a
large set of nearby galaxies. We cross-index these with integrated CO
fluxes from \citet{YOUNG95}, \citet{HELFER03}, and \citet{LEROY09} to derive $\Sigma_{\rm mol}$
assuming that CO emission covers the same area as H$\alpha$. From the combination of these data 
we have disk-average $ \Sigma_{\rm SFR}$ and $\Sigma_{\rm mol}$ estimates for $41$ galaxies.

\citet{SAINTONGE12}, following \citet{SAINTONGE11B}, present the COLDGASS survey, which
obtained integrated molecular gas mass and SFRs for $366$ galaxies with $M_* > 10^{10}$~M$_\odot$,
$215$ with secure CO detections. This large survey represents the best sample of integrated galaxy
measurements to date. To convert to surface densities, we take the area of the star-forming disk in these galaxies to be
$0.75~r_{25}$. \citet{SAINTONGE12} derive their SFRs from SED modeling that yields results close to
what one would obtain converting the UV+IR luminosity directly to a SFR. This yields higher SFRs 
than our approach for matched measurements. Comparing galaxies with matched stellar mass or molecular
gas content, we find the offset to be $\approx 0.19$~dex, a factor of $\approx 1.55$. This agrees well with what
one would expect accounting for our subtraction of an IR cirrus with magnitude $\approx 1.2$ and our 
24$\mu$m coefficient, which is $\approx 1.2$ lower than what one would adopt to match a bolometric
TIR SFR indicator (see L12 for calculations and discussion).

\citet{LEROY05} combined new data with measurements by \citet{YOUNG95}, \citet{ELFHAG96}, 
\citet{TAYLOR98}, \citet{BOKER03}, and \citet{MURGIA02} to compare $\Sigma_{\rm SFR}$ 
and $\Sigma_{\rm mol}$ for  individual $\sim 30$--$50\arcsec$ pointings in a wide sample of 
nearby galaxies. They estimate $\Sigma_{\rm SFR}$ from the 20cm radio continuum \citep[][]{CONDON92}.  
These low-resolution pointings typically cover several kpc, a larger area than our resolution elements but less than an average over 
a whole galaxy disk.

\citet{WONG02}, \citet{SCHUSTER07}, and \citet{CROSTHWAITE07}
presented radial profiles of $\Sigma_{\rm mol}$ and $\Sigma_{\rm SFR}$
for several nearby galaxies. \citet{WONG02} targeted 7 nearby spirals,
using H$\alpha$ to calculate $\Sigma_{\rm SFR}$. \citet{SCHUSTER07}
targeted M51 and derived $\Sigma_{\rm SFR}$ from 20-cm radio continuum
to estimate $\Sigma_{\rm SFR}$. We only present the \cite{WONG02} and
\citet{SCHUSTER07} profiles down to $\Sigma_{\rm mol} \approx
5$~M$_\odot$~pc$^{-2}$, below which we consider them somewhat unreliable. \citet
{CROSTHWAITE07} targeted NGC~6946 and used IR emission to estimate $\Sigma_{\rm SFR}$.

\citet{KENNICUTT07}, \citet{BLANC09}, \citet{RAHMAN11}, and \citet{RAHMAN12} targeted
small regions, similar to B08 and the work presented here. \citet{KENNICUTT07}
focused on luminous regions in M51, mainly in the spirals
arms. They infer $\Sigma_{\rm SFR}$ from a combination of \halpha\ and
24\,\micron\ emission. \citet{RAHMAN11} explored a range of
methodologies. We focus on their most robust measurements, drawn from
bright regions in NGC\,4254 with $\Sigma_{\rm SFR}$ from a combination
of NUV and 24\,\micron\ emission. \citet{RAHMAN12} extended this work to consider the full set of 
CARMA STING galaxies, using only the 24$\mu$m emission 
with a nonlinear calibration to infer $\Sigma_{\rm SFR}$. \citet{BLANC09} studied the central
$4.1\times4.1$\,kpc$^{2}$ of M51, deriving $\Sigma_{\rm SFR}$ from
H$\alpha$ spectroscopy corrected using the Balmer decrement.

\section{$\Sigma_{\rm SFR}$-$\Sigma_{\rm mol}$ Scaling Relations: First Order Constancy of $\tau_{\rm dep}^{\rm mol}$}
\label{sec:scaling}

We estimate $\Sigma_{\rm SFR}$ and $\Sigma_{\rm mol}$ for $\sim 14,500$ points in $30$ nearby galaxies.
In this section we analyze these data in the context of a traditional ``star formation law'' scaling relation
(\S \ref{sec:background}). We show the data distribution in $\Sigma_{\rm SFR}$-$\Sigma_{\rm mol}$ parameter 
space (\S\,\ref{sec:combined}) and examine how this depends on methodology (\S \ref{sec:tracers}). Using a Monte Carlo 
technique based on that of \citet{BLANC09}, we consider the best fit power-law to the ensemble data and individual 
galaxies (\S \ref{sec:index}). We compare our results to a broad sample of literature data (\S \ref{sec:lit}).

\subsection{Combined Measurement}
\label{sec:combined}

\begin{figure*}[]
\plotone{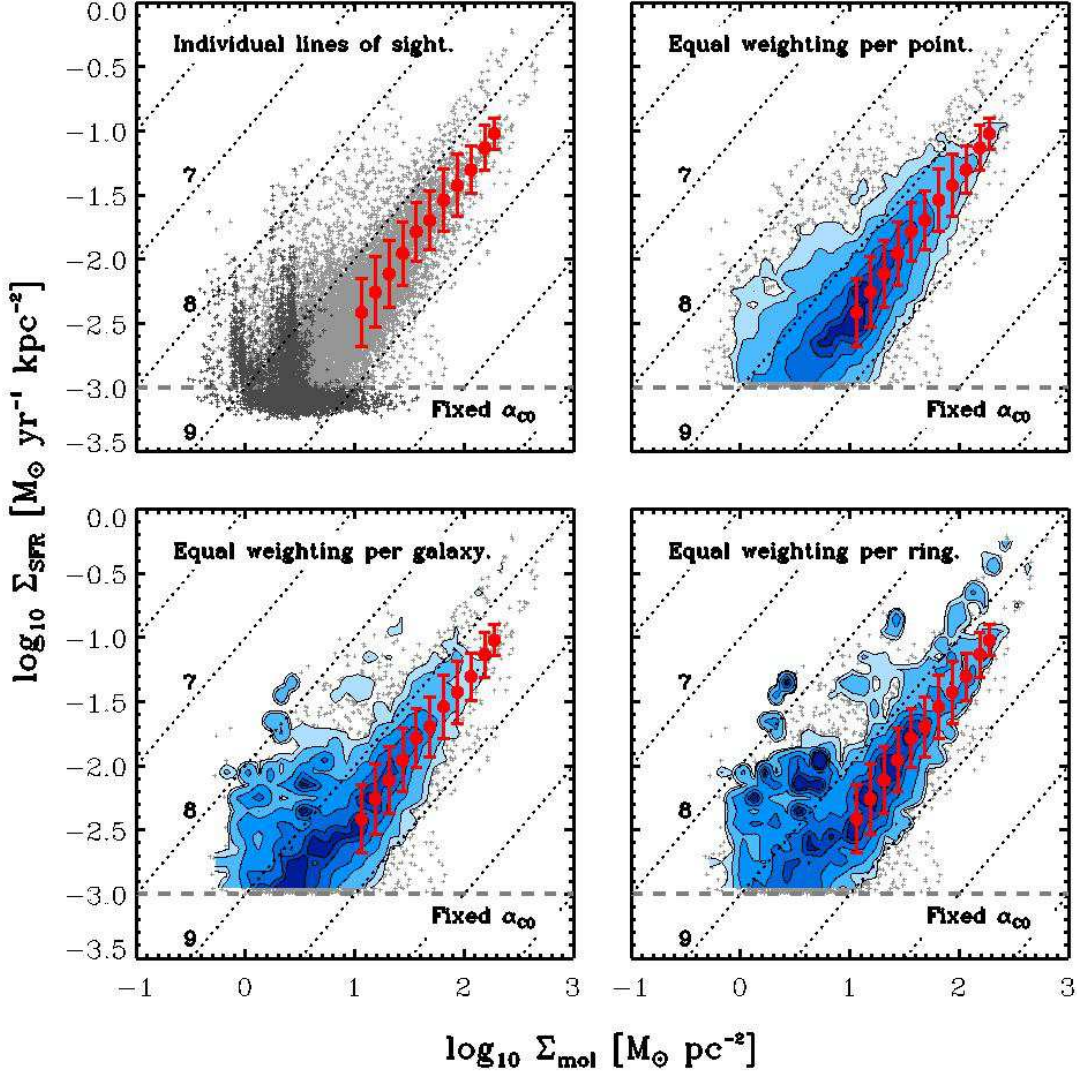}
\caption{Star formation rate surface density, \sigsfr , estimated from H$\alpha$+24$\mu$m 
emission, as a function of molecular gas surface density, \sightwo , derived from CO(2-1) emission 
for 30 nearby disk galaxies. The top left panel shows individual points (dark gray points show upper limits) 
with the running median and standard deviation indicated by red points and error bars. The red points 
with error bars from the first panel appear in all four panels to allow easy comparison. Dotted lines indicated
fixed H$_2$ depletion times; the number indicates $\log_{10} \tau_{\rm Dep}$ in yr. The top 
right panel shows the density of the data in the top left panel. In the bottom panels we vary the weighting 
used to derive data density. The bottom left panel gives equal weight to each galaxy. The bottom 
right panel gives equal weight to each galaxy and each radial bin.}
\label{fig:combined}
\end{figure*}

\begin{deluxetable}{lccc}
\tablecaption{$\tau_{\rm dep}^{\rm mol}$ at 1~kpc Resolution} 
\tablehead{
\colhead{Tracer} & 
\colhead{Median $\tau_{\rm dep}^{\rm mol}$} & 
\colhead{Scatter} & 
\colhead{$r_{\rm corr}$} \\
\colhead{} & 
\colhead{[Gyr]} & 
\colhead{[$1\sigma$ dex]} & 
\colhead{$\left( \Sigma_{\rm mol}, \Sigma_{\rm SFR} \right)$}
}
\startdata
\multicolumn{4}{c}{Weighting as equal each ...} \\
\hline
line-of-sight & & \\
... fixed $\alpha_{\rm CO}$ & 2.2 & 0.28 & $0.72 \pm 0.02$ \\
... $\Sigma=100$ $\alpha_{\rm CO}$ & 2.6 & 0.26 & $0.75 \pm 0.01$ \\
... $\Sigma=50$ $\alpha_{\rm CO}$ & 3.1 & 0.28 & $0.70 \pm 0.01$ \\
galaxy & & \\
... fixed $\alpha_{\rm CO}$ & 1.3 & 0.32 & $0.67 \pm 0.19$\\
... ... only $M_* > 10^{10}$~M$_\odot$ & 1.7 & 0.21 & $0.87 \pm 0.18$\\
... ... only $M_* < 10^{10}$~M$_\odot$ & 0.4 & 0.29 & $0.53 \pm 0.37$\\
... $\Sigma=100$ $\alpha_{\rm CO}$ & 1.8 & 0.20 & $0.89 \pm 0.17$\\
... ... only $M_* > 10^{10}$~M$_\odot$ & 2.0 & 0.13 & $0.95 \pm 0.25$\\
... ... only $M_* < 10^{10}$~M$_\odot$ & 1.1 & 0.26 & $0.87 \pm 0.35$\\
... $\Sigma=50$ $\alpha_{\rm CO}$ & 2.4 & 0.26 & $0.90 \pm 0.19$ \\
... ... only $M_* > 10^{10}$~M$_\odot$ & 2.1 & 0.31 & $0.94 \pm 0.21$\\
... ... only $M_* < 10^{10}$~M$_\odot$ & 2.7 & 0.21 & $0.78 \pm 0.44$\\
\hline
\multicolumn{4}{c}{Tracing $\Sigma_{\rm SFR}$ with ...} \\
\multicolumn{4}{c}{(weighting lines-of-sight equally)} \\
\hline
H$\alpha$+24$\mu$m & & \\
... {\bf best estimate} & 2.2 & 0.28 & $0.72\pm0.02$ \\
... no cirrus & 2.0 & 0.22 & $0.79\pm0.02$ \\
... double cirrus & 3.0 & 0.37 & $0.62 \pm 0.02$ \\
FUV+24$\mu$m & & \\
... {\bf best estimate} & 2.2 & 0.27 & $0.72 \pm 0.02$ \\
... no cirrus & 1.9 & 0.21 & $0.81 \pm 0.02$ \\
... double cirrus & 3.2 & 0.39 & $0.58 \pm 0.02$ \\
H$\alpha$ + $1$~mag & 2.1 & 0.30 & $0.66 \pm 0.02$ 
\enddata
\tablecomments{Median molecular gas depletion time, scatter, and correlation
between $\Sigma_{\rm SFR}$ and $\Sigma_{\rm mo}$ in our sample. Line-of-sight
averages treat each kpc-resolution line of sight as equal. Galaxy averages refer to 
$\tau_{\rm dep}^{\rm mol} = \left<\Sigma_{\rm mol}\right>/\left<\Sigma_{\rm SFR}\right>$ inside 
$0.75~r_{25}$ for each galaxy. Unless otherwise noted, we calculate $\Sigma_{\rm mol}$
using fixed $\alpha_{\rm CO}$ and $\Sigma_{\rm SFR}$ from H$\alpha+24\mu$m. Quoted error 
bars on $\tau_{\rm dep}^{\rm mol}$ report $1\sigma$ scatter, uncertainties on the rank correlation arise from randomly repairing data.}
\label{tab:combined}
\end{deluxetable}

Figure \ref{fig:combined} compares $\Sigma_{\rm SFR}$, estimated from
H$\alpha$+24$\mu$m, and $\Sigma_{\rm mol}$ at 1~kpc resolution for our
whole sample. Individual kpc resolution lines of sight appear as gray
points and the red points show the median $\Sigma_{\rm SFR}$ and
standard deviation after binning the data by $\Sigma_{\rm mol}$. In the top right
panel and bottom row, blue contours show data density adopting
different weightings. The top right panel gives identical weight to
each line of sight, treating each kpc$^2$ as equal regardless of
location. The bottom left panel gives equal weight to each galaxy and
so weights measurements from small galaxies with little area more than
measurements from large galaxies. The bottom right panel treats each
radial ring in each galaxy equally, and so gives more weight to the
central parts of galaxies than their outer regions. Dashed lines here
and throughout this paper indicate fixed $\tau_{\rm dep}^{\rm mol}$ and a
horizontal line indicates the limit of our $\Sigma_{\rm SFR}$
measurements. In the top left panel, dark points show measurements
where one quantity is an upper limit. Table \ref{tab:combined}
summarizes key values from the plots in this section.

The top rows of Figure \ref{fig:combined} and Table \ref{tab:combined}
show the good correspondence between $\Sigma_{\rm SFR}$ and $\Sigma_{\rm mol}$ 
that we have previously found in the HERACLES sample \citep[B08,L08,][]{SCHRUBA11,BIGIEL11}.
Our dynamic range at 1~kpc resolution spans from $\Sigma_{\rm SFR} \sim 10^{-3}$ to
$10^{-1}$~M$_\odot$~yr$^{-1}$~kpc$^{-2}$ and $\Sigma_{\rm mol}$ from a few
to $100$~M$_\odot$~pc$^{-2}$. Across this range, $\Sigma_{\rm mol}$ and
$\Sigma_{\rm SFR}$ correlate well, exhibiting a Spearman rank correlation
coefficient $\gtrsim 0.7$ for most tracers and weightings. This quantifies the tight,
one-to-one relationship visible by eye in the top row.

The median $\tau_{\rm dep}^{\rm mol}$ weighting each line of sight equally is
$2.2$~Gyr with a scatter of $0.3$ dex, a factor of two. The absolute value of the median
$\tau_{\rm dep}^{\rm mol}$, i.e., the scale of the $x$- and $y$-axes in Figure \ref{fig:combined}, 
depends on the calibration of our SFR tracer and CO-to-H$_2$
conversion factor. Each remains uncertain at the $30$--$50\%$ level and we suggest 
that an overall uncertainty of 60\% on the absolute value of $\tau_{\rm dep}^{\rm mol}$
represents a realistic, if somewhat conservative, value. Our $\Sigma_{\rm mol}$ and
$\Sigma_{\rm SFR}$ estimates can be compared internally with much better accuracy than this 
\citep[L12,][]{SANDSTROM12}, so we suggest that this uncertainty be viewed as an overall 
scaling of our results.

The bottom row in Figure \ref{fig:combined} begins to reveal the deviations from 
a simple one-to-one scaling that will be the subject of Section \ref{sec:tdep_vary}. 
Weighting all galaxies equally (bottom left panel) reveals a significant population of 
low $\Sigma_{\rm mol}$, high $\Sigma_{\rm SFR}$, low $\tau_{\rm dep}^{\rm mol}$ data. 
This drives the median depletion time for the sample from $\approx 2.2$~Gyr, weighting by line-of-sight,
to $\approx 1.3$~Gyr, weighting by galaxy. In Section \ref{sec:global} we show that these
low apparent $\tau_{\rm dep}^{\rm mol}$ originate from low-mass, low-metallicity systems 
\citep[see also][]{SCHRUBA11,KRUMHOLZ11,SCHRUBA12}. Because of their small size, these systems do not
contribute many data compared to large, metal-rich spirals. Therefore, they only weakly influence the 
overall data distribution seen in the top row. We examine $\tau_{\rm dep}^{\rm mol}$ as a function of 
host galaxy properties and local conditions in Sections \ref{sec:global} and \ref{sec:local}.
In the appendix we present $\Sigma_{\rm SFR}-\Sigma_{\rm mol}$ relations for individual galaxies
(see also Table \ref{tab:integrated}), allowing the reader to see how Figure \ref{fig:combined} emerges 
from the superposition of individual systems (see also Section \ref{sec:index}).

Weighting radial rings equally (bottom right panel) highlights these same 
low $\tau_{\rm dep}^{\rm mol}$-low $\Sigma_{\rm mol}$ galaxies and brings out an additional low
$\tau_{\rm dep}^{\rm mol}$ population at higher $\Sigma_{\rm mol}$. These point emerge
because the radial weighting emphasizes points in the central parts of galaxies relative to their
outskirts. We show in Section \ref{sec:centers} that the central regions of many of our targets
exhibit enhanced efficiency compare to their disks. As will small galaxies, these central regions
contribute only a tiny fraction of the area in our survey and thus exert little impact on the plots
in the top row.

Figure \ref{fig:combined} thus illustrates our main conclusions: a first order
simple linear correlation between $\Sigma_{\rm SFR}$ and $\Sigma_{\rm mol}$ and
real second order variations. It also illustrates the limitation of considering only
$\Sigma_{\rm SFR}$-$\Sigma_{\rm mol}$ parameter space to elicit these second-order
variations. Metallicity, dust-to-gas ratio, and position in a galaxy all play
key roles but are not encoded in this plot, leading to double-valued $\Sigma_{\rm SFR}$
at fixed $\Sigma_{\rm mol}$ in some regimes. We explore these systematic variations 
in $\tau_{\rm dep}^{\rm mol}$ and motivate our explanations throughout the rest of the paper.

\subsection{Relationship for Different SFR and Molecular Gas Tracers}
\label{sec:tracers}

Figure \ref{fig:combined} shows our best-estimate $\Sigma_{\rm SFR}$ and $\Sigma_{\rm mol}$ 
computed from fixed $\alpha_{\rm CO}$. Many approaches exist to estimate each quantity 
\citep[see references in][L12]{LEROY11} and the recent literature includes many claims about
the effect of physical parameter estimation on the relation between $\Sigma_{\rm SFR}$ and $\Sigma_{\rm mol}$.
In this section, we explore the effects of varying our approach to estimate $\Sigma_{\rm SFR}$ and $\Sigma_{\rm mol}$.

\subsubsection{Choice of SFR Tracer}
\label{sec:choicesfr}

\begin{figure*}[]
\plotone{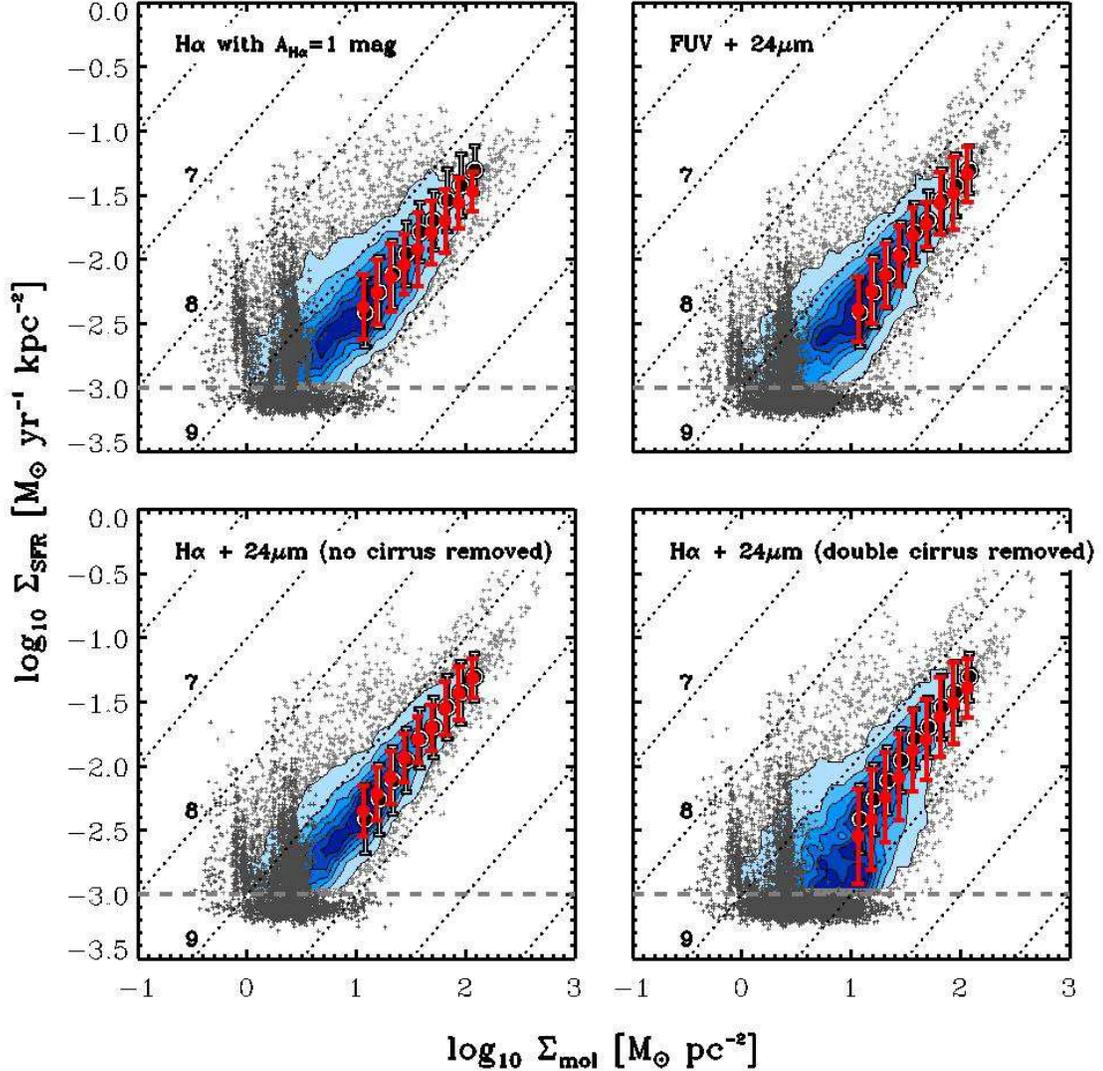}
\caption{Star formation rate surface density, \sigsfr , estimated using different tracers as a function of molecular gas surface
density, \sightwo . Light gray points show individual measurements, dark gray points indicate upper limits, contours show data density, 
and red points show the running median and standard deviation using each tracer. For comparison, black-and-white points show the bins from
the top right panel of Figure \ref{fig:combined}. We show $\Sigma_{\rm SFR}$ estimated from: ({\em top left}) H$\alpha$ 
assuming 1 magnitude of extinction, ({\em top right}) a combination of FUV and 24$\mu$m emission (L12), ({\em bottom 
left}) H$\alpha$ combined with 24$\mu$m emission with no IR cirrus correction applied to the 24$\mu$m, ({\em bottom right}) H$\alpha$ 
combined with 24$\mu$m with double our IR cirrus correction applied to the 24$\mu$m.}
\label{fig:by_sfr}
\end{figure*}

Figure \ref{fig:by_sfr} and the lower part of Table \ref{tab:combined} report the 
results of varying our approach to trace the SFR. We show $\Sigma_{\rm SFR}$ estimated from only H$\alpha$, with 
a fixed, typical $A_{\rm H\alpha} = 1$~mag (top left), along with results combining FUV, instead of H$\alpha$, 
with 24$\mu$m emission (top right). We also show the results of varying the approach to the 
IR cirrus. Our best-estimate $\Sigma_{\rm SFR}$ combines H$\alpha$ or FUV with 24$\mu$m after correcting the 
24$\mu$m emission for contamination by an IR cirrus following L12. We illustrate the impact of this correction by 
plotting results for two limiting cases of IR cirrus correction: no cirrus subtraction (bottom left) and removing double 
our best cirrus estimate (bottom right), which we consider a maximum reasonable correction. Data density contours in 
Figure \ref{fig:by_sfr} weight each point equally and the large black points indicate the original binned results from 
Figure \ref{fig:combined}. 

The top left panel of Figure \ref{fig:by_sfr} and Table \ref{tab:combined} show
that the basic relationship between $\Sigma_{\rm SFR}$ and $\Sigma_{\rm mol}$ persists 
even when we derive $\Sigma_{\rm SFR}$ from H$\alpha$ alone. The median $\tau_{\rm dep}^{\rm mol}$
and scatter using only H$\alpha$ resemble what we find for our best estimate and 
the correlation between H$\alpha$ and CO appears only moderately
weaker than for the hybrid SFR tracer. It also appears moderately flatter than relations that incorporate IR
emission as we underestimate extinction in the central parts of galaxies (\S \ref{sec:index}). Inasmuch 
as H$\alpha$ represents an unambiguous tracer
of recent star formation, the top left panel in Figure \ref{fig:by_sfr} demonstrates that
subtle biases in the treatment of IR emission, e.g., 24$\mu$m emission tracing the ISM rather than
recent star formation, do not drive our results.

The top right panel shows $\Sigma_{\rm SFR}$ traced by FUV+24$\mu$m emission. The 
distribution agrees well with what we found using H$\alpha$+24$\mu$m, as do the median
and scatter in $\tau_{\rm dep}^{\rm mol}$. The agreement of
FUV+24$\mu$m and H$\alpha$ with our best estimate H$\alpha$+24$\mu$m occur
partially because we have designed our SFR tracers to yield self-consistent results (L12). However,
that procedure considered only the overall normalization and did not require the detailed agreement 
we see comparing Figures \ref{fig:combined} and \ref{fig:by_sfr}.

In the bottom row, we vary our approach to the infrared cirrus. By default, we correct 
the 24$\mu$m map for infrared cirrus following L12. The bottom left panel shows
the results of applying no cirrus subtraction, while in the bottom right
panel we double our cirrus subtraction. Turning off the cirrus subtraction yields median
$\tau_{\rm dep}^{\rm mol}$ $\approx 10\%$ shorter than our best estimate with
notably lower scatter and our strongest observed correlation. The tighter 
correlation reflects the fact that the relationship between 24$\mu$m and CO emission 
is the strongest in the data \citep[see also][]{SCHRUBA11}. $\Sigma_{\rm SFR}$ tracers 
that more heavily emphasize 24$\mu$m exhibit the strongest correlation with $\Sigma_{\rm mol}$ 
traced by CO.

Doubling the cirrus subtraction leads to an $\approx 25\%$ longer $\tau_{\rm dep}^{\rm mol}$,
larger scatter, and a mildly weaker correlation between $\Sigma_{\rm SFR}$ and $\Sigma_{\rm mol}$.
This partially reflects uncertainty in the cirrus calculation, which relies on model fits to observed
data. It also reflects the deemphasis of 24$\mu$m emission, which exhibits a very tight correspondence to
CO, in favor of H$\alpha$, which still exhibits a good correspondence but with more scatter. The fraction
of data that have upper only limits for $\Sigma_{\rm SFR}$ also increases, so that extending this analysis to lower surface density
will require improved data and methodology. 

Thus we observe subtle variations in the relation between $\Sigma_{\rm SFR}$ and $\Sigma_{\rm mol}$
depending on the exact treatment of 24$\mu$m emission, including up to a $\approx 50\%$ variation in
$\tau_{\rm Dep}^{\rm mol}$ across the full plausible range of cirrus treatments. However, our main results of
a simple correspondence between $\Sigma_{\rm mol}$ and $\Sigma_{\rm SFR}$ hold even
when we omit IR data from the analysis. Note that this conclusion relies on the assumption that
H$\alpha$ emission traces recent star formation. If a substantial fraction of H$\alpha$ emission arises 
from sources other than recent star formation or if the mean free path of an ionizing photon regularly exceeds one of our
kpc-sized resolution elements, then this general agreement may break down \citep[][see discussion in L12]{RAHMAN11,LIU11}.
These more exotic situations aside, overall Figure \ref{fig:by_sfr} and Table \ref{tab:combined} show  good qualitative and 
quantitative agreement among different approaches to estimate $\Sigma_{\rm SFR}$. We will find the 
same when fitting the data in Section \ref{sec:index}.

Throughout the rest of the paper, we adopt H$\alpha$+24$\mu$m, corrected for 
the effects of a 24$\mu$m cirrus, as our single, best estimate of $\Sigma_{\rm SFR}$. L12 justify 
this choice and we refer to that paper for more discussion.

\subsubsection{Choice of CO Line}
\label{sec:choiceh2}

\begin{figure}[]
\plotone{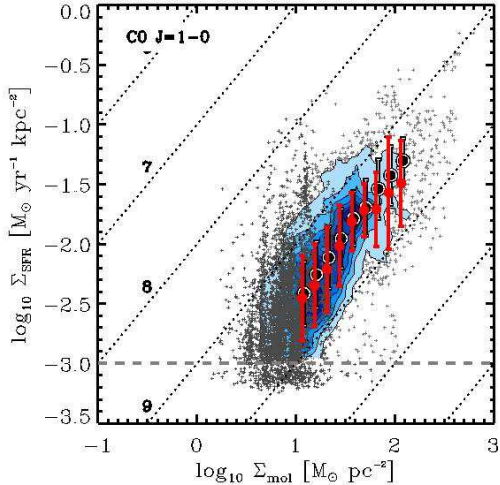}
\caption{Star formation rate surface density, \sigsfr , as a function
  of molecular gas surface density, \sightwo , here estimated from
  literature CO(1-0) data (Section \ref{sec:data}). Gray points show individual
  measurements, contours indicate data density, and red points show
  the running median and standard deviation using each tracer. For comparison,
  large black-and-white circles show the same bins from Figure
  \ref{fig:combined}.}
\label{fig:by_co}
\end{figure}

HERACLES consists of maps of CO (2-1) emission, which we use to
estimate the distribution of H$_2$. CO (1-0) has been more
commonly used to trace the distribution of H$_2$ \footnote{
We emphasize that \citet{SANDSTROM12} demonstrate the 
ability of CO (2-1) to robustly trace molecular gas in our sample
(Section \ref{sec:param})}. CO (1-0)
maps exist for a subset of our targets \citep{HELFER03,KUNO07}. Though
these do not have the same overall quality as the HERACLES maps, we use them
to assess the impact of our choice of molecular gas tracer. Figure \ref{fig:by_co} plots
$\Sigma_{\rm SFR}$ as a function of $\Sigma_{\rm mol}$ estimated from
literature CO(1-0) data. We allow repeats, so that if
\citet{HELFER03} and \citet{KUNO07} each mapped a galaxy we include
each data set in Figure \ref{fig:by_co}. 

Overall, results for CO (2-1) and CO (1-0) agree fairly well. The CO (1-0) data 
tend to yield higher $\tau_{\rm dep}^{\rm mol}$. This is exclusively a product of the
\citeauthor{KUNO07} data. In the overlap of our sample and the
\citeauthor{KUNO07} data, the \citeauthor{KUNO07} data yield median $\tau_{\rm
  Dep}^{\rm mol} \approx 3.3$~Gyr. Our data yield median $\tau_{\rm
  Dep}^{\rm mol} \approx 2.3$~Gyr for the same points. However, for the
overlap with BIMA SONG \citep{HELFER03}, the BIMA SONG CO (1-0) data
give median $\tau_{\rm dep}^{\rm mol} \approx 2.1$~Gyr. Over the same
points, HERACLES implies $\tau_{\rm dep}^{\rm mol} \approx
2.2$~Gyr. The disagreement between our CO (2-1) data and
CO (1-0) data thus appears no larger than the disagreement
among published CO (1-0) data sets.

\subsubsection{CO-to-H$_2$ Conversion Factor}
\label{sec:alphaco}

\begin{figure*}[]
\plotone{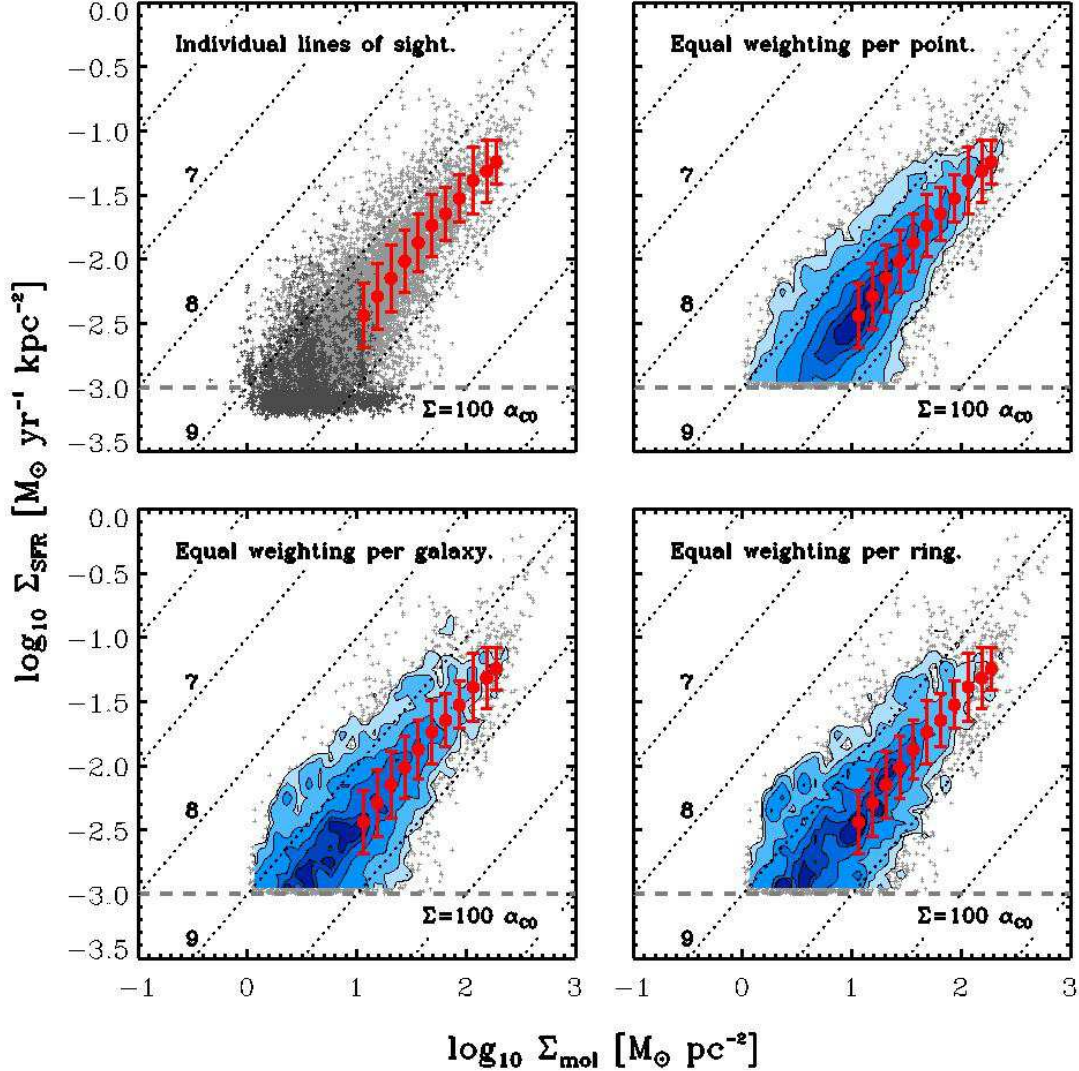}
\caption{Star formation rate surface density, $\Sigma_{\rm SFR}$, as a function of
molecular gas surface density, $\Sigma_{\rm mol}$, here estimated from CO (2-1)
data combined with our ``$\Sigma=100$'' conversion factor $\alpha_{\rm CO}$, which 
reflects variations in the dust-to-gas ratio and the central depressions in $\alpha_{\rm CO}$
found by \citet{SANDSTROM12}. Panels and annotations are as Figure \ref{fig:combined}.}
\label{fig:combine_alpha}
\end{figure*}

In Figures \ref{fig:combined} - \ref{fig:by_co} we adopt a fixed CO-to-H$_2$ conversion, 
$\alpha_{\rm CO}$. This assumption may be too coarse given the wide range of 
metallicities, dust-to-gas ratios, and central CO concentrations in our
targets. Figure \ref{fig:combine_alpha} shows $\Sigma_{\rm SFR}$ and $\Sigma_{\rm mol}$
after the application of our ``$\Sigma=100$'' conversion factor (Section \ref{sec:param}), which 
attempts to account for the presence of CO-poor envelopes of molecular clouds and central CO
depressions. This conversion factor assumes that all CO emission originates from clouds with 
surface densities of $100$~M$_\odot$~pc$^{-2}$ with PDR structure like that described in
\citet{WOLFIRE10}. The resulting dependence of $\alpha_{\rm CO}$ on metallicity approximates the 
current consensus \citep{WOLFIRE10,GLOVER11,LEROY11,FELDMANN12,
NARAYANAN12}, but note that this remains highly uncertain because of limited observational constraints.
We also report results for the ``$\Sigma=50$'' conversion factor in Table \ref{tab:combined}.
This conversion factor makes the more aggressive assumption that a large amount of molecular emission
emerges from weakly shielded parts of clouds, rendering $\alpha_{\rm CO}$ very sensitive to the
dust-to-gas ratio.

Applying the ``$\Sigma=100$'' $\alpha_{\rm CO}$ to the data in Figure \ref{fig:combined} yields
Figure \ref{fig:combine_alpha}. The top rows, which show the bulk distribution of the data, appear qualitatively similar 
in the two plots, though Table \ref{tab:combined} and close inspection of the plots do indicate that the normalization 
of the $\Sigma_{\rm SFR}$-$\Sigma_{\rm mol}$ relation changes between the two plots. The median $\tau_{\rm dep}^{\rm mol}$
weighting each line of sight equally rises from $2.2$~Gyr to $2.6$~Gyr with the application of the variable conversion factor.

The most dramatic contrast between Figures \ref{fig:combined} and \ref{fig:combine_alpha} appears in the bottom rows. 
Many of the low $\tau_{\rm dep}^{\rm mol}$ (high $\Sigma_{\rm SFR}$) data in Figure \ref{fig:combined} arise from
small galaxies with low dust-to-gas ratios. With the application of a variable conversion factor, our estimate of $\Sigma_{\rm mol}$ 
in these galaxies moves to higher values while $\Sigma_{\rm SFR}$ remains constant. The result, visible in the bottom rows
of Figure \ref{fig:combine_alpha}, is that data from these low-mass, low metallicity galaxies now overlap the other points, forming a 
(more) continuous single $\Sigma_{\rm SFR}$-$\Sigma_{\rm mol}$ trend. Table \ref{tab:combined} reports that the scatter
among galaxies drops from $\approx 0.3$~dex to $\approx 0.2$~dex with the application of the ``$\Sigma=100$'' conversion factor,
with the systematic difference in $\tau_{\rm dep}^{\rm mol}$ between high and low mass galaxies reduced from a factor of $\sim 4$ to a 
factor of $\sim 2$.

Thus, as we will see in Section \ref{sec:tdep_vary}, application of a dust-to-gas ratio-dependent conversion factor 
CO-to-H$_2$ conversion factor does affect the derived $\Sigma_{\rm SFR}$-$\Sigma_{\rm mol}$ relation, with the sense of moving many points with low 
apparent $\tau_{\rm dep}^{\rm mol}$ into closer agreement with the distribution defined by large galaxies. This scenario of a rapidly 
varying conversion factor and a weakly varying $\tau_{\rm dep}^{\rm mol}$ has been discussed in the context of the Small 
Magellanic Cloud by \citet{BOLATTO11} and in a theoretical context by \citet{KRUMHOLZ11}. They interpret weak variations of
$\tau_{\rm dep}^{\rm mol}$ but strong variations of $\alpha_{\rm CO}$ as evidence that the requisite preconditions for star
formation more closely resemble those for H$_2$ formation than those required for a high CO abundance.

In \S \ref{sec:tdep_vary}, we will show that the even more  aggressive ``$\Sigma=50$'' case may in fact explain {\em most} apparent 
galaxy-wide variations in $\tau_{\rm dep}^{\rm mol}$, but note that while the ''$\Sigma=50$'' conversion factor may offer an explanation 
for our observed trends, it also requires that a substantial fraction of the integrated CO emission from galaxies arise from relatively low 
$A_V$ regions.

The other portion of the ``$\Sigma=100$'' conversion factor,  the $\alpha_{\rm CO}$ depressions in the central parts of galaxies,
affect too few data to be prominent in Figure \ref{fig:combine_alpha}. However, one can see many individual low $\tau_{\rm dep}^{\rm mol}$ 
points at moderate $\Sigma_{\rm mol}$ on close inspection. We return to this point in \S \ref{sec:centers}.

\subsection{Power Law Index}
\label{sec:index}

\begin{figure*}[]
\plottwo{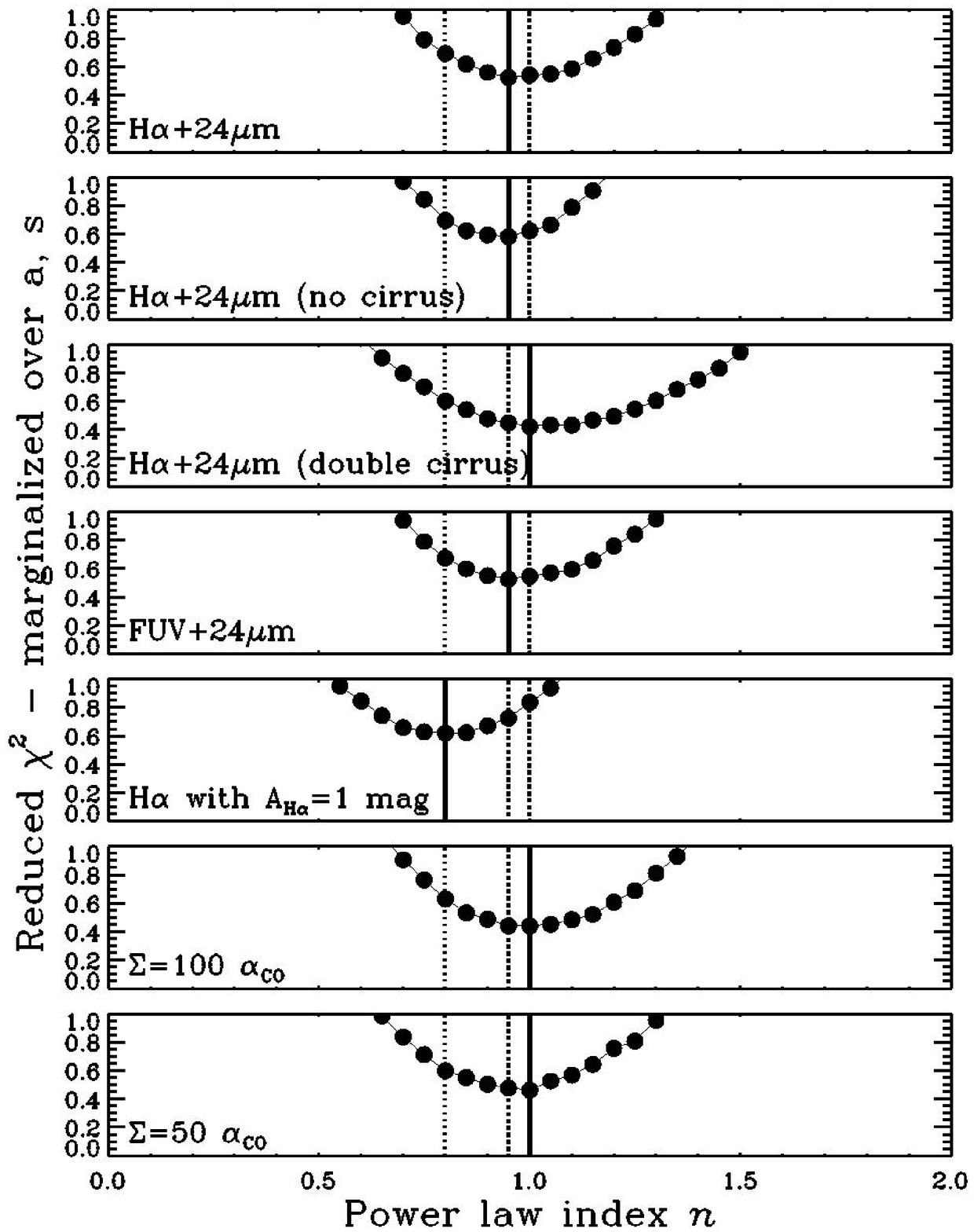}{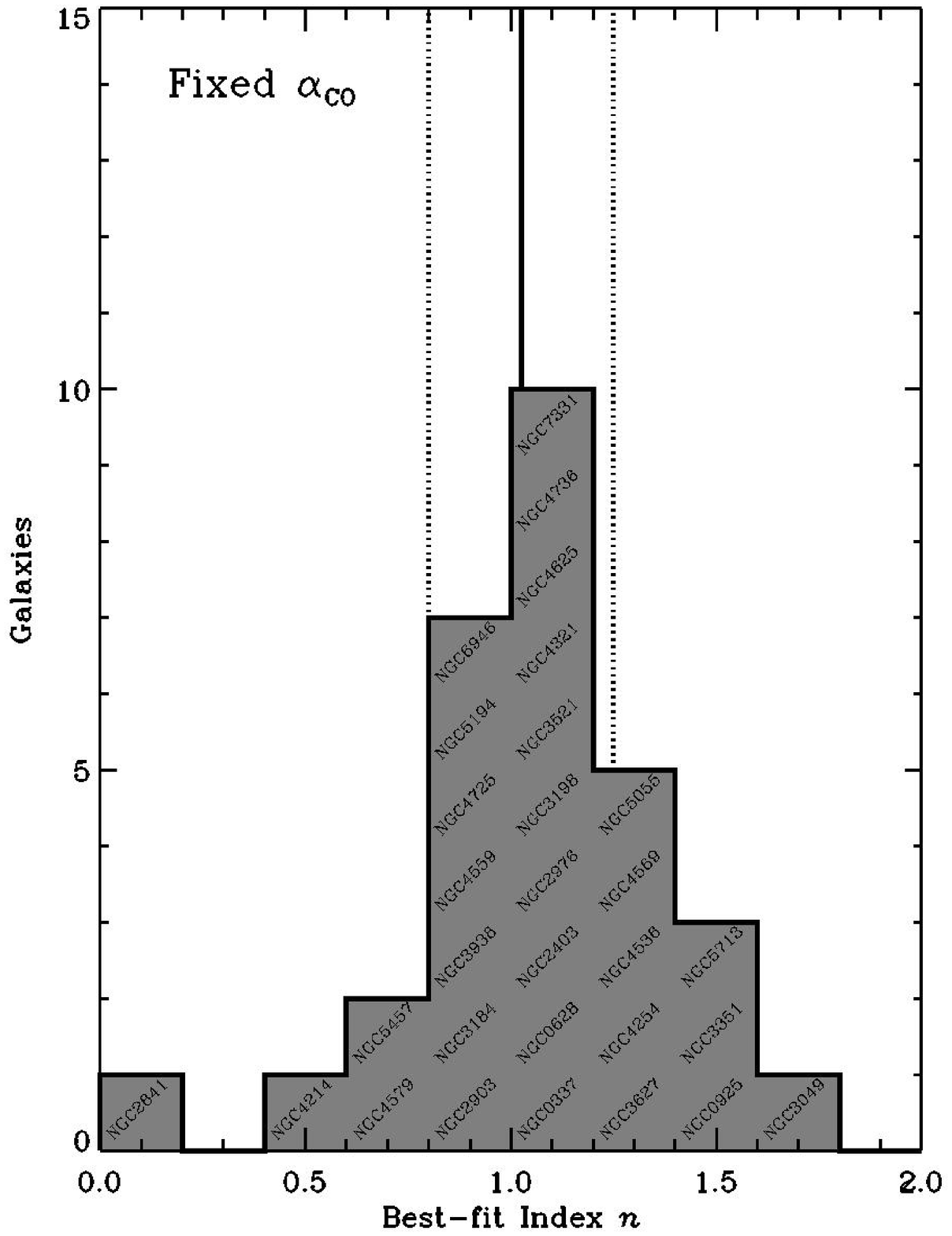}
\caption{Results of power-law fitting for ({\em left}) our combined data set and
({\em right}) individual galaxies. {\em Left:} Goodness of fit as a function of power law index, $n$, for
fits to our entire ensemble of data where $\Sigma_{\rm mol} > 5$~M$_\odot$~pc$^{-2}$. The fitting 
uses a modified version of the Monte Carlo technique outlined by \citet{BLANC09}. From top to bottom, 
we show results for fixed $\alpha_{\rm CO}$ using our best H$\alpha$+24$\mu$m 
and FUV+24$\mu$m estimates; H$\alpha$+24$\mu$m with no cirrus removed 
or double the best-estimate cirrus removed; and H$\alpha$ with an assumed fixed 1~magnitude of 
extinction. We also show results for our ``$\Sigma=100$'' and ``$\Sigma=50$'' conversion factors. 
Vertical lines indicate where $n$ reaches its minimum value, bold for the value in that panel, dotted for 
other panels. Each case shows a broad minimum in $\chi^2$ with a best-fit value near $n \approx 1 \pm 0.2$.
{\em Right:} Best fit power-law index, $n$, for individual galaxies using H$\alpha$+24$\mu$m and a
fixed CO-to-H$_2$ conversion factor. The distribution of best-fit $n$ for individual galaxies peaks
near $n \approx 1$ with most galaxies in the range $n \approx 0.8$--$1.2$ but outliers up to $n \gtrsim 1.5$.
}
\label{fig:fit}
\end{figure*}

\begin{deluxetable*}{lccc}
\tablecaption{Results of Monte Carlo Fitting to Equation \ref{eq:powlaw}} 
\tablehead{
\colhead{Tracer} & 
\colhead{$\log_{10} a$} & 
\colhead{$n$} & 
\colhead{$s$} \\
& 
(M$_\odot$~yr$^{-1}$~kpc$^{-2}$) & 
& (dex) 
}
\startdata
H$\alpha$+24$\mu$m & $-2.40 \pm 0.09$ & $0.95 \pm 0.15 $ & $0.30 \pm 0.05$  \\
$...$ no cirrus removed & $-2.35 \pm 0.08$ & $0.95 \pm 0.13$ & $0.225\pm0.05$ \\
$...$ double cirrus removed & $-2.50 \pm 0.10$ & $1.00 \pm 0.17$ & $0.375\pm0.05$\\
FUV+24$\mu$m & $-2.40 \pm 0.09$ & $0.95 \pm0.16$ & $0.30\pm0.05$ \\
H$\alpha$ with $A_{\rm H\alpha} = 1$ mag & $-2.35 \pm 0.09$ & $0.80 \pm0.14$ & $0.275\pm0.06$ \\
\hline
\\
$\Sigma = 100$ $\alpha_{\rm CO}$ and H$\alpha$+24$\mu$m & $-2.45 \pm 0.09$ & $1.0 \pm 0.15$ & $0.30 \pm 0.05$ \\
$\Sigma = 50$ $\alpha_{\rm CO}$ and H$\alpha$+24$\mu$m & $-2.35 \pm 0.09$ & $0.90 \pm 0.15$ & $0.30 \pm 0.05$
\enddata
\label{tab:fit}
\tablecomments{Results of Monte Carlo fitting to Equation \ref{eq:powlaw} for different combinations of $\Sigma_{\rm SFR}$ and $\Sigma_{\rm mol}$ tracers. Column (1) reports the tracer used; in the top five rows we vary the $\Sigma_{\rm SFR}$ tracer while adopting fixed $\alpha_{\rm CO}$. The last two rows adopt our best $\Sigma_{\rm SFR}$ estimate, H$\alpha$+24$\mu$m, and vary the adopted conversion factor. Columns (2)--(4) report the best-fit coefficient at $\Sigma_{\rm mol} = 10$~M$_\odot$~pc$^{-2}$, the power law index, and the intrinsic scatter. We quote uncertainties from the Monte Carlo simulations described in the appendix.}
\end{deluxetable*}

Studies of the star formation-gas connection in galaxies have treated the relationship as a power law and focused on the index of this power
law. While this single parameter undoubtedly makes for easy shorthand, the fixation on this parameter
obscures environmental factors other than $\Sigma_{\rm gas}$. Recent observations offer good evidence that $\Sigma_{\rm SFR}$ relates
to $\Sigma_{\rm HI}$ and $\Sigma_{\rm mol}$ in fundamentally different ways
\citep[L08, B08][]{WONG02,SCHRUBA11} and that $\Sigma_{\rm SFR}$ is even a multivalued function of $\Sigma_{\rm mol}$ \citep[][and this paper]{DADDI10,GENZEL10,SCHRUBA11,SAINTONGE11B,SAINTONGE12}.

Despite the shortcomings of this approach, we consider the best-fit index in our data as a useful, or 
at least expected, point of comparison to previous studies. We derive best-fit relations for our ensemble of measurements and individual galaxies. We fit a relation with three 
parameters: a normalization, $a$, power-law index, $n$, and intrinsic, log-normally distributed scatter with RMS magnitude $s$. Then

\begin{equation}
\label{eq:powlaw}
\Sigma_{\rm SFR} = a \left(\frac{\Sigma_{\rm gas}}{10~{\rm M}_\odot~{\rm pc}^2} \right)^{n}
\end{equation}

\noindent with data intrinsically scattered by $s$. We derive the best-fit $a$, $n$, and $s$ using a 
Monte Carlo approach based on the work of \citet{BLANC09}. This  resembles Hess diagram fitting 
used for optical color-magnitude diagrams. It includes observational uncertainties, upper limits, and 
intrinsic scatter in the relation. This approach also avoids important biases that can easily arise fitting scaling relations 
to noisy, bivariate data. We illustrate these biases, which affect many commonly adopted approaches,  in 
the appendix \citep[see also][]{BLANC09} and note that they can easily shift the derived index by a 
few tenths for realistic data distributions.

Following \citet{BLANC09} we grid our data, deriving a two-dimensional image of data density in 
regularly-spaced cells in $\log_{10} \Sigma_{\rm SFR}$-$\log_{10} \Sigma_{\rm mol}$ space. Unlike 
\citet{BLANC09} we work in logarithmic space. This gives us a better ability to resolve the 
distribution of our data, but forces a coarser approach to upper limits. We treat upper limits by essentially 
creating an ``upper limit row" along the $\Sigma_{\rm SFR}$ axis. In detail, we adopt the following approach:

\begin{enumerate}

\item We exclude all data with $\Sigma_{\rm mol} < 5$~M$_\odot$~pc$^{-2}$. This gives us a data 
set with a well-defined $x$-axis.

\item We generate Monte Carlo data sets for a wide range of $a$, $n$, and $s$ in the following way. 
We take our observed $\Sigma_{\rm mol}$ to represent the true physical distribution. We draw 100,000 data 
points from this distribution (allowing repeats) for each combination of $a$, $n$, 
and $s$. We derive $\Sigma_{\rm SFR}$ for each of these points. We then apply the expected 
uncertainty to $\Sigma_{\rm mol}$ (the statistical uncertainty from HERACLES) and $\Sigma_
{\rm SFR}$ (0.15 dex). We grid these data in $\log_{10} \Sigma_{\rm SFR}$-$\log_{10} \Sigma_{\rm 
mol}$ space, using cells $0.125$~dex wide in both dimensions. We treat this grid as the expected probability 
distribution function for those underlying parameters $a$, $n$, and $s$.

\item We grid our observed data in $\log_{10} \Sigma_{\rm SFR}$-$\log_{10} \Sigma_{\rm mol}$ 
using the same grid on which we derived probability distribution functions. We create different grids for 
each set of $\Sigma_{\rm SFR}$ estimates.

\item We compare our gridded data to the Monte Carlo realization for each $a$, $n$, $s$ 
combination and calculate a goodness-of-fit estimate, which we here loosely refer to as $\chi^2$. 
After re-normalizing the Monte Carlo grid to have the same amount of data as the observed grid, we calculate:

\begin{equation}
\label{eq:gof}
\chi^2 = \sum\limits_i \frac{(N_{\rm obs}^i - N_{\rm model}^i)^2}{N_{\rm model}^i}~.
\end{equation}

\noindent where the sum runs across all cells, $i$, $N_{\rm obs}$ refers to the observed number of 
data in the grid cell, and $N_{\rm model}$ refers to the expected number of data in that cell given $a$, $n$, and $s$ and 
our observational uncertainties. The goodness-of-fit statistic is thus analogous to $\chi^2$ 
calculated for the case of Poisson noise in each cell. Points with only upper limits on $\Sigma_{\rm 
SFR}$ (where $\Sigma_{\rm SFR} < 10^{-3}$~M$_\odot$~yr$^{-1}$~kpc$^{-2}$) are included in the calculation. 
These have an associated $\Sigma_{\rm mol}$ value but all upper limits are treated as having the same $\Sigma_{\rm SFR}$.

\end{enumerate}

We apply this method to our ensemble of data, repeating the exercise for each SFR tracer discussed in Section \ref{sec:tracers}
and for our fixed, ``$\Sigma=100$'', and ``$\Sigma=50$'' conversion factors. We also fit each galaxy on its own\footnote{Due to
the lower density of points, we use 0.2-dex cell sizes and require only $10,000$ points to populate the theoretical distribution.}. 
Figure \ref{fig:fit} and Table \ref{tab:fit} report our fits to the combined data set. Figure \ref{fig:fit} plots the approximate reduced 
$\chi^2$ as a function of power law index, marginalizing over $a$ and $s$. We observe  clear minima in the range $n = 0.5$--$1.5$ for 
all SFR tracers. The appendix presents a Monte Carlo treatment that considers a number of effects: robustness to removal of individual 
data or galaxies, statistical noise, calibration (gain) uncertainties for each data set, and exact choice of fitting methodology. We quote 
uncertainties derived from this Monte Carlo treatment in Table \ref{tab:fit}.

The fits in Table \ref{tab:fit} suggest a power law with $n \approx 1 \pm 0.15$, intrinsic scatter of a factor of $\approx 2$ 
(0.3~dex), and $\tau_{\rm dep}^{\rm mol} \approx 2.5$~Gyr at $\Sigma_{\rm mol} = 10$~M$_\odot$~pc$^{-2}$. The 
slope remains consistent with a linear relation between H$_2$ and star formation \citep[B08,][]{BIGIEL11} or with 
the weakly super-linear slope of \citet{GENZEL10} or \citet{DADDI10}, though note that our 1-kpc scale does not precisely match 
their observations. The mild difference between the best-fit coefficient, $a$, and the median
$\tau_{\rm dep}^{\rm mol}$ reported in Table \ref{tab:combined} reflect the inadequacy of the power law to capture the full distribution
of the data.

The choice of SFR tracer affects the fit, but offers more of a refinement than a qualitative shift in these 
conclusions. The sense of the shifts resemble those seen in Section \ref{sec:tracers}. Replacing FUV for H$\alpha$ as the 
unobscured tracer has minimal effect. Using only H$\alpha$ to estimate SFR yields a slightly shallower slope. Based on the 
observed H$\alpha$-to-IR ratio, extinction increases with increasing $\Sigma_{\rm SFR}$ \citep[see plots in][and L12]{PRESCOTT07}. By 
assuming a fixed $A_{\rm H\alpha}$ we would expect to underestimate $\Sigma_{\rm SFR}$ at the high end 
and overestimate it at low end, somewhat  ``tilting" the relationship to shallower slope. If we do not remove 
any cirrus from the data, working only with the measured 24$\mu$m emission, the scatter in the 
relation diminishes to less that $0.2$~dex. This underscores the point that it is the tight observed correlation 
between CO and 24$\mu$m emission that drives much of the recent work on this topic 
\citep[see more discussion in][]{RAHMAN11,SCHRUBA11,LIU11} so that SFR tracers that emphasize 24$\mu$m data tend to yield the 
tightest relations. Conversely, increasing the cirrus removed leads to a somewhat longer overall $\tau_{\rm dep}^{\rm mol}$
with larger intrinsic scatter. Adjusting the conversion factor exerts only a mild impact on the fit because largest corrections
apply to small galaxies and often to low apparent $\Sigma_{\rm mol}$ regions. These significant variations to a small
subset of the data do not drive substantial variations in the fit.

The simple nearly linear scaling given by our fits could result from the superposition of a varied set of distinct relations for individual
galaxies \citep[see][]{SCHRUBA11}. In the right hand panel of Figure \ref{fig:fit} we show that indeed the best-fit index
for individual galaxies exhibits significant scatter. We find a median $1.05$ but best-fit values span $\sim 0.5$--$1.5$. We report best-fit
indices for individual galaxies in the appendix and stress two general conclusions here. First, we see variation in index from
galaxy to galaxy, but the $67\%$ range is still $0.8$--$1.25$, consistent with the idea that to first order the molecular gas supply
regulates the star formation distribution and in sharp contrast to the steep indices relating $\Sigma_{\rm SFR}$ to atomic gas \citet{BIGIEL08,
SCHRUBA11}. Second, the fact that these galaxy-to-galaxy variations wash out into Figure \ref{fig:combined} implies that while 
$\tau_{\rm dep}^{\rm mol}$ may correlate with $\Sigma_{\rm mol}$ within an individual galaxy, simply knowing
 $\Sigma_{\rm mol}$ at 1~kpc resolution with no other knowledge of local conditions or host galaxy does not allow one to 
 predict $\tau_{\rm dep}^{\rm mol}$ better than simply adopting a median $\tau_{\rm dep}^{\rm mol}$. That is, in the absence of 
 knowledge of other conditions, $\Sigma_{\rm mol}$ is not a good predictor of the molecular gas depletion time.

\subsection{Comparison to Literature Data}
\label{sec:lit}

\begin{figure*}[]
\plotone{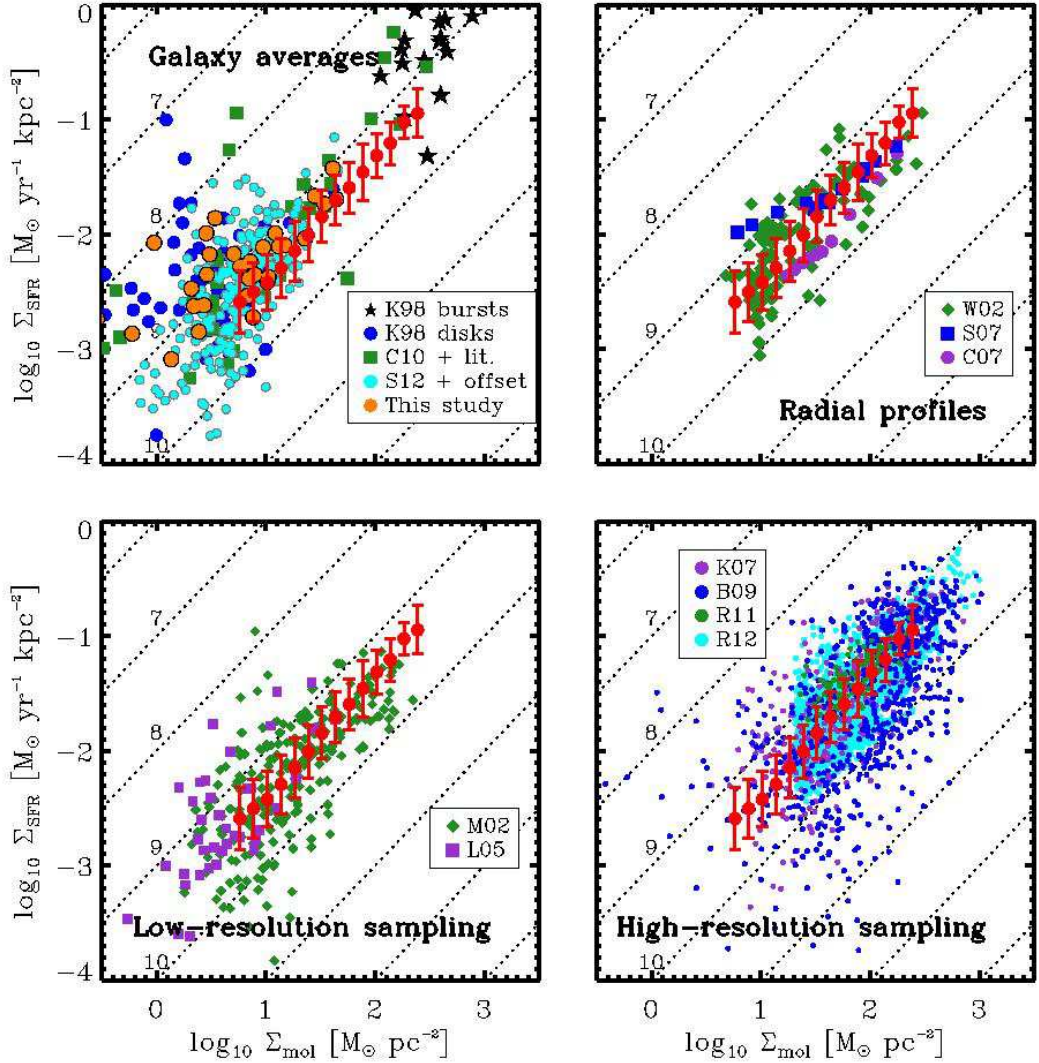}
\caption{\sigsfr\ versus \sightwo\ for our results and literature measurements. The type of literature measurement varies with panel:
the top left shows galaxy averages, the top right shows radial profiles, the lower left shows sparse, coarse ($\sim 1\arcmin$) 
sampling of galaxy disks, and the lower right shows fine sampling ($\lesssim 15\arcsec$). In all four panels, red points show a 
running median and standard deviation (error bars) for our kpc-scale measurements (Figure \ref{fig:combined}) using H$\alpha$+24$\mu$m 
and CO (2-1). Orange circles in the top left panel show integrated measurements for our data. Literature measurements are described in Section 
\ref{sec:data}; references: K98 --- \citet{KENNICUTT98B}, C10 + lit. --- \citet{CALZETTI10} combined with literature CO, 
S12 --- \citet{SAINTONGE12} with offset to bring SFR estimates into agreement (Section \ref{sec:param}),
W02 --- \citet{WONG02}, S07 --- \citet{SCHUSTER07}, C07 --- \citet{CROSTHWAITE07}, M02 --- \citet{MURGIA02}, 
L05 --- \citet{LEROY05}, K07 --- \citet{KENNICUTT07}, B09 --- \citet{BLANC09}, R11 --- \citet{RAHMAN11}, R12 --- 
\citet{RAHMAN12}.}
\label{fig:lit}
\end{figure*}

\begin{deluxetable}{lcc}
\tablecaption{Comparison to Literature} 
\tablehead{
\colhead{Galaxy} & 
\colhead{Median $\tau_{\rm dep}^{\rm mol}$} &
\colhead{Scatter} \\
\colhead{} & 
\colhead{[Gyr]} &
\colhead{[dex]}
}
\startdata
This study ... & \\
... weighting by measurement & 2.2 & 0.28 \\
... weighting by galaxy & 1.3 & 0.32 \\
Median of literature ... & \\
... weighting by measurement & 2.7\tablenotemark{a} & 0.36\\
... weighting by study & 2.0 & 0.23 \\
\hline
\multicolumn{2}{c}{Disk Averages} \\  
\hline
\citet{SAINTONGE12} & 0.7 & 0.37 \\
... offset SFR\tablenotemark{b} & 1.1 & 0.37 \\
\citet{CALZETTI10} + literature CO\tablenotemark{c} & 1.5 & 0.38 \\
\citet{KENNICUTT98A} ... & \\
... disks & 1.1 & 0.46 \\
... starbursts & 0.4 & 0.39 \\
\hline
\multicolumn{2}{c}{Radial Profiles} \\
\hline
\citet{SCHUSTER07} & 2.0 & 0.12\tablenotemark{d} \\
\citet{CROSTHWAITE07} & 4.4 & 0.06\tablenotemark{d} \\
\citet{WONG02} & 2.0 & 0.36 \\
\hline
\multicolumn{2}{c}{Low Resolution} \\
\hline
\citet{LEROY05} & 2.1 & 0.33 \\
\citet{MURGIA02} & 2.8 & 0.41 \\
\hline
\multicolumn{2}{c}{High Resolution} \\
\hline
\citet{RAHMAN12} & 2.9 & 0.37 \\
\citet{RAHMAN11} & 1.6 & 0.15\tablenotemark{d} \\
\citet{BLANC09} & 3.2 & 0.64\tablenotemark{d} \\
\citet{KENNICUTT07} & 2.2 & 0.37\tablenotemark{d}
\enddata
\tablecomments{Average molecular gas depletion time, in Gyr, for matched assumptions --- $\xco = 2 \times 10^{20}$~\xcounits , a Kroupa IMF, and including helium in the gas estimate. The left column gives the study, with the list broken down by sampling approach and the right column reports the median molecular gas depletion time in that study. Error bars report the $1\sigma$ scatter, in dex, for each study.}
\tablenotetext{a}{Dominated by \citet{RAHMAN12}. Without \citet{RAHMAN12} median is 2.1~Gyr.}
\tablenotetext{b}{SFR estimate offset to match our estimates (Section \ref{sec:param}). }
\tablenotetext{c}{CO from \citet{YOUNG95} and \citet{HELFER03}.}
\tablenotetext{d}{Study considered a singe galaxy. Others combine multiple galaxies.}
\label{tab:lit}
\end{deluxetable}

Many studies have assessed the relationship between gas and star formation in nearby galaxies 
(Section \ref{sec:intro}). Figure \ref{fig:lit} and Table \ref{tab:lit} compare our measurements to a 
compilation of these studies \citep[see Section \ref{sec:litdata} and][]{BIGIEL11}. We adjust each set of 
measurements to share our adopted CO-to-H$_2$ conversion factor and stellar initial mass function.

Table \ref{tab:lit} gives $\tau_{\rm dep}^{\rm mol}$ by study. These span from 0.4 for the \citet{KENNICUTT98A} 
starbursts to 4.4~Gyr for the study of NGC~6946 by \citet{CROSTHWAITE07}. Considering 
all measurements equally, the median literature $\tau_{\rm dep}^{\rm mol}$ is 2.7~Gyr, which drops to 2.1~Gyr if 
we exclude the large data set of \citet{RAHMAN12}, which otherwise dominates the statistics. Treating each
study as a single independent measurement, the median is $\tau_{\rm dep}^{\rm mol} = 2.0$~Gyr. These 
are in good agreement with the estimates of this study (Table \ref{tab:combined}) $\tau_{\rm dep}^{\rm mol} \approx 2.2$~Gyr 
weighting all measurements equally. The scatter among individual literature data is $\approx 0.36$~dex and from
study to study the scatter is $\approx 0.23$ dex.

Figure \ref{fig:lit} shows a more detailed comparison between our measurements and individual 
literature data. We separate the literature studies according to the scale sampled. The top left panel 
shows measurements where one point corresponds to one galaxy. The top right panel shows data from studies 
that measure azimuthally averaged $\Sigma_{\rm mol}$ and $\Sigma_{\rm SFR}$ in a series of  concentric tilted 
rings. The bottom left panel shows data for individual pointings with comparatively poor angular resolution, $\approx 40$--$60\arcsec$. 
The bottom right panel shows studies that obtain high-angular-resolution sampling of each target. In each panel 
we plot the running median and standard deviation for our data, binned by $\Sigma_{\rm mol}$, as red points.

The final three panels of Figure \ref{fig:lit} demonstrate excellent agreement between our data and 
previous studies that resolve the disks of galaxies \citep[see also][]{BIGIEL11}. This agreement may not be 
surprising given that our study shares targets with many of these literature studies, which also heavily overlap 
one another. Nonetheless, we show here that repeated measurements of the distribution of $\Sigma_{\rm SFR}$ 
and $\Sigma_{\rm mol}$ in the nearest star-forming spiral galaxies mostly cover the same part of parameter space 
regardless of exact methodology. Uncertainty in interpretation and fitting techniques have clouded this basic agreement 
in where the data lie. Given our basic approach to physical parameter estimation, there appears to be overall agreement 
for a typical $\tau_{\rm dep}^{\rm mol} = 1$--$3$~Gyr in local disk galaxies.

The first panel of the Figure \ref{fig:lit} looks qualitatively different from the other three. This panel 
shows galaxy-integrated measurements, so that one point is one galaxy. These tend to scatter from overlapping our 
data up to lower $\tau_{\rm dep}^{\rm mol}$ at low $\Sigma_{\rm mol}$ and comparatively high $\Sigma_{\rm SFR}$. This 
same effect appears in Table \ref{tab:lit} as low values of $\tau_{\rm dep}^{\rm mol}$ for studies that focus on 
measurements of whole galaxies. Our synthesis of literature CO and SFR measurements yields $\tau_{\rm dep}^{\rm mol} = 1.3$~Gyr 
while for the \citet{KENNICUTT98B} disk galaxies the median $\tau_{\rm dep}^{\rm mol} = 1.1$~Gyr. Treating our own 
sample as a set of integrated measurements we find a similar value, $\tau_{\rm dep}^{\rm mol} = 1.3$~Gyr. 

This disk-integrated $\tau_{\rm dep}^{\rm mol}$ is significantly shorter than the $\tau_{\rm dep}^{\rm mol}$ that 
we measure treating each point equally. We noted this effect in Section \ref{sec:combined}. It arises because 
weighting each galaxy equally emphasizes small, low-mass, low SFR galaxies relative to large, massive galaxies. These 
low mass galaxies have less physical area than large disks, so that they do not affect the ensemble of measurements much.  However 
these small galaxies do exhibit short apparent $\tau_{\rm dep}^{\rm mol}$ and when given equal weight they 
drive the median $\tau_{\rm dep}^{\rm mol}$ down by a factor of $\sim 2$. We explore this and other systematic variations
in $\tau_{\rm dep}^{\rm mol}$ in the second part of this paper.

\section{Systematic Second-Order Variations in $\tau_{\rm dep}^{\rm mol}$: Global Correlations, Efficient Galaxy Centers, and Correlated Scatter}
\label{sec:tdep_vary}

In Section \ref{sec:scaling} we demonstrate that our ensemble of data can be described to first order by a roughly linear relation between
$\Sigma_{\rm SFR}$ and $\Sigma_{\rm mol}$ with a slope corresponding to a typical depletion time $\tau_{\rm dep}^{\rm mol} \approx 2.2$~Gyr
with a factor of two scatter from line of sight to line of sight. However, we also show that the apparent uniformity of $\tau_{\rm dep}^{\rm mol}$ 
results at least partially from the emphasis that our approach places on the disks of large, star-forming galaxies. These contribute most of the 
area in our sample. When we apply weightings that emphasize small galaxies or the inner parts of galaxies, we observe departures from this simple
picture. 

In this section, we explore these variations. We examine the dependence of disk-average $\tau_{\rm dep}^{\rm mol}$ on integrated 
galaxy properties (\S \ref{sec:global}) and the dependence of $\tau_{\rm dep}^{\rm mol}$  on local physical conditions (\S \ref{sec:local}).
We highlight apparent variations in $\tau_{\rm dep}^{\rm mol}$ as a function of galaxy mass, metallicity, and dust-to-gas ratio and we discuss
the CO-to-H$_2$ conversion factor as a potential cause. In Section \ref{sec:centers} we contrast the central regions of our targets with the disks and
show evidence for a shift to more efficient star formation in galaxy centers, perhaps indicative of a transition between ``disk'' and ``starburst'' modes of star 
formation. Finally we examine the scale-dependence of scatter in $\tau_{\rm dep}^{\rm mol}$ (\S \ref{sec:scatter}) to show that undiagnosed 
systematic variations in $\tau_{\rm dep}^{\rm mol}$ persist in our data, reflecting either large-scale synchronization of star formation or 
real correlated efficiency variations spanning the disks of our targets.

\subsection{Galaxy-to-Galaxy Variations in $\tau_{\rm dep}^{\rm mol}$}
\label{sec:global}

\begin{figure*}[]
\epsscale{1.0}
\plotone{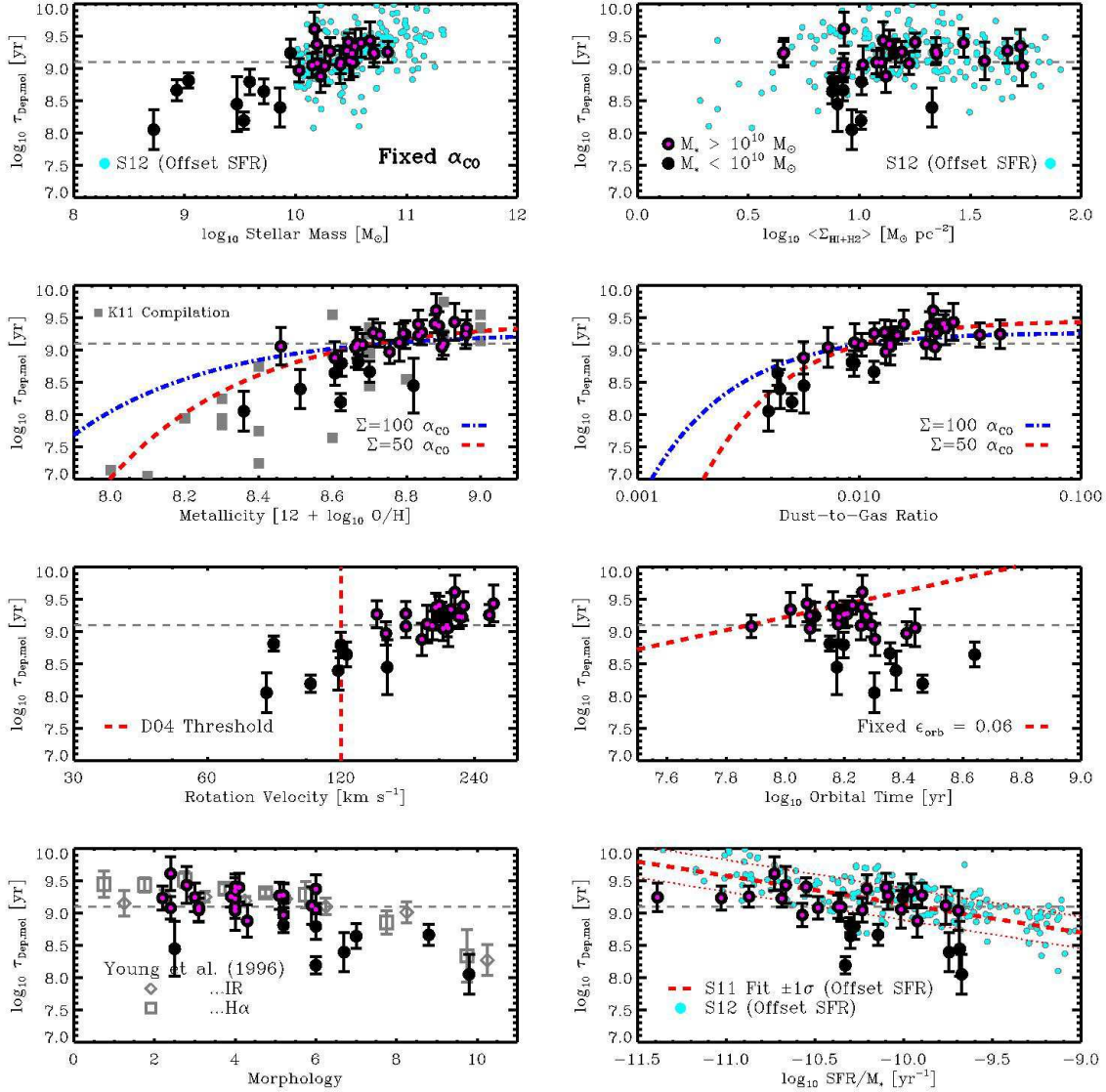}
\caption{Integrated $\tau_{\rm dep}^{\rm mol}$ ($\Sigma_{\rm mol}/\Sigma_{\rm SFR}$) within 
$0.75~r_{25}$ and standard deviation (error bars) in $\tau_{\rm Dep}^{\rm mol}$ over individual lines of sight 
plotted against integrated galaxy properties: stellar mass, average gas mass surface 
density, metallicity, dust-to-gas mass ratio, rotation velocity at $r_{25}$, orbital period at 0.4~$r_{25}$, morphology, 
and specific star formation rate ($\left<\Sigma_{\rm SFR} \right> /\left< \Sigma_* \right>$). Literature data appear 
in gray \citep[Y96, K11, and S12 refer to ][]{YOUNG96,KRUMHOLZ11,SAINTONGE12} and fits or predictions 
appear as curves: \citet{WOLFIRE10} conversion factors for $\Sigma = 50$ and $100$~M$_\odot$~pc$^{-2}$, the 
rotation-velocity threshold associated with the suppression of dust lanes by \citet[][D04]{DALCANTON04}, a fixed 
efficiency per orbital period, and the $\tau_{\rm dep}^{\rm mol}$ vs SSFR fit by \citet[][S11]
{SAINTONGE11B}. Low mass, metal poor, late type galaxies show significantly lower $\tau_{\rm dep}^{\rm mol}$ (high SFR-to-CO ratios) than 
their massive, early type counterparts, but these trends persist with smaller dynamic range even among massive 
galaxies (we mark galaxies with $M_* > 10^{10}$ M$_\odot$ using magenta dots in all panels).}
\label{fig:tdep_global}
\end{figure*}

\begin{deluxetable}{lccc}
\tablecaption{Rank Correlation of $\tau_{\rm dep}^{\rm mol}$ with Galaxy-Average Properties} 
\tablehead{
\colhead{Quantity} & 
\colhead{Fixed $\alpha_{\rm CO}$} & 
\colhead{$\Sigma = 100$ $\alpha_{\rm CO}$} & 
\colhead{$\Sigma = 50$ $\alpha_{\rm CO}$} \\
}
\startdata
Stellar mass & $+0.72 \pm 0.18$ & $+0.47 \pm 0.19$ & $-0.07 \pm 0.19$ \\
... M$_* > 10^{10}$~M$_\odot$ & $-0.37 \pm 0.20$ & $-0.15 \pm 0.24$ &$+0.02 \pm 0.22$ \\
\\
$\left< \Sigma_{\rm HI + H2}\right>$ & $+0.40 \pm 0.19$ & $+0.64 \pm 0.18$ & $+0.49 \pm 0.20$ \\
... M$_* > 10^{10}$~M$_\odot$ & $+0.19 \pm 0.20$ & $+0.60 \pm 0.22$ & $+0.60 \pm 0.25$ \\
\\
Metallicity & $+0.73 \pm 0.21$ & $+0.44 \pm 0.19$ & $-0.14 \pm 0.19$ \\
... M$_* > 10^{10}$~M$_\odot$ & $+0.59 \pm 0.20$ & $0.16 \pm 0.23$ & $-0.21 \pm 0.22$ \\
\\
Dust-to-gas ratio & $+0.81 \pm 0.19$ & $+0.28 \pm 0.19$ & $-0.49 \pm 0.19$ \\
... M$_* > 10^{10}$~M$_\odot$ & $+0.57 \pm 0.20$ & $-0.24 \pm 0.22$ & $-0.72 \pm 0.23$ \\
\\
Rotation velocity & $+0.74 \pm 0.20$ & $+0.47 \pm 0.20$ & $-0.11 \pm 0.23$ \\
... M$_* > 10^{10}$~M$_\odot$ & $-0.40 \pm 0.24$ & $+0.02 \pm 0.026$ & $-0.24 \pm 0.23$ \\
\\
Orbital time & $-0.45 \pm 0.21$ & $-0.36 \pm 0.19$ & $-0.12 \pm 0.20$ \\
... M$_* > 10^{10}$~M$_\odot$ & $-0.32 \pm 0.24$ & $-0.21 \pm 0.23$ & $-0.03 \pm 0.22$ \\
\\
Morphology & $-0.49 \pm 0.19$ & $-0.30 \pm 0.19$ & $-0.03 \pm 0.20$ \\
... M$_* > 10^{10}$~M$_\odot$ & $-0.16 \pm 0.23$ & $+0.03 \pm 0.23$ & $+0.16 \pm 0.22$ \\
\\
SFR/M$_*$ & $-0.38 \pm 0.19$ & $-0.06 \pm 0.19$ & $0.32 \pm 0.19$ \\ 
... M$_* > 10^{10}$~M$_\odot$ & $-0.27 \pm 0.22$ & $+0.16\pm 0.23$ & $0.34 \pm 0.25$ \\
\enddata
\tablecomments{Rank correlation coefficient relating the molecular gas depletion time, $\tau_{\rm
Dep}^{\rm mol}$ to galaxy-average properties. Quoted uncertainties report the $1\sigma$ scatter of the correlation coefficient about 0 obtained by randomly repairing the
data. The columns give results for different assumptions about the CO-to-H$_2$ conversion factor.
We report results for all galaxies and only high mass galaxies, $M_* > 10^{10}$~M$_\odot$. {\em The average molecular gas depletion time is 
strongly covariant with galaxy average properties. This covariance can be reduced but not removed by 
the application of a D/G-dependent CO-to-H$_2$ conversion factor.}}
\label{tab:tdep_global}
\end{deluxetable}

\begin{figure*}[]
\epsscale{1.0}
\plotone{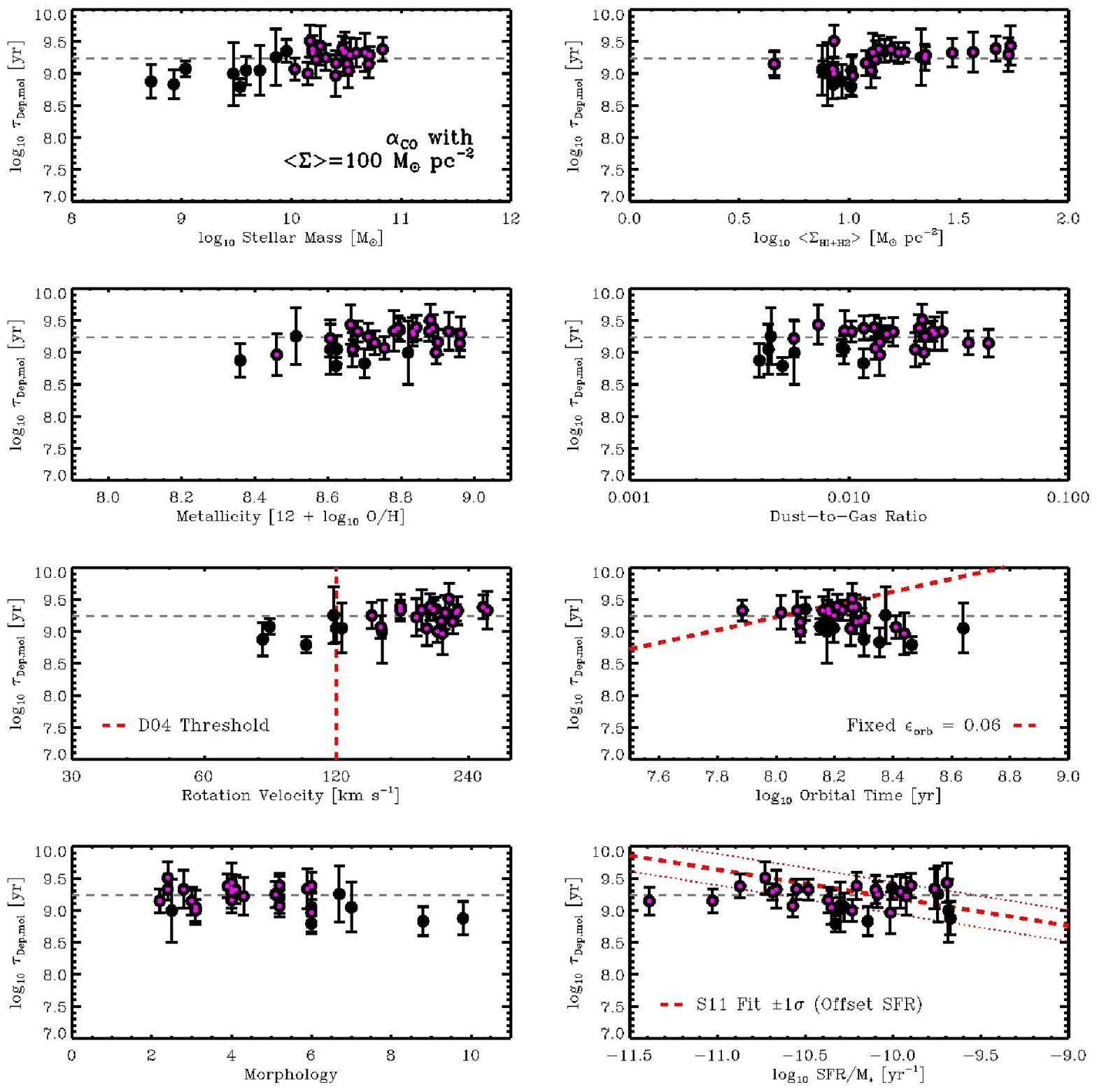}
\caption{Same as Figure \ref{fig:tdep_global} but {\em after} application of our variable ``$
\Sigma=100$'' CO-to-H$_2$ conversion factor. Most of the galaxy-scale variations in $\tau_{\rm dep}^{\rm mol}$ visible 
in Figure \ref{fig:tdep_global} can largely be explained by  this dust-to-gas ratio-
dependent conversion factor, though significant galaxy-to-galaxy scatter remains. A dashed horizontal line in each panel shows 
the median $\tau_{\rm dep}^{\rm mol}$ across our sample and we reproduce several comparison lines from Figure \ref{fig:tdep_global}.}
\label{fig:tdep_global_corr}
\end{figure*}

Our comparison with the literature reveals the same effect that we saw in Section \ref{sec:combined}: 
when we weight by host galaxy instead of by individual kpc-scale measurement, clear variations
in $\tau_{\rm dep}^{\rm mol}$ become visible. In particular, a large population of short $\tau_{\rm dep}^{\rm mol}$
data emerges in the bottom left panel of Figure \ref{fig:combined} and the top left panel of Figure \ref{fig:lit}. 
In Figure \ref{fig:tdep_global} we investigate the physical origin of these differences. Adopting a fixed
$\alpha_{\rm CO}$, we plot $\tau_{\rm dep}^{\rm mol}$ for each galaxy in our sample as a function of a series 
of galaxy properties: stellar mass, disk-average total (\hi\ + H$_2$) gas surface density, metallicity, dust-to-gas ratio, rotation velocity, 
average orbital time (weighted by $\Sigma_{\rm mol}$), morphological type, and specific star 
formation rate (sSFR=$SFR/M_*$). $\tau_{\rm dep}^{\rm mol}$ correlates with many of these quantities 
with the overall sense that {\em low mass, low metallicity, late-type, high sSFR galaxies exhibit 
shorter apparent molecular gas depletion times than massive, metal-rich, earlier type disk galaxies.} 

Such trends have been observed in various ways before. In Figure \ref{fig:tdep_global}, gray points 
show the approximate $\tau_{\rm dep}^{\rm mol}$ as a function of morphology as inferred by \citet{YOUNG96} 
from comparing the FCRAO Extragalactic CO Survey with H$\alpha$ (squares) and IR (diamonds) data. 
They found later-type galaxies to exhibit mildly enhanced star formation efficiencies. We also plot $\tau_{\rm dep}^{\rm mol}$ 
derived by \citet{KRUMHOLZ11} combining literature SFR and CO data (including HERACLES measurements), who
showed a strong trend of SFR/CO decreasing with increasing metallicity \citep[see also][]{SCHRUBA12}. Both literature trends agree well with our own data
and here we present the first direct comparison to the dust-to-gas ratio, the quantity thought to link
metallicity to both CO emission and the star formation process \citep[e.g.,][]{WOLFIRE10,KRUMHOLZ11}.

\citet{SAINTONGE11B} and \citet{SAINTONGE12} found a clear relation between $\tau_{\rm dep}^{\rm mol}$ and
$M_*$ and identified sSFR as the global galaxy property most directly linked to $\tau_{\rm dep}^{\rm mol}$. 
We plot their data, corrected to bring our SFR estimates into agreement in the top row and bottom right panel.
In the bottom right panel,  we plot their fit of $\tau_{\rm dep}^{\rm mol}$, corrected for different treatments of helium.
We also  indicate the rotation velocity threshold below which \citet{DALCANTON04} noted the disappearance of 
prominent dust lanes in edge-on disk galaxies. The existence of such dust lanes 
should indicate the presence of a dusty, dense ISM of the type that might host molecular material and star formation.
Though \citet{DALCANTON04} make no clear prediction for $\tau_{\rm dep}^{\rm mol}$ we note that there is a generally
lower $\tau_{\rm dep}^{\rm mol}$ below the threshold than above; at the very least, the lack of abundant dust shielding may
drive both observations. This contrasts with the finding of \citet{WATSON12} who do not find such a trend with circular velocity.

Table \ref{tab:tdep_global} quantifies the relationship between $\tau_{\rm dep}^{\rm mol}$ and these other quantities using 
the Spearman rank correlation coefficient. $\tau_{\rm dep}^{\rm mol}$ correlates with stellar mass, metallicity, dust-to-gas ratio,
and rotation velocity at $>3\sigma$ significance. Weaker ($\sim 2\sigma$) correlations link $\tau_{\rm dep}^{\rm mol}$ to gas surface density, 
orbital time, morphology, and specific star formation rate. The trends with stellar mass and rotation velocity appear continuous, without clear 
thresholds. Indeed they still emerge, though at lower significance, even if we consider only high mass galaxies. 

The sense of the correlation with gas surface density is that we find the shortest depletion times for galaxies below 
$\left<\Sigma_{\rm HI+H2}\right> \sim 10$~M$_\odot$~pc$^{-2}$. These systems still yield the lowest $\tau_{\rm dep}^{\rm mol}$ regardless
of our treatment of $\alpha_{\rm CO}$ (see below). In our sample, these low-$\left<\Sigma_{\rm HI+H2}\right>$ hold a more of their
gas in the atomic phase than systems with high $\left<\Sigma_{\rm HI+H2}\right>$. Therefore, the trend we see amounts to longer 
depletion times in gas-rich, molecule dominated systems. However, note that our high $\left< \Sigma_{\rm HI+H2} \right>$ systems are gas-rich spirals. Our 
sample does not include truly extreme systems like the local U/LIRGs. Many of these systems have high gas surface densities but very {\em low} $\tau_{\rm dep}^{\rm mol}$ 
\citep[e.g.,][]{SANDERS96}, so that if they were added to Figure \ref{fig:tdep_global}, they would not extend the trend that we
see. We do see similar efficient starbursts {\em within} our targets (\S \ref{sec:centers}), but the effect remains mostly confined to the central parts of galaxies 
and does not propagate to these galaxy-scale measurements. This behavior may be somewhat evident in the \citet{SAINTONGE12} sample, but not our data.

The relation between sSFR and $\tau_{\rm dep}^{\rm mol}$ does not appear as strong as other trends in our sample, contrary to the 
finding of \citet{SAINTONGE11B}. This may simply reflect our sample's lack  of massive early type galaxies with low sSFR or U/LIRGs with 
high sSFR. Where we do overlap \citet{SAINTONGE12}, the agreement between the samples appears very good after our adjustment for different 
approaches to SFR estimation (\S \ref{sec:data}). The most substantive observational disagreement between our results and those of \citet{SAINTONGE11B} is that we 
find $\tau_{\rm dep}^{\rm mol}$ to correlate with metallicity and dust-to-gas ratio, even for relatively high mass galaxies, while their tests revealed no such correlations.

{\em Dust-to-Gas Ratio and Conversion Factor:} We emphasize that the correlations in Figure \ref{fig:tdep_global} relate {\em apparent} $\tau_{\rm  dep}^{\rm mol}$ to 
integrated galaxy properties. The second row of Figure \ref{fig:tdep_global} shows strong trends in $\tau_{\rm dep}^{\rm mol}$ as a function of metallicity and dust-to-gas ratio and
these trends raise a fundamental issue regarding the interpretation of Figure \ref{fig:tdep_global}. There is good evidence that the 
CO-to-H$_2$ conversion factor increases with decreasing metallicity and dust-to-gas ratio \citep[see references in][]{ISRAEL97,LEROY11,BOLATTO11}. 
Metallicity and dust-to-gas ratio vary with the stellar and dynamical mass of a galaxy, so that the correlations in Figure \ref{fig:tdep_global} and
Table \ref{tab:tdep_global} may reflect either true variation in the efficiency with which molecular gas forms stars, variations in
the CO-to-H$_2$ conversion factor, or a mixture of the two.

The blue and red curves in the second row show the dependence of $\alpha_{\rm CO}$ on metallicity or dust-to-gas ratio for 
the ``$\Sigma=100$'' (blue) and ``$\Sigma=50$'' (red) models described in Section \ref{sec:data} assuming a fixed molecular gas depletion time. 
The third and fourth columns in Table \ref{tab:tdep_global} report the correlations between $\tau_{\rm dep}^{\rm mol}$ with global properties after 
application of these conversion factors\footnote{Note that the conversion factors are calculated from the dust-to-gas ratio profiles and then applied to 
the CO maps to derive the integrated H$_2$ masses and that these conversion factors include central $\alpha_{\rm CO}$ depressions, so that the mapping 
of conversion factor to the average dust-to-gas ratio is not perfect.}. The red curve, representing the more extreme assumption that CO emission arises from
low surface density, $\Sigma=50$~M$_\odot$~pc$^{-2}$, clouds fits the dust-to-gas ratio trend well. Neither trend is steep enough
to fit the metallicity data (left panel, second row), but given the large uncertainties in the metallicities \citep[often several 0.1s of a dex, e.g.,][]{MOUSTAKAS10}, 
the ``$\Sigma=50$'' curve is not a bad match. 

Comparison of these two curves and the second row of Figure \ref{fig:tdep_global} highlights several technical points 
also found in the recent literature. First, the conversion factor needed to impose a fixed $\tau_{\rm dep}^{\rm mol}$
depends more sharply on metallicity than the predictions from many theoretical models and some observational determinations 
\citep[see plots in][]{KRUMHOLZ11,SCHRUBA12,GENZEL12}. Second the left and right panels of the second row do
not perfectly agree. The dust-to-gas ratio derived from {\em Spitzer} observations of low metallicity regions appears somewhat 
lower than one would predict from a linear scaling of their metallicity \citep[][]{MUNOZMATEOS09B}. Because 
both the observed and predicted trends flatten at high D/G and high metallicities, the zero points (solar values) used to compute
the theoretical curves might be adjusted to yield better agreement between metallicity and D/G or to better match one theoretical curve
or the other to the data. That is, given the uncertainties in the absolute measurements of both metallicity and D/G, one can 
``slide'' the relative positions of the theoretical curves and the D/G and metallicity determinations left and right. As a result,
the differences between the ``$\Sigma=50$'' and ``$\Sigma=100$'' curves should not be overemphasized.

These details aside, the key questions are: to what degree can conversion factor variations explain the global trends seen
in Figure \ref{fig:tdep_global} and are the adopted conversion factors reasonable? In Figure \ref{fig:tdep_global_corr} we repeat 
Figure \ref{fig:tdep_global} but adopt our ``$\Sigma=100$'' conversion factor rather than fixed $\alpha_{\rm CO}$. Many correlations
of $\tau_{\rm dep}^{\rm mol}$ with local properties appear weaker with the ``$\Sigma=100$'' $\alpha_{\rm CO}$ than with a fixed conversion 
factor. The more extreme ``$\Sigma=50$'' case removes almost all variation with stellar mass and metallicity, but reverses the trend
with dust-to-gas ratio. Thus, the ``$\Sigma=50$'' may represent the simplest single explanation for the bulk of the variation in Figure \ref{fig:tdep_global}.
``$\Sigma=100$'' corresponds to a conservative $\alpha_{\rm CO}$ estimate that reflects present conventional wisdom. It lessens the strength of many
correlations but (marginally) significant variations in $\tau_{\rm dep}^{\rm mol}$ with other properties still exist when using this conversion factor.

\subsubsection{Conversion Factors and ISM Structure} 

Both of our conversion factors assume a simple, universal structure for the molecular ISM beneath our kpc resolution. In actuality, the typical surface density of 
GMCs, the universality of this value, and the balance of GMCs and more diffuse molecular 
material remain poorly constrained by observations. Much previous work places typical surface densities of Milky Way GMCs at $\sim 150$~M$_\odot$~pc$^{-2}$ 
\citep{SOLOMON87,ROMANDUVAL10B}. High resolution observations of the inner parts of nearby spirals (NGC~5194, NGC~6946) yield similar surface 
densities \citep[][Hughes et al., in review; Colombo et al., submitted]{DONOVANMEYER12}. Other studies find lower surface densities, even in the same 
systems: e.g., studies of the Milky Way ring by \citet{HEYER09}, the disk of NGC~6946 \citep{REBOLLEDO12}, the LMC \citep{FUKUI08,HUGHES10,WONG11}, or
a sample of Local Group galaxies \citep{BOLATTO08} all find typical GMC surface densities $\Sigma \sim 50$--$100$~M$_\odot$~pc$^{-2}$. Some of
these differences may be attributed to methodology, but Hughes et al. (in review) decisively demonstrate that the structure of the molecular ISM does vary 
substantially with environment via a carefully matched comparison M51, M33, and the LMC. They show substantial differences in the probability distribution
function of CO emission at $\sim 50$~pc resolution both within and among these galaxies.

Given these uncertainties, the best way to read the ``$\Sigma=50$'' conversion factor is a case where $\alpha_{\rm CO}$ depends sharply
on the dust-to-gas ratio because most CO emission to come from low-extinction lines of sight, $A_V \sim 1$--$2$ at Milky Way dust-to-gas ratios. In addition 
to the mixed evidence on surface densities of whole populations, maps of very local clouds \citep[as in][]{HEIDERMAN10,LADA10} do tend to find significant mass 
in low-A$_V$ components. Furthermore, the contribution of diffuse sight lines \citep[e.g., those studied by][]{LISZT10} to the integrated CO emission from a large part of 
a galaxy remains poorly known, but may easily be several 10s of percent in the Milky Way (H. Liszt, private communication).

A detailed investigation of GMC structure is beyond the scope of this paper. The key points are the following: (1) the basic structure of the molecular ISM remains
uncertain at a level that significantly affects our interpretation of integrated CO emission, (2) the structure of the molecular ISM does vary substantially with environment,
and (3) observations do appear to admit the possibility of substantial CO emission from low-A$_V$ lines of sight or unbound clouds. Given these caveats, we must view
our adopted conversion factors as more schematic than exact, so that subtle differences between the ``$\Sigma=50$'' and ``$\Sigma=100$'' case should not be over-interpreted.
We can say from Table \ref{tab:tdep_global} and Figure \ref{fig:tdep_global} and \ref{fig:tdep_global_corr} that the conversion factor can explain many of the largest
systematic variations in $\tau_{\rm dep}^{\rm mol}$ across our data, but probably not all of them. High mass, high surface density galaxies probably are less efficient at
forming stars from their molecular reservoir (i.e., high $\tau_{\rm dep}^{\rm mol}$) than low mass galaxies.

\subsubsection{Low Efficiency (High $\tau_{\rm dep}^{\rm mol}$) in High Mass Galaxies?} 

Our best guess is that weak correlation exists relating $\tau_{\rm dep}^{\rm mol}$ to galaxy mass and average surface density, reflecting a factor of $\sim 2$ increase 
in $\tau_{\rm dep}^{\rm mol}$ moving from low ($M_* < 10^{10}$~M$_\odot$) to high ($M_* > 10^{10}$~M$_\odot$) mass galaxies. Several natural explanations 
exist for such trends and a combination of these are almost certainly at play. Suppression of star formation in molecular gas by dynamical effects appears evident 
from high resolution observations of M51 (Meidt et al. submitted) and in some barred spiral galaxies \citep[see discussion in][]{JOGEE05}. These large-scale dynamical 
effects may suppress the ability of molecular gas to collapse into gravitationally bound, star-forming clouds. 

This represents a subset of a more general effect: in regions of high pressure and high gas surface density, the ISM becomes increasingly molecular
\citep[][L08]{BLITZ06}. A diffuse molecular ISM, meaning unbound material and the low density outskirts of bound clouds, may represent a reservoir of non-star-forming molecular
gas \citep[though note that even the bulk of bound, molecular material does not directly participate in star-formation; see references in][]{LADA12}. In low mass, low molecular
fraction galaxies such gas, if present, will often be in the atomic phase. Higher mass galaxies will have both a higher molecular fraction and stronger dynamical effects such as shear
and streaming motions working at suppressing the creation of bound gas. See \citet[][]{SAINTONGE12} for discussion of analogous effects.

This schematic picture leads to several predictions and is clouded by several complications. Both a prediction and a complication is that in this sketch, the physical state of molecular gas
varies systematically as a function of galaxy mass. This should manifest via several direction observables such as the CO-to-HCN or CO-to-HCO+ ratio, which will trace the
fraction of dense molecular gas, the ratio of $^{13}$CO-to-$^{12}$CO, tracing the optical depth of the gas, and the ratio of CO rotational transitions, which trace a complex
combination of density, temperature, and deviation from local thermodynamic equilibrium. An immediate complication from such variations is that changes in
internal conditions propagate to $\alpha_{\rm CO}$ variations that are distinct from the dust-shielding effects accounted in our adopted conversion factors \citep[e.g., see
discussion in][]{MALONEY88}. A second, weak prediction would be a general increase in $\tau_{\rm dep}^{\rm mol}$ with increasing molecular fraction, at least up to a certain
extent --- starbursts exhibit both low $\tau_{\rm dep}^{\rm mol}$ and high molecular fractions. Systematic trends in $\tau_{\rm dep}^{\rm mol}$ with molecular fraction 
were not immediately clear in L08; we search for them in our present data set in the next section. A similar mixed picture arises from dynamical effects: one might expect
to see high molecular fractions but low efficiencies in regions where shear or streaming motions suppress bound cloud formation. However, \citet{FOYLE10} found little
or no evidence for $\tau_{\rm dep}^{\rm mol}$ variations between spiral arms and the surrounding material. Again, a similar case of competing effects comes into play,
given that spiral arms are also invoked as a mechanism to collect inefficient, low $\tau_{\rm dep}^{\rm mol}$ gas into bound, high efficiency objects \citep[e.g.,][]{TAN00,KODA09}. 
Similarly, bars may both suppress inflowing gas and concentrate material into nuclear starbursts \citep[e.g.,][]{SAKAMOTO99,JOGEE05,SHETH05}.

Though competing effects cloud a simple interpretation of the data, the path forward here remains relatively clear: systematic measurements of the internal 
conditions in the molecular gas represent a critical next step and our knowledge of the CO-to-H$_2$ conversion factor must improve to further refine our 
understanding of how $\tau_{\rm dep}^{\rm mol}$ depends on physical conditions. More immediately, we need to understand which local conditions drive 
galaxy-averaged trends seen in this section. Finally, we emphasize that while these details are essential to a complete understanding of star formation in galaxies, 
they appear less critical to explain Figure \ref{fig:tdep_global} than the CO-to-H$_2$ conversion factor and represent, in some sense, a factor of $\sim 2$ level correction
to a constant $\tau_{\rm dep}^{\rm mol}$. 

\subsection{Local Variations in $\tau_{\rm dep}^{\rm mol}$}
\label{sec:local}

\begin{deluxetable*}{lccc}
\tablecaption{Rank Correlation of $\tau_{\rm dep}^{\rm mol}$ with Local Conditions} 
\tablehead{
\colhead{Quantity} & 
\colhead{Fixed $\alpha_{\rm CO}$} & 
\colhead{$\Sigma = 100$ $\alpha_{\rm CO}$} & 
\colhead{$\Sigma = 50$ $\alpha_{\rm CO}$} \\
}
\startdata
$\Sigma_*$ & $-0.03 \pm 0.02$ & $-0.01 \pm 0.02$ & $-0.08 \pm 0.02$ \\
$\Sigma_{\rm mol}$ & $-0.06 \pm 0.02$ & $-0.04 \pm 0.02$ & $+0.15 \pm 0.02$ \\
$f_{\rm mol} = \Sigma_{\rm mol} / (\Sigma_{\rm mol} + \Sigma_{\rm HI})$ & $+0.04 \pm 0.03$ & $+0.01 \pm 0.02$ & $+0.04 \pm 0.02$ \\
Dust-to-gas ratio & $+0.39 \pm 0.02$ & $+0.09 \pm 0.02$ & $-0.23 \pm 0.02$ \\
Orbital Time & $-0.05 \pm 0.02$ & $+0.07 \pm 0.02$ & $+0.12 \pm 0.02$\\
Galactocentric Radius ($r/r_{25}$) & $-0.05 \pm 0.02$ & $+0.09 \pm 0.02$ & $+0.13 \pm 0.02$ \\
\enddata
\tablecomments{Rank correlation between $\tau_{\rm dep}^{\rm mol}$ for individual kpc-resolution lines of sight
and local conditions for regions where we are reasonably complete. Uncertainties give the $1\sigma$ scatter
of the rank correlation about zero obtained when randomly repairing the data and accounting for an oversampling 
factor of $4$. The three columns report results for different assumed CO-to-H$_2$ conversion factors (\S \ref{sec:data}).
We report results only for the range over which $\lesssim 30\%$ of our data are limits. This is approximately:
$\Sigma_* > 40$~M$_\odot$~pc$^{-2}$, $\Sigma_{mol} > 4$~M$_\odot$~pc$^{-2}$, $f_{\rm mol} > 0.5$, 
$D/G > 0.006$, $\tau_{\rm orb} < 0.23$~Gyr, and $r_{\rm gal} < 0.55~r_{25}$. A strong correlation with the dust-to-gas
ratio is evident for fixed $\alpha_{\rm CO}$. For a variable $\alpha_{\rm CO}$ weak correlations.}
\label{tab:tdep_local}
\end{deluxetable*}

\begin{figure*}[]
\plotone{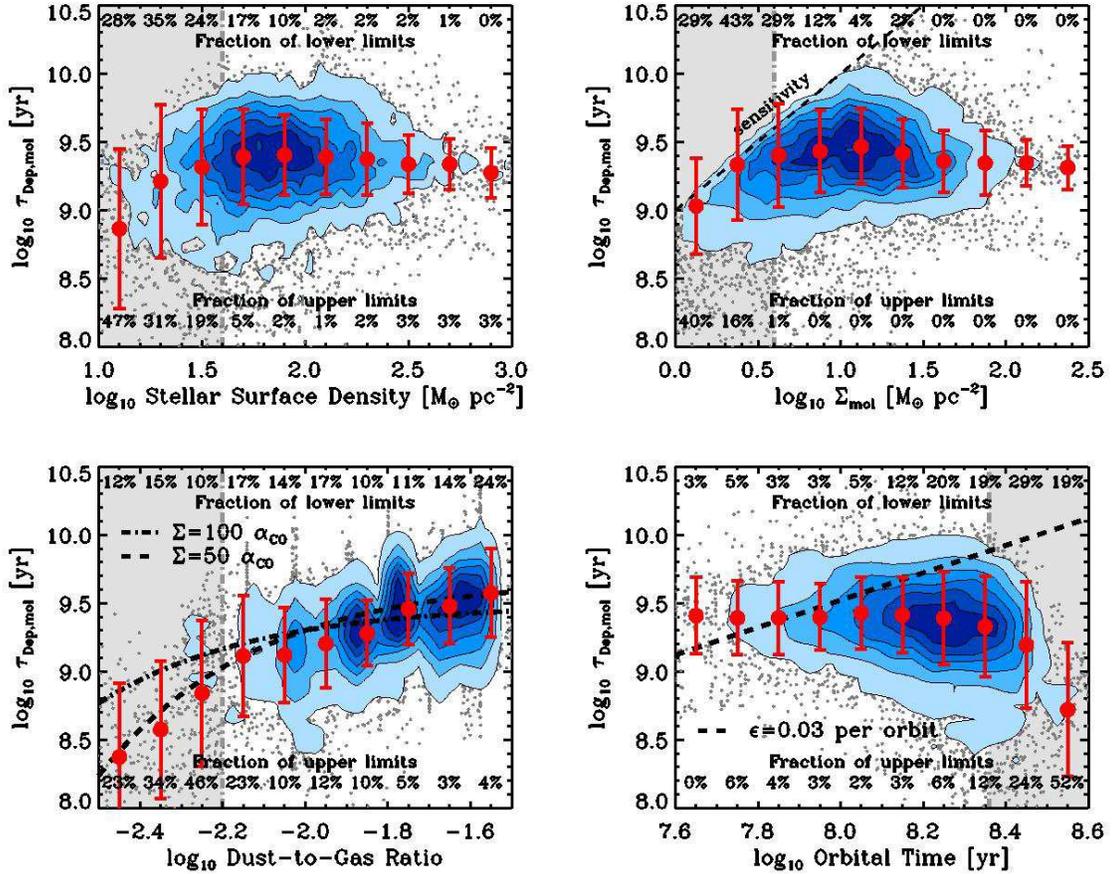}
\caption{Local $\tau_{\rm dep}^{\rm mol}=\Sigma_{\rm mol}/\Sigma_{\rm SFR}$ for fixed $\alpha_{\rm CO}$
as a function stellar mass surface density (top left), H$_2$ surface density (top right), dust-to-gas ratio (bottom left), and orbital 
period (bottom right) for a fixed $\alpha_{\rm CO}$. Gray points show individual lines of 
sight and shaded blue contours show density of data. Red points show the median (including relevant 
upper and lower limits) and standard deviation in $\log_{10} \tau_{\rm dep}^{\rm mol}$ binned by the $x$-axis.
Percentages along the top and bottom indicate the fraction of lower and upper limits in each bin. Gray
regions indicate where $\gtrsim 30\%$ of our $\tau_{\rm dep}^{\rm mol}$ estimates are limits so that completeness
represents a serious concern. The dashed curve in the lower left panel shows the expected relation 
for an $\alpha_{\rm CO}$-dependent $D/G$ and the dashed line in the lower right panel shows the expectation 
for a fixed fraction of molecular converted to stars per orbital period.}
\label{fig:tdep_local}
\end{figure*}

\begin{figure*}[]
\plotone{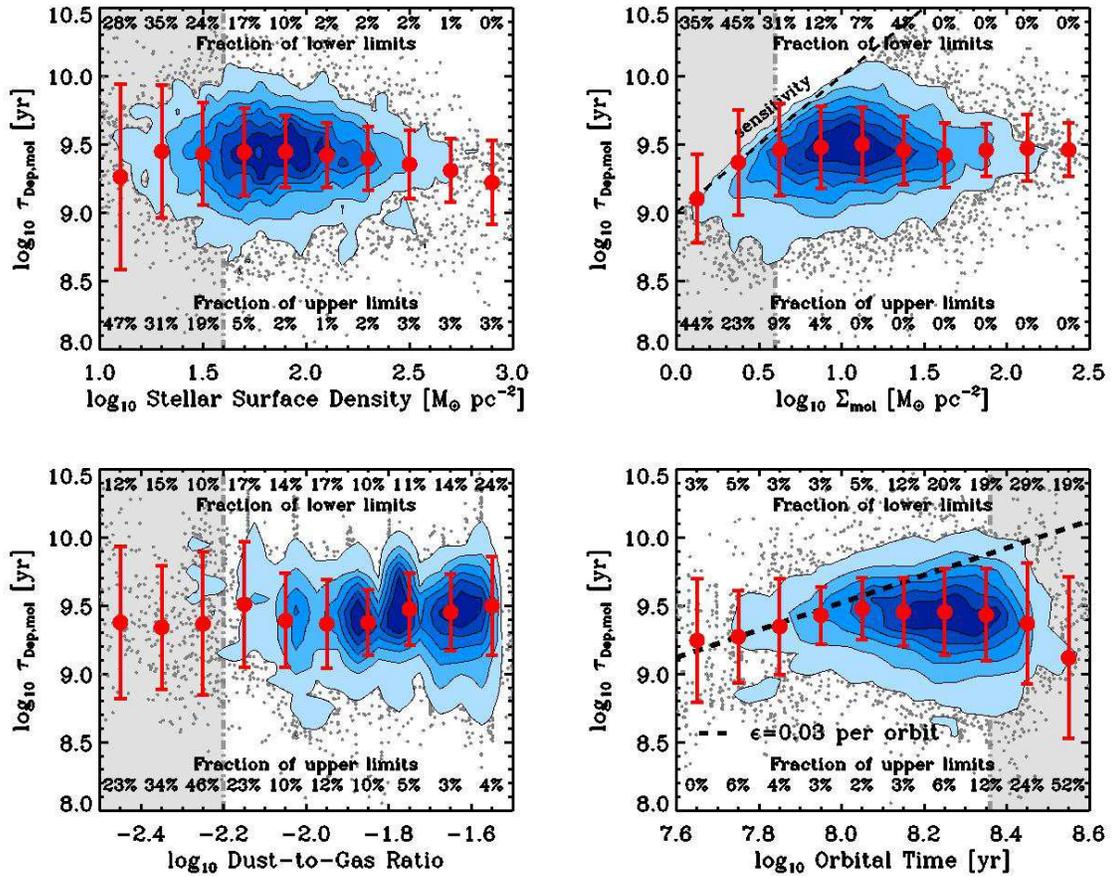}
\caption{As Figure \ref{fig:tdep_local} but {\em after} application of our ``$\Sigma=100$''  conversion 
factor following \citet[][Equation 
\ref{eq:alphaco}]{WOLFIRE10}. The conversion factor correction removes the strongest visible 
trend in the data.}
\label{fig:tdep_local_corr}
\end{figure*}

\begin{figure*}[]
\plottwo{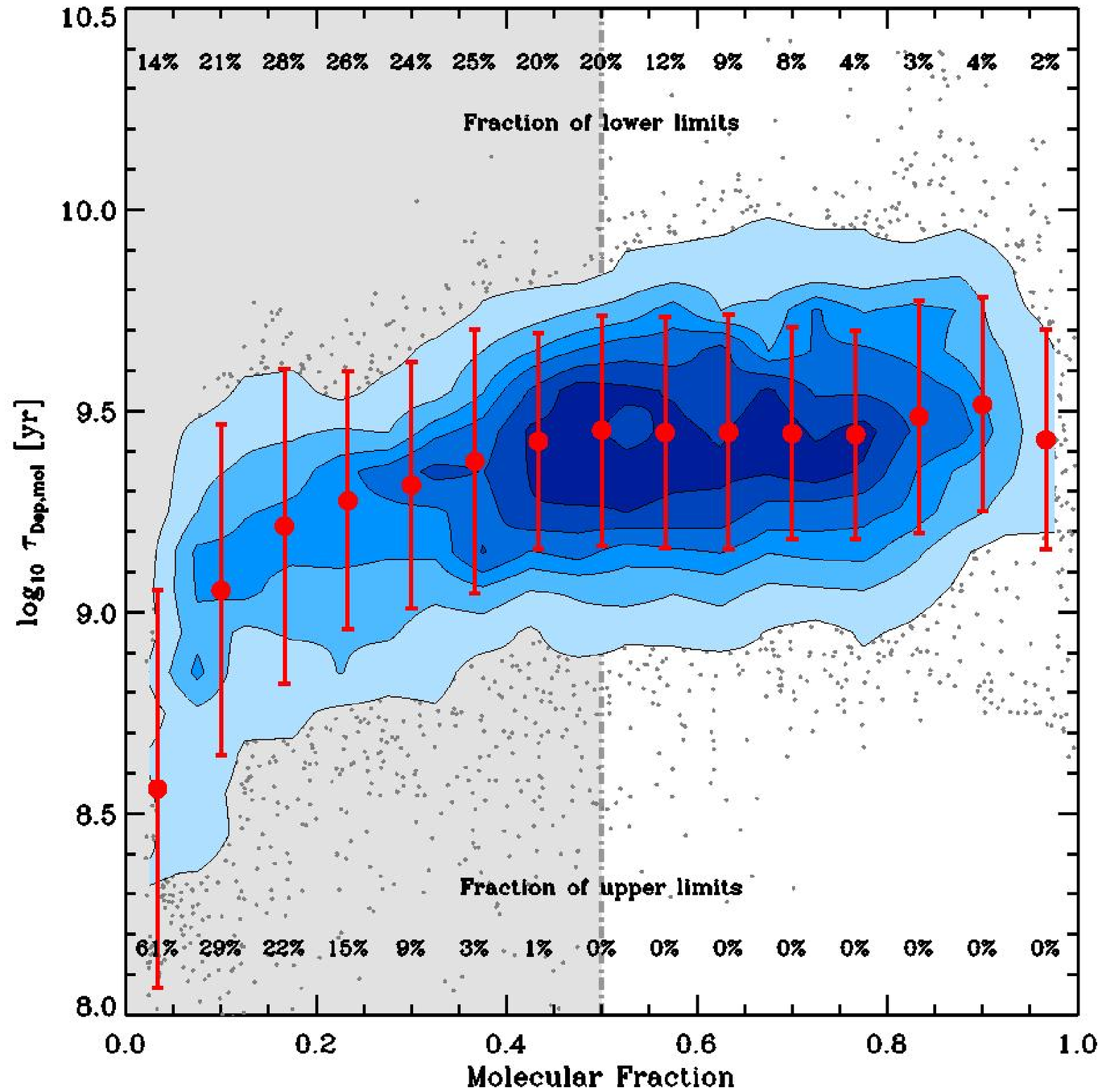}{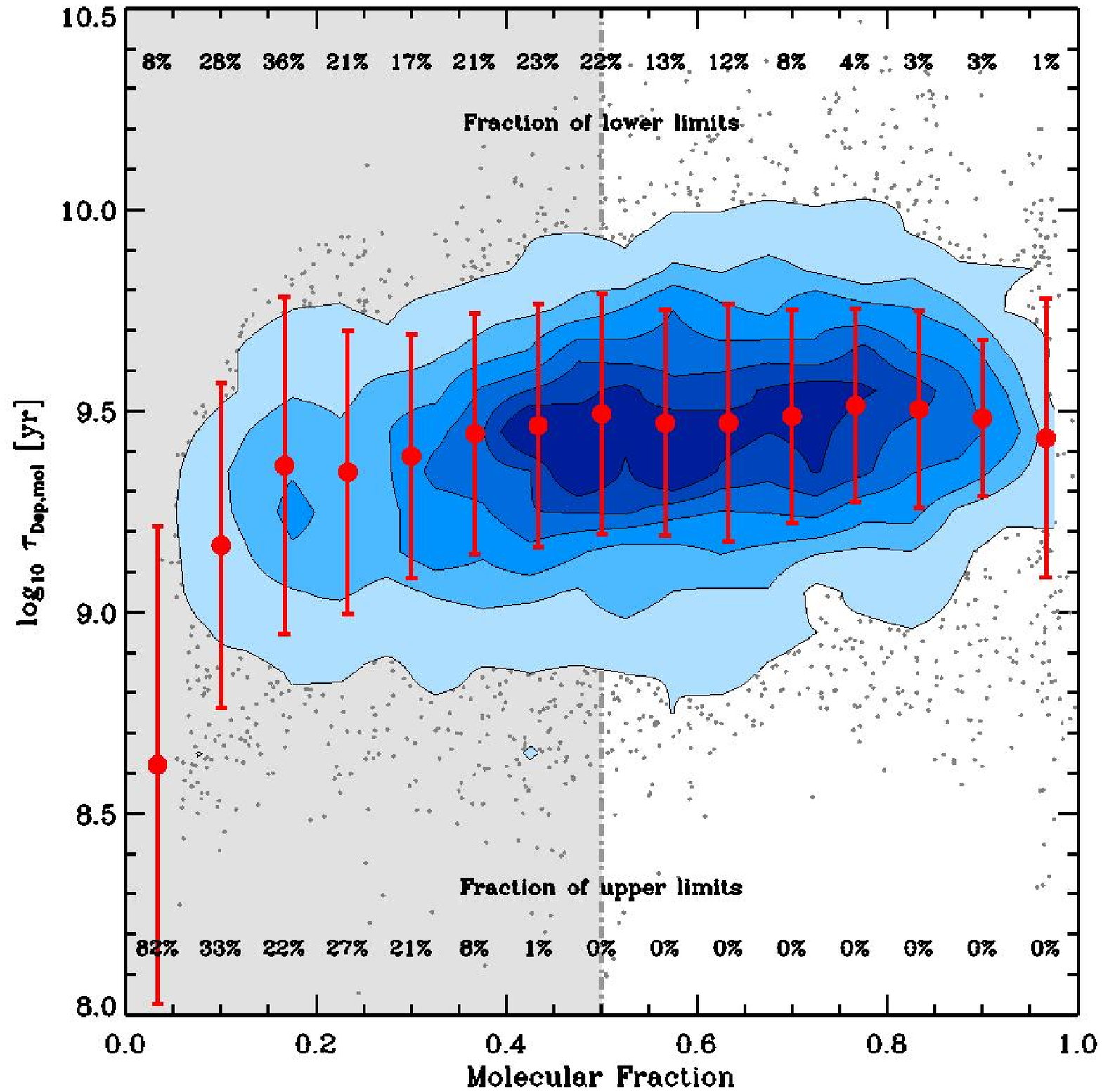}
\caption{As Figure \ref{fig:tdep_local} but showing the molecular gas depletion time, $\tau_{\rm dep}^{\rm mol}$, as 
a function of the molecular fraction for ({\em left}) a fixed $\alpha_{\rm CO}$ and ({\em right}) our ``$\Sigma=100$'' $\alpha_{\rm CO}$.}
\label{fig:tdep_fmol}
\end{figure*}

HERACLES resolves our targets, so that we can investigate the dependence of the local $\tau_{\rm dep}^{\rm mol}$ 
on local conditions. As Figure \ref{fig:tdep_global} shows, galaxy averaged properties have a high degree of
covariance. Examining local conditions may help break this degeneracy.  Figure \ref{fig:tdep_local} plots the data 
density (blue contours) for fixed $\alpha_{\rm CO}$ as a function of local (kpc-scale) conditions: ({\em top left})  stellar mass surface 
density, $\Sigma_*$, ({\em top right}) H$_2$ surface density, $\Sigma_{\rm mol}$, ({\em bottom left}) dust-to-gas ratio, and ({\em bottom 
right}) orbital time. In red we show the median and standard deviation for data binned by the 
abscissa. These medians incorporate the fraction of upper and lower limits in the bin, which appear as percentages 
running along the top and bottom of each plot. The gray region in each plot indicates where the percentage
of limits exceed $\sim 30\%$, indicating significant issues with completeness.

Table \ref{tab:tdep_local} reports rank correlations between $\tau_{\rm dep}^{\rm mol}$ and local physical conditions. 
We restrict these calculations to the region of approximate completeness. This minimizes biases, but
prevents us from probing very low surface densities. Binning and profile work will allow us to extend these analyses in 
future work \citep[][]{SCHRUBA11}. Figure \ref{fig:tdep_local} shows little systematic variation $\tau_{\rm dep}^{\rm mol}$ as a 
function of $\Sigma_*$ and $\Sigma_{\rm mol}$, only a weak tendency to have with shorter $\tau_{\rm dep}^{\rm mol}$ at the highest surface 
densities (for constant $\alpha_{\rm CO}$). These trends have the opposite sense of those expected based on our examination
of galaxy average properties, where high mass, high $\Sigma_{\rm HI+H2}$ systems showed {\em longer} depletion times
than low mass systems.

We observe a more significant relation between $\tau_{\rm dep}^{\rm mol}$ and $D/G$, one that becomes even stronger when 
limits are factored in. The functional form  of the local trend in $\tau_{\rm Dep}^{\rm mol}$ as a function of $D/G$ closely matches 
the trend seen for whole galaxies. To some degree this reflects the fact that the $D/G$, unlike $\Sigma_*$ or $\Sigma_{\rm mol}$ remains
relatively constant across the disks of many of our targets (see Appendix). By contrast even a galaxy with very high total $M_*$ or 
$M_{\rm mol}$ will have a wide range of $\Sigma_*$ and $\Sigma_{\rm mol}$.

Figure \ref{fig:tdep_local_corr} shows $\tau_{\rm dep}^{\rm mol}$ as a function of the same local conditions after applying 
our ``$\Sigma=100$'' conversion factor. The correspondence between $\tau_{\rm dep}^{\rm mol}$ and $D/G$ mostly vanishes
if we adopt this conversion factor. Meanwhile the weak correlations of $\tau_{\rm dep}^{\rm mol}$ with $\Sigma_*$ and $\Sigma_{\rm mol}$ remain
weak.

For fixed $\alpha_{\rm CO}$, Figure \ref{fig:tdep_local} does not support the idea that a fixed fraction of gas is converted to stars each orbital 
time. In the bottom right panel the red bins do not match the dashed line, which shows a fixed fraction of gas converted to stars 
per orbital time. Adopting the ``$\Sigma=100$'' conversion factor changes this picture somewhat, as the central $\alpha_{\rm CO}$ depressions lead to shorter
$\tau_{\rm dep}^{\rm mol}$ in regions with short orbital times. With the ``$\Sigma=100$'' $\alpha_{\rm CO}$, a fixed efficiency per orbital time
becomes a reasonable description of the data below $\tau_{\rm orb} \sim 100$~Myr. Most of our data have longer $\tau_{\rm orb}$ than this, so that this
statement relates mostly to the inner parts of galaxies. Regardless of $\alpha_{\rm CO}$, a fixed efficiency per orbital time does not appear to
describe most of our data. Instead, this may be reasonable description of integrated galaxies across a wide range of 
luminosities \citep{KENNICUTT98B,DADDI10,GENZEL10} or relevant to inner regions where the orbital timescale becomes comparable to the internal
dynamical time of bound clouds. Treatment of $\alpha_{\rm CO}$ makes a large difference to the results in this plot, highlighting the need
for improved constraints in the inner disks of galaxies where variations in physical condition presumably dominate $\alpha_{\rm CO}$ variations.

In Section \ref{sec:global} we discuss the idea that the conversion of the diffuse, unbound ISM from atomic to molecular may lead to the high
$\tau_{\rm dep}^{\rm mol}$ found in massive galaxies. The local drivers of such trends are not immediately obvious from Figure \ref{fig:tdep_local}.
In Figure \ref{fig:tdep_fmol} we directly plot $\tau_{\rm dep}^{\rm mol}$ as function of the local molecular fraction, $f_{\rm mol} = \Sigma_{\rm mol} / (\Sigma_{\rm mol} + 
\Sigma_{\rm HI})$. For $f_{\rm mol} > 0.5$ molecular gas dominates the ISM mass budget. We do not know whether this molecular gas
is organized into bound, star-forming clouds or diffuse, inert material but if high $f_{\rm mol}$ does correspond to a higher fraction of
diffuse molecular material, we might expect to observe a general increase in $\tau_{\rm dep}^{\rm mol}$ as $f_{\rm mol}$ increases. As Figure \ref{fig:tdep_fmol} shows,
our completeness severely limits this calculation, restricting us to $f_{\rm mol} \gtrsim 0.5$. Above this value regardless of how we treat $\alpha_{\rm CO}$ we
find little or no correlation of $\tau_{\rm dep}^{\rm mol}$ with $f_{\rm mol}$.

Comparison of $\tau_{\rm dep}^{\rm mol}$ to local conditions thus reveals the same strong trend with $D/G$ observed for galaxy-average properties but 
only weak trends with other parameters, including the molecular fraction. We observe a suggestion of decreased $\tau_{\rm dep}^{\rm mol}$ at high $\Sigma_*$
or high $\Sigma_{\rm mol}$ and, after apply depressed $\alpha_{\rm CO}$ in galaxy centers, we find a weak correlation of orbital time and $\tau_{\rm dep}^{\rm mol}$
for short orbital times. 

We emphasize that these represent our broad-brush results. Our database will allow deeper exploration via detailed analysis of individual galaxies, deep profiles, varying
weighting and normalization, and stacking \citep{SCHRUBA11}. Kinematic analysis \citep[e.g.,][]{TAN00} and the inclusion of outer disks \citep[e.g.,][]{BIGIEL10} should yield 
the lever arms to better understand the impact of local conditions on star formation in molecular gas. Indeed, as we discuss in the next section, significant peripheral 
evidence points to the existence of significant environmental dependencies of $\tau_{\rm dep}^{\rm mol}$. Such effects are not immediately evident from the simple
tests that we carry out here, however. Beyond the clear correlation of $\tau_{\rm dep}^{\rm mol}$ with $D/G$, which we interpret as likely due to conversion factor effects, 
the absence of ``smoking gun'' correlations represents the main first-order result of this section.

\subsection{Spatial Correlation of $\tau_{\rm dep}^{\rm mol}$}
\label{sec:scatter}

\begin{figure}[]
\plotone{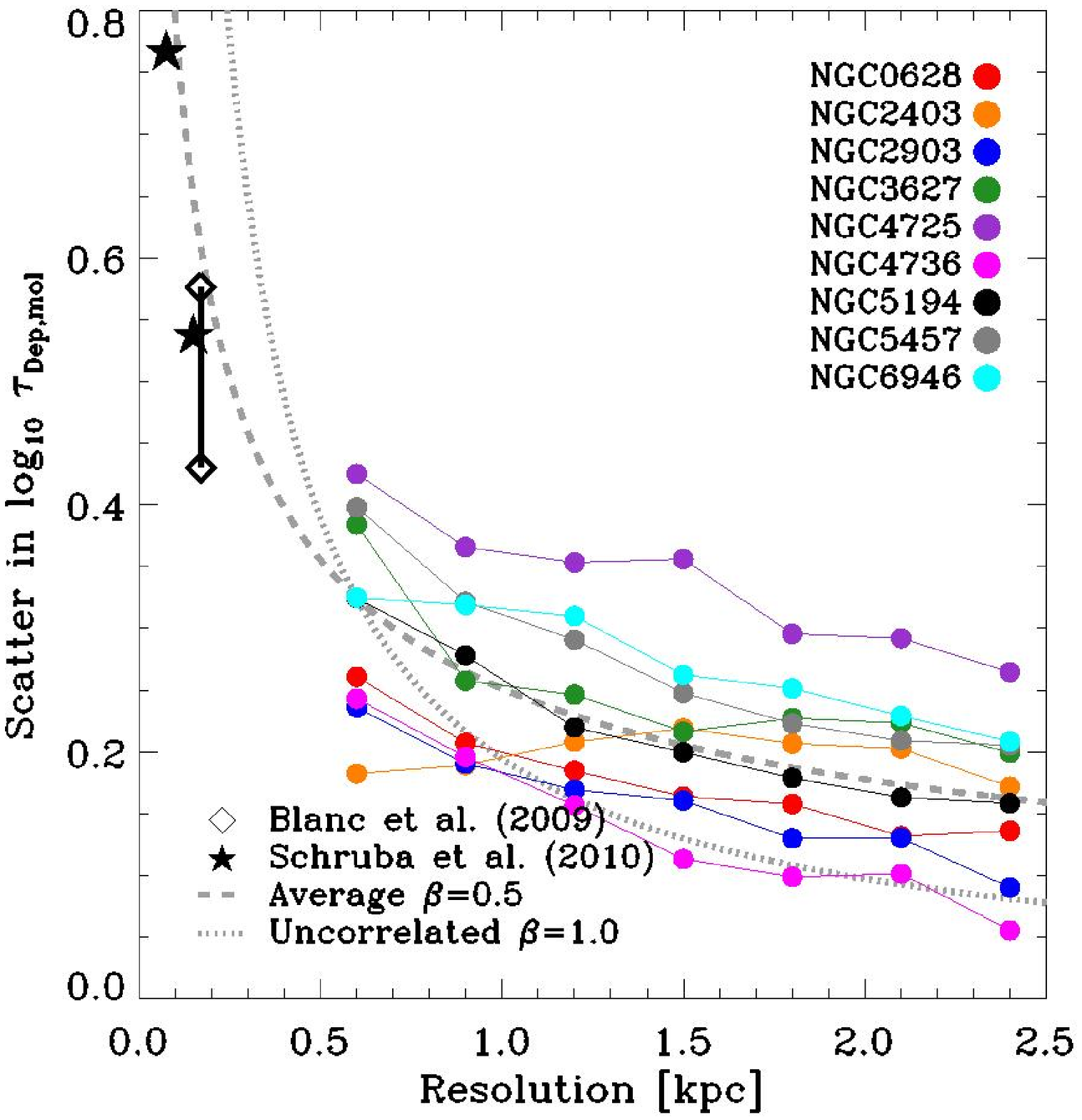}
\caption{Evidence for spatial correlation among $\tau_{\rm dep}^{\rm mol}$ within 
galaxies. We plot RMS scatter in $\log_{\rm 10} \tau_{\rm dep}^{\rm mol}$ ($y$-axis) as a
function of spatial resolution ($x$-axis) for a subset of nearby, large galaxies. The scatter in $\tau_
{\rm dep}^{\rm mol}$ changes more slowly as a function of resolution than one would expect 
averaging uncorrelated data. Extrapolating the trend that we observe to much higher resolution yields 
reasonable consistency with (scarce) high resolution literature data \citep{BLANC09, SCHRUBA10}. We 
illustrate the median trend, $\beta = 0.5$, while $\beta = 1$ would be expected
for a disk filled with uncorrelated star-forming regions. This low $\beta$ reflects systematic variations 
in $\tau_{\rm dep}^{\rm mol}$on intermediate scales, either due to synchronization of star formation on intermediate scales \citep
[e.g., see][]{FELDMANN11} or systematic but still undiagnosed environment dependencies of the efficiency of star formation.}
\label{fig:scatter}
\end{figure}

\begin{deluxetable}{lc}
\tablecaption{Scale Dependence of Scatter in $\tau_{\rm dep}^{\rm mol}$} 
\tablehead{
\colhead{Galaxy} & \colhead{Averaging Index, $\beta$} \\
}
\startdata
NGC~0628 & 0.5 (0.2--0.6) \\
NGC~2403 & 0.0 (0.0--0.3) \\
NGC~2903 & 0.6  (0.4--0.6) \\
NGC~3627 & 0.5 (0.4--0.6) \\
NGC~4725 & 0.3 (0.3--0.4) \\
NGC~4736 & 0.8 (0.1--0.8) \\
NGC~5194 & 0.6 (0.5--0.6) \\
NGC~5457 & 0.5 (0.1--0.5) \\
NGC~6946 & 0.3 (0.2--0.4) 
\enddata
\label{tab:scaledep}
\tablecomments{Averaging index, $\beta$, for well-resolved galaxies. We quote the best-fit $\beta$, 
defined in Equation \ref{eq:scaledep}, estimated from a $\chi^2$ minimization using 
H$\alpha$+24$\mu$m and our variable ``$\Sigma = 100$'' $\alpha_{\rm CO}$. In parenthesis, we give the range of values measured 
as we vary SFR tracer and change adopted $\alpha_{\rm CO}$. For comparison, we expect $\beta 
\approx 1$ for the case of uncorrelated averaging in a thin disk.}
\end{deluxetable}

We have examined the explicit dependence of $\tau_{\rm dep}^{\rm mol}$ on various host galaxy properties and 
local physical conditions. We find many systematic dependencies on host galaxy, but the picture relating
$\tau_{\rm dep}^{\rm mol}$ to local conditions remains more ambiguous. Considering the
scatter in $\tau_{\rm dep}^{\rm mol}$ as a function of scale offers another way to approach this problem. For
star formation uncorrelated on the scale of individual regions in a two-dimensional disk, there is a clear
analytic expectation for the functional form of this averaging. Deviations from this scaling can reveal the
degree to which adjacent regions share the same $\tau_{\rm dep}^{\rm mol}$, or at least appear synchronized.

At small, $\sim 10$--$50$~pc, scales observations of the Milky Way and the nearest galaxies resolve 
$\Sigma_{\rm SFR}$ and $\Sigma_{\rm mol}$ into discrete star forming regions, clouds, and clusters. These 
individual elements have distinct ages and evolutionary sequences \citep[e.g.][]
{KAWAMURA09,FUKUI10} so that the ratio of molecular gas to stars in a region and the emission of 
star formation rate tracers both evolve as a function of time (see discussion and plots in L12). As a result, scaling relations between SFR and H$_2$, which 
capture the time-averaged relation between gas and star formation, emerge only after averaging together many distinct regions \citep[see][]
{SCHRUBA10,FELDMANN11}. 

In a disk with fixed $\tau_{\rm dep}^{\rm mol}$ if these individual regions form stars independently then
we will expect the scatter in $\tau_{\rm dep}^{\rm mol}$ over part of the galaxy to go as $\sqrt{N}^{-1}$, where
$N$ is the number of star-forming region in that part of the galaxy. For a region of extent $l$ in a smooth disk,
$N \propto l^2$, so that the expectation for the ``uncorrelated case'' is $\sigma \propto l^{-1}$.

Deviations from this scaling will emerge if $\tau_{\rm dep}^{\rm mol}$ varies systematically on large scales across
the disk. In that case a high (or low) $\tau_{\rm dep}^{\rm mol}$ in one region is likely to be reflected with 
a similar $\tau_{\rm dep}^{\rm mol}$ in the adjacent regions. Thus, if $\tau_{\rm dep}^{\rm mol}$ is correlated between 
two adjacent regions, we expect a weaker dependence of the scatter in $\tau_{\rm dep}^{\rm mol}$ on scale. We would
expect this to occur in the case that $\tau_{\rm dep}^{\rm mol}$ has real, but still undiagnosed, dependence on local
physical conditions. Moreover large scale dynamical effects like spiral density waves, bars, supernova explosions, may 
synchronize the star formation process on scales larger than a single cloud.

With $\sim$ kpc resolution, HERACLES offers limited ability to measure the scale dependence of scatter over a large
dynamic range, but we have identified a subset of large, nearby galaxies where we can measure the scatter in 
$\tau_{\rm dep}^{\rm mol}$ at linear resolutions from 0.6--2.4~kpc. These are labeled as our ``multi-scale'' sample in
Table \ref{tab:sample}. We use this sample to measure the scatter in $\log_{10} \tau_{\rm dep}^{\rm mol}$ as a function
of linear resolution. To do this, we convolve each galaxy in our ``multi-scale'' sample (Table \ref{tab:sample}) to have 
linear resolution 0.6--2.4~kpc and measure the RMS scatter in $\log_{10} \tau_{\rm dep}^{\rm mol}$ 
across the galaxy at each resolution. Doing so, we make no correction for inclination, so that this 
exercise consists of placing targets at larger and larger distances.

Figure \ref{fig:scatter} plots the results of this exercise for each resolved galaxy. We 
show $\sigma$, the RMS scatter in $\log_{10} \tau_{\rm dep}^{\rm mol}$ as a function of 
spatial resolution. The figure shows results calculated using the ``$\Sigma=100$'' $\alpha_
{\rm CO}$, which removes a significant internal gradient in $\tau_{\rm dep}^{\rm mol}$ from NGC~5457. 
For comparison we plot the scatter in $\tau_{\rm dep}^{\rm mol}$ measured at high resolution 
in M33\footnote{We infer the scatter from their measurements of CO-to-H$\alpha$+24$\mu$m 
near H$\alpha$ peaks and CO peaks.} by \citet[][stars]{SCHRUBA10} and M51 \citep[][diamonds]
{BLANC09}.

Figure \ref{fig:scatter} shows a steady increase in scatter with improving linear resolution. 
We characterize this scale dependence of the scatter via a power law,
 
\begin{equation}
\label{eq:scaledep}
\sigma \left( l \right) = \sigma_{600} \left( \frac{l}{600~{\rm pc}} \right)^{-\beta}~
\end{equation}

\noindent where $l$ represents the spatial resolution, $\sigma_{600}$ is the 
scatter in $\tau_{\rm dep}^{\rm mol}$ at 600~pc resolution and the power-law 
index $\beta$ measures the rate at which changing the resolution changes the 
measured scatter in $\tau_{\rm dep}^{\rm mol}$. We report the best-fit $\beta$ for 
each multiscale target in Table \ref{tab:scaledep}. As described above, we expect $\beta = 1$ for 
uncorrelated star formation in a disk.

In most cases the best-fit averaging index, $\beta$, is $\sim 0.5$, significantly less than the $\beta = 1$ 
expected for uncorrelated, fixed-efficiency star formation. Based on the previous sections,
we expect that this reflect real $\tau_{\rm dep}^{\rm mol}$ variations or correlated systematic uncertainties
in our physical parameter estimation (e.g., undiagnosed $\alpha_{\rm CO}$ variations). That is, this 
is another way to see the subtle but real systematic variations in $\tau_{\rm dep}^{\rm mol}$ considered
in the last two sections. Alternatively, Figure \ref{fig:scatter} and Table \ref{tab:scaledep} could reflect a
high  degree of synchronization, with adjacent regions likely to be at the same stage of the star formation 
process and thus show similar ratios of star formation tracers to CO emission. Either synchronization or real $\tau_{\rm dep}^{\rm mol}$ 
variations might be achieved by dynamical phenomena at small scales and systematic efficiency 
variations may arise from dependence on local conditions that we have yet to identify. 

This weak, $\beta \sim 0.5$, scale dependence agrees with predictions based on numerical simulations by \citet{FELDMANN11}, who discuss $\beta$ in 
terms of dimensionality. As an example, $\beta = 0.5$ would be consistent with the degree of correlation 
expected by spiral arms or any other phenomenon that synchronizes star formation along one dimension, 
though in the simulations of \citet{FELDMANN11} it arises more generally. The low $\beta$ also agrees 
qualitatively with numerous observations of highly structured star formation on scales $\lesssim 1$~kpc in nearby 
galaxies; for example, see the recent synthesis by \citet[][]{ELMEGREEN11}.

From the perspective of galactic-scale star formation, the key point from this calculation is that the 
scatter in $\tau_{\rm dep}^{\rm mol}$ at kpc scales depends on processes operating at scales larger than that
of individual star-forming regions. That is, key information on the distribution of star formation in 
galaxies remains to be extracted from comparison of maps of $\Sigma_{\rm SFR}$ and $\Sigma_{\rm mol}$. This 
represents another manifestation of the overall theme of this section, that real second order variations
in $\tau_{\rm dep}^{\rm mol}$ do appear visible in our sample. Future 
investigation of HERACLES and similar surveys will allow tests of the degree to which
this correlation can be attributed to systematic variations in $\tau_{\rm dep}^{\rm mol}$ as a function 
of yet-unexplored local conditions or to physical parameter estimation. Simultaneously, observations with high spatial 
dynamic range will allow more a detailed diagnosis of the scale-dependence of $\tau_{\rm dep}^{\rm mol}$.

\subsection{Enhanced Efficiency in Galaxy Centers}
\label{sec:centers}

\begin{figure*}[]
\plottwo{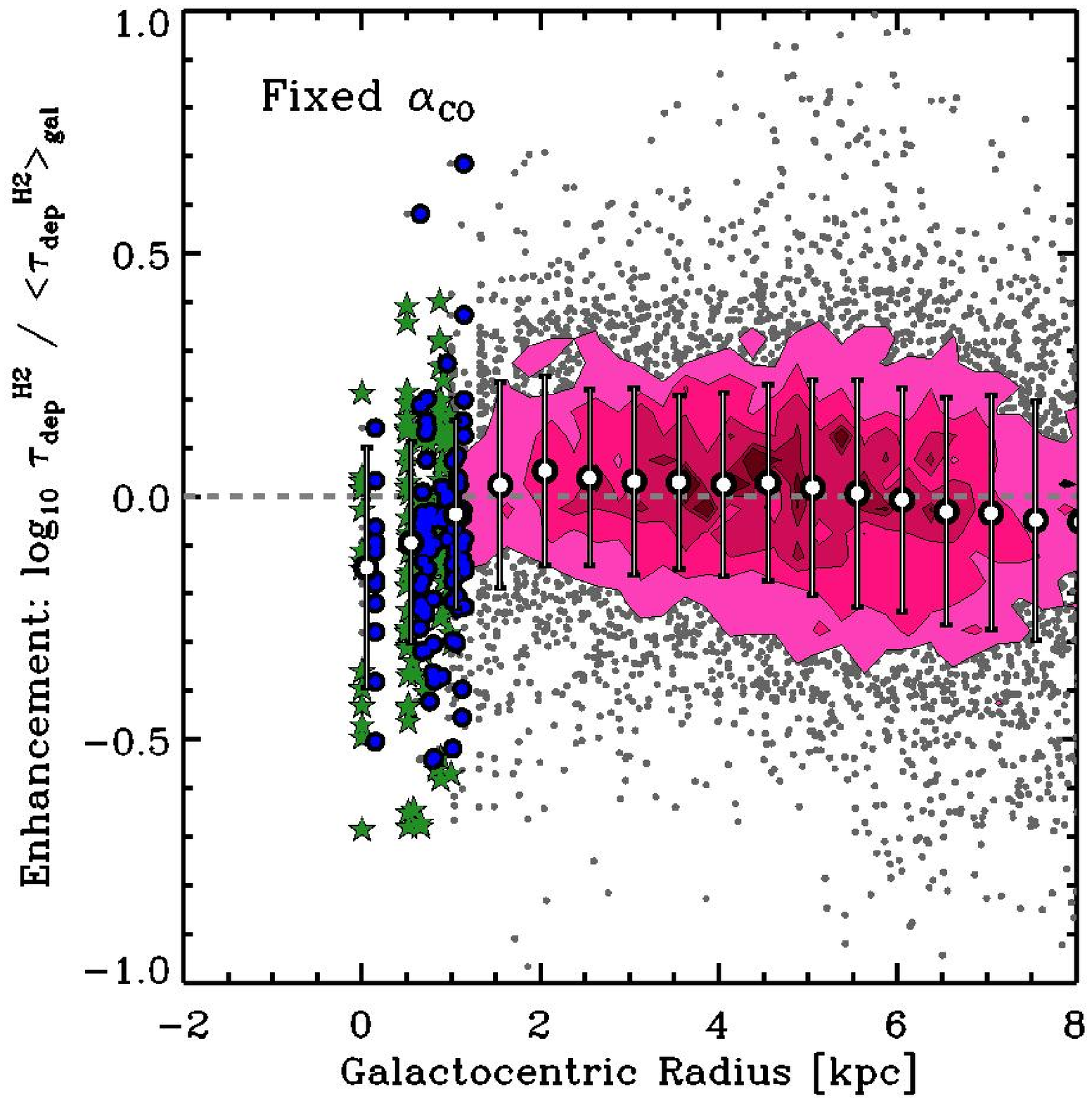}{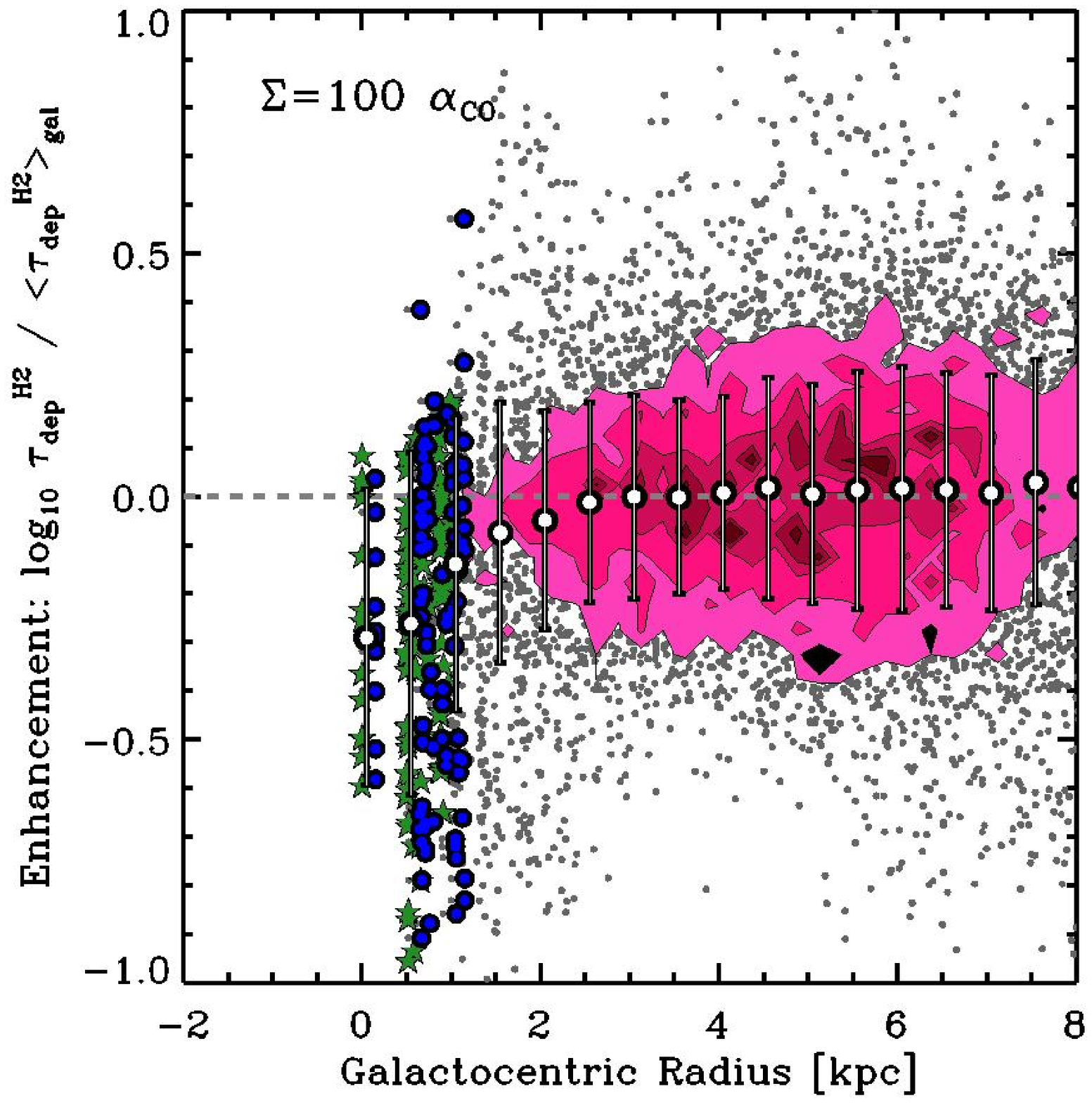}
\plottwo{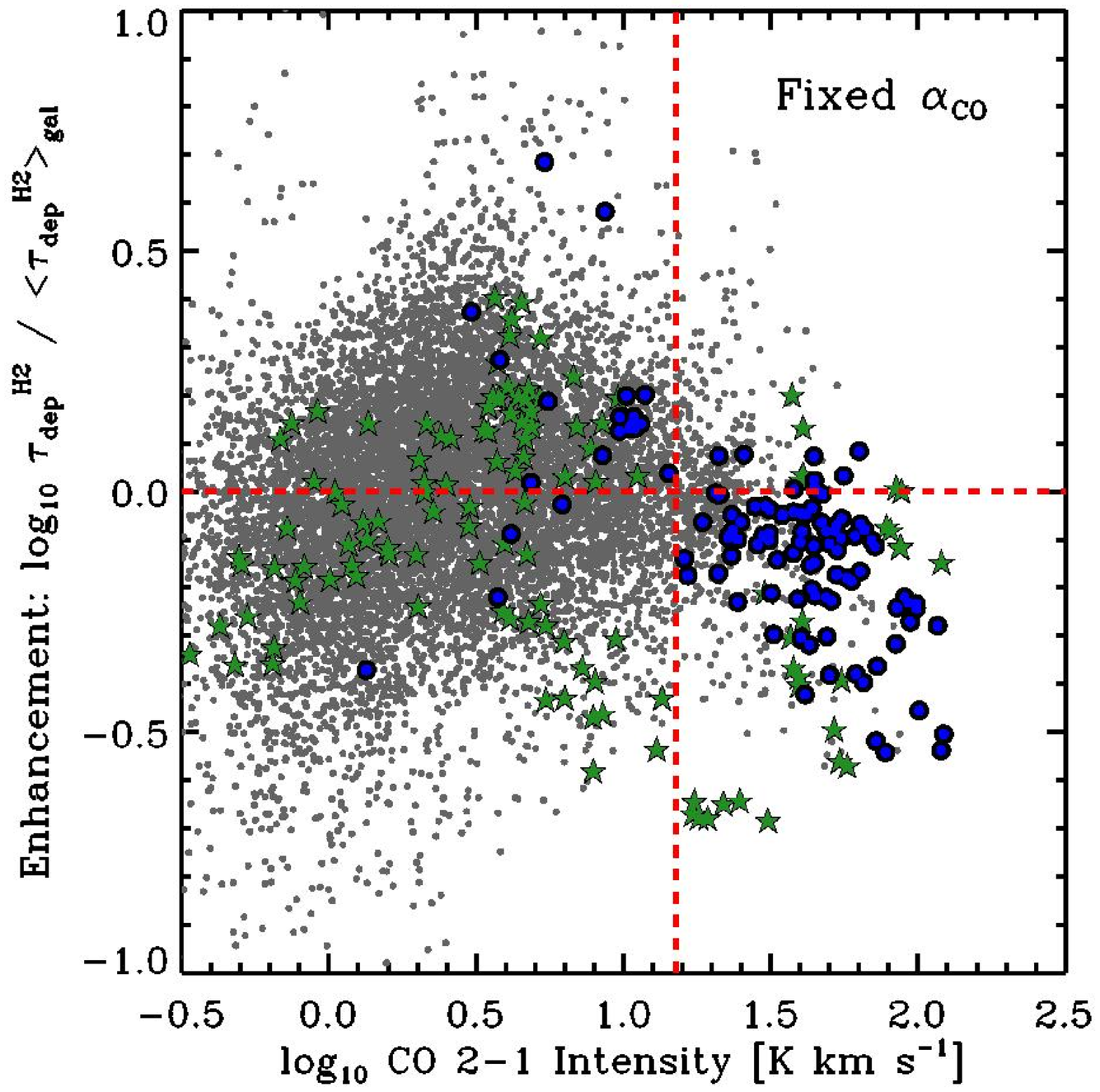}{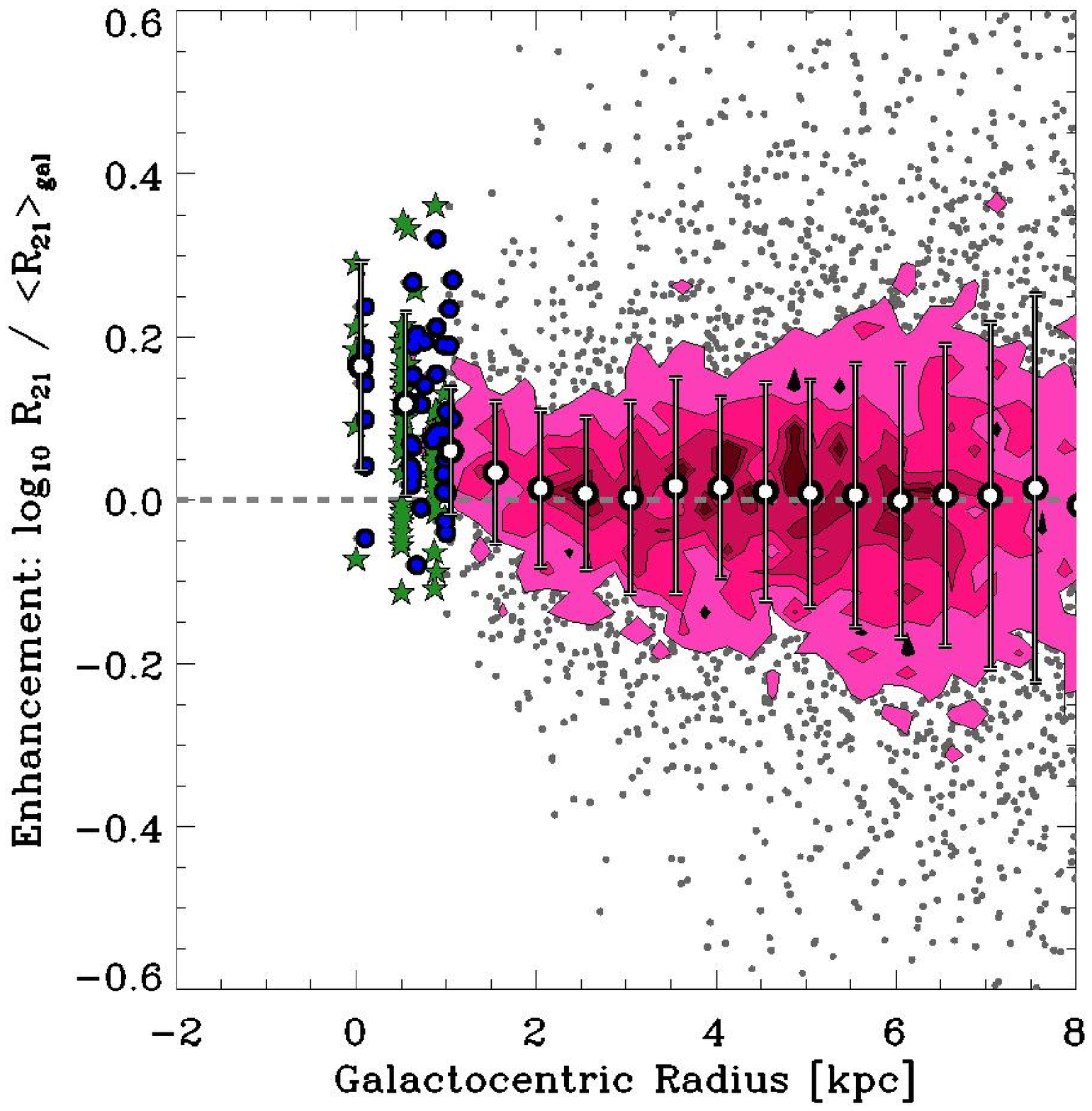}
\caption{Enhanced efficiency in galaxy centers. We plot $\tau_{\rm dep}^{\rm mol}$ normalized to 
the galaxy average as a function of galactocentric radius for a fixed $\alpha_{\rm CO}$ (top left) and
a ``$\Sigma=100$'' conversion factor (top right). We also plot enhancement in $\tau_{\rm dep}^{\rm mol}$
as a function of CO (2-1) intensity (bottom left) and show the enhancement in the CO(2-1) to CO(1-0) line
ratio relative to the disk average as a function of radius. Individual gray points show kpc-resolution lines
of sight; contours show data density; black-and-white points and error bars show median and RMS scatter
in binned data. Blue points and green stars show data from the inner kpc of our targets --- blue points
show systems with a spectroscopic classification indicating the likely presence of an AGN \citep{MOUSTAKAS10},
green points show star formation dominated nuclei. Galaxy centers show shorter depletion times and
enhanced line ratios relative to galaxy disks, indicative of more excited, more efficiently star-forming gas.
The effect appears even stronger once we account for variations in $\alpha_{\rm CO}$ (Sandstorm et al.)
and essentially all CO-bright nuclear regions show some level of enhancement.}
\label{fig:tdep_center}
\end{figure*}

Our sample does not contain any true galaxy-wide starbursts, but many of our targets do host nuclear
concentrations of star formation and gas \citep[][]{HELFER03}, so that the inner kpc of our targets
represents a subsample intermediate between normal disk galaxies and starbursts. These 
regions represent only a small fraction of the area in our targets, so exert a negligible impact on
the ensemble of data seen in Figure \ref{fig:combined} but they probe an important part of parameter 
space --- high $\Sigma_{\rm mol}$, high $\Sigma_{\rm SFR}$ --- and so we explicitly consider them 
in Figure \ref{fig:tdep_center}. We plot the enhancement in $\tau_{\rm dep}^{\rm mol}$, calculated 
by dividing $\tau_{\rm dep}^{\rm mol}$ for each point by the average for that galaxy, and mark
points from the inner kpc of our targets with large symbols: blue dots for galaxies with some 
spectroscopic indication of an AGN \citep[][or NED]{MOUSTAKAS10} and green stars show star-formation
dominated regions.

The top left panel of Figure \ref{fig:tdep_center} shows enhancement in $\tau_{\rm dep}^{\rm mol}$ for a
fixed $\alpha_{\rm CO}$. We find systematically lower $\tau_{\rm dep}^{\rm mol}$ in the centers of our
targets, both AGN and starbursts. We find this shorter $\tau_{\rm dep}^{\rm mol}$ using a fixed $\alpha_{\rm CO}$. \citet{SANDSTROM12} 
find evidence for systematically lower conversion factors in the central parts of our sample. The top right panel of Figure \ref{fig:tdep_center}
shows the results of applying these central corrections, as part of our ``$\Sigma=100$'' conversion factor. 
$\tau_{\rm dep}^{\rm mol}$ become even shorter in the central regions compared to the disks. Figure \ref{fig:tdep_center} thus 
exhibits one of the clearest systematic effects in our sample: the central regions of our targets tend to have significantly 
shorter $\tau_{\rm dep}^{\rm mol}$ than the ensemble of disk regions. Central $\tau_{\rm dep}^{\rm mol}$ are  $0.8$ times 
the disk value with $\pm 0.2$~dex ($1\sigma$ scatter) for fixed $\alpha_{\rm CO}$. The median central-to-disk ratio drops to 
$0.6$ with $\pm 0.35$~dex scatter our ``$\Sigma=100$''' case. 

This apparently real shortening of $\tau_{\rm dep}^{\rm mol}$ coincides with an increase in the CO(2-1) to CO(1-0) ratio, 
indicative of more excited gas. The bottom right panel shows that this lower $\tau_{\rm dep}^{\rm mol}$
coincides with higher CO(2-1) to CO(1-0) ratios \citep[calculated with comparison to][]{KUNO07}. The central regions of
our targets show systematically enhanced CO(2-1)/CO(1-0) compared to the disks. This excitement presumably
reflects the same changing physical conditions that drive the lower conversion factors found by \citet{SANDSTROM12},
underscoring that nuclear gas concentrations represent a distinct physical regime from galaxy disks. Molecular gas in these regions 
gives off more CO emission, appears more excited, and forms stars more rapidly than molecular gas further out in the
disks of galaxies.

Not all nuclear regions exhibit shorter $\tau_{\rm dep}^{\rm mol}$. A sufficient but not necessary condition to find such 
enhancements in our data is: (1) to lie within the central kpc of a target and (2) have CO(2-1) intensity $\gtrsim 15$~K~kms$^{-1}$. We 
indicate this criteria using a red line in the bottom left panel of Figure \ref{fig:tdep_center}. This does not precisely equate 
to a $\Sigma_{\rm mol}$ threshold because of ambiguities introduced by $\alpha_{\rm CO}$; the figure shows that central parts of galaxies
with bright CO emission tend to show lower $\tau_{\rm dep}^{\rm mol}$.

\citet{DADDI10} and \citet{GENZEL10} argue for multiple ``modes'' or ``sequences'' of star formation, in which the shorter dynamical
time and higher density in starburst galaxies lead to shorter depletion times at fixed $\Sigma_{\rm mol}$. Figure \ref{fig:tdep_center} supports 
this idea inasmuch as it shows that physical conditions other than gas surface density play an important role determining $\tau_{\rm dep}^{\rm mol}$. 
The population of low $\tau_{\rm dep}^{\rm mol}$ in Figure \ref{fig:tdep_center} appear to be set by factors other than kpc-scale $\Sigma_{\rm mol}$ alone:
first, identifying these lines of sight requires knowing that the points line in the centers of our targets; second, $\tau_{\rm dep}^{\rm mol}$ exhibits a
wide range, presumably set by other physical parameters.  Figure \ref{fig:tdep_center} does not support the idea of two cleanly distinguished sequences, suggesting
instead that a continuum of $\tau_{\rm dep}^{\rm mol}$ exist in the central regions of our targets; we plot the histogram of central $\tau_{\rm dep}^{\rm mol}$
in Section \ref{sec:summary}.

What drives these enhancements? Beam dilution certainly plays some role. Our kpc resolution will average out nuclear gas concentrations
with surface densities well in excess of the disks of our targets but small spatial extent \citep[e.g.,][]{JOGEE05}. We expect that high pressure, driven
by the deep potential well in the central parts of galaxies, also plays a role driving gas to higher densities. This effect is seen in our 
own Galaxy \citep{OKA01} and others \citep{ROSOLOWSKY05}.
\citet{DADDI10} and \citet{GENZEL10} suggest that the shorter dynamical timescales in starburst galaxies may also play a key role and
our data may offer some tentative support for this (Figure \ref{fig:tdep_local_corr}). For this paper, the key result is that the shorter $\tau_{\rm dep}^{\rm mol}$ do exist, 
so that the inner parts of disk galaxies represent a kind of ``transition regime'' between the disks of spirals and galaxy-wide starbursts. Follow-up interferometry and
spectroscopy contrasting physical conditions in the nuclear regions with those in disks will yield more insight.

\section{Discussion and Conclusions}
\label{sec:summary}

\begin{figure*}[]
\plotone{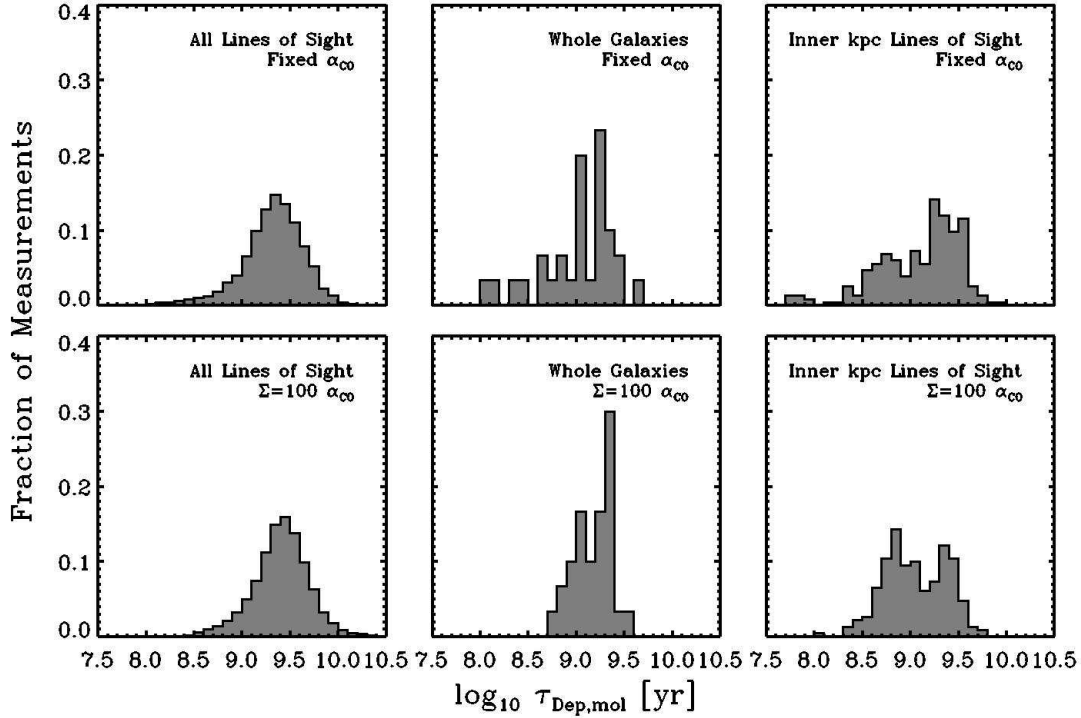}
\caption{Distribution of molecular gas depletion time, $\tau_{\rm dep}^{\rm mol}$, in HERACLES.
We show histograms for ({\em left}) all lines of sight at kpc resolution, ({\em middle}) whole galaxies, 
and ({\em right}) lines of sight in the central kpc of our targets. The top row shows results for fixed $\alpha_{\rm CO}$ 
while the bottom row applies our ``$\Sigma=100$'' conversion factor, which accounts for variable dust shielding and central 
$\alpha_{\rm CO}$ depressions. The distribution of $\tau_{\rm dep}^{\rm mol}$ treating lines of sight equally appears lognormal, with 
a depletion time $1$--$3$~Gyr, depending on the adopted conversion factor, and $\lesssim 0.3$~dex scatter, though low-$\tau_{\rm 
Dep}^{\rm mol}$  excursions from the lognormal distribution are already visible in this panel. The distribution of $\tau_{\rm  Dep}^{\rm mol}$ for whole 
galaxies shows more scatter for a fixed $\alpha_{\rm CO}$, reflecting systematic trends in which low-mass, 
low-metallicity, high sSFR galaxies show faint CO relative to SFR tracers. After accounting for varying $\alpha_{\rm CO}$, there 
is less scatter  among galaxies but some spread remains. Meanwhile the inner kpc of our targets shows a much broader, much less normal distribution, with 
both low and high $\tau_{\rm dep}^{\rm mol}$ common. Correcting for the central $\alpha_{\rm CO}$ depressions observed by 
\citet{SANDSTROM12} and expected from observed line ratio variations exacerbates this effect, but it is clearly present even adopting fixed $\alpha_{\rm CO}$. Thus, 
$\tau_{\rm dep}^{\rm mol}$ appears constant to first order but clear systematic second-order variations do emerge as a function of local and galaxy scale 
conditions in a broad sample.}
\label{fig:histogram}
\end{figure*}

We study the relationship between molecular gas surface density, $\Sigma_{\rm mol}$, as traced by CO (2-1) emission 
and SFR surface density, \sigsfr --- traced by combinations of H$\alpha $, FUV, and IR emission --- at 
1~kpc resolution across the optical disks of 30 nearby spiral galaxies. Broadly, we demonstrate two 
conclusions, which we illustrate using histograms of $\tau_{\rm dep}^{\rm mol}$ in Figure \ref{fig:histogram}: a first-order 
simple correspondence between $\Sigma_{\rm mol}$ and $\Sigma_{\rm SFR}$ and
second-order systematic variations in the apparent molecular gas depletion time, $\tau_{\rm dep}^{\rm mol}$, including
lower values in nuclear starbursts and low, mass low-metallicity galaxies. Some, but not all, of these variations
may be explained by invoking a dependence of the CO-to-H$_2$ conversion factor, $\alpha_{\rm CO}$, on
dust-to-gas ratio

First, molecular gas and star formation are tightly correlated in both individual galaxies and over 
individual kpc-resolution lines of sight. Their ratio, the molecular gas depletion time, $\tau_{\rm dep}^{\rm mol}$ 
appears to first-order constant across the disks of massive, large star-forming galaxies, with a median value
$\tau_{\rm dep}^{\rm mol} = 2.2$~Gyr, a scatter of $\approx 0.3$ dex, and systematic uncertainty $\sim 60\%$ (see the 
nearly lognormal histogram in the left panels of Figure \ref{fig:histogram}). We arrive
at this conclusion testing a wide range of methodologies to trace the recent star formation rate and distribution of molecular gas. This 
includes the interchange of H$\alpha$ and FUV emission, varying treatments of the infrared ``cirrus'', substitution 
of literature CO (1-0) for our CO (2-1) maps, and adoption of variable CO-to-H$_2$ conversion factors. We demonstrate 
that our measurements agree with a large collection of literature data from the last 
decade (Section \ref{sec:lit}), with all data occupying a common region of $\Sigma_{\rm mol}$-$
\Sigma_{\rm SFR}$ parameter space. Our data reinforce and extend a consensus for a ``large disk galaxy'' value 
of $\tau_{\rm dep}^{\rm mol} \approx 2$~Gyr for matched assumptions about the CO-to-H$_2$
conversion factor and stellar IMF with $\approx 0.3$~dex scatter.

Adopting a forward-modeling approach similar to that of \citet{BLANC09}, we derive a best-fit power 
law index of $N \approx 1 \pm 0.2$ for $\Sigma_{\rm SFR} \propto \Sigma_{\rm mol}^{N}$ (Section \ref
{sec:index}) for a fixed $\alpha_{\rm CO}$ and show that individual galaxies exhibit a distribution of $N$ with
$N$ mostly in the range 0.8--1.2. However, we stress the inadequacy of a power law to capture 
important changes in physical conditions other than $\Sigma_{\rm mol}$ and the substantial uncertainty in such fits. 
Reinforcing the conclusions of \citet{BLANC09}, we caution that the commonly used combination of 
``sigma-clipping" and bivariate fitting has the potential to substantially bias results (see Appendix), yielding seemingly 
discordant values of $N$ even when the data substantially agree.

Our second major conclusion is that with a broad sample spanning a wide range of physical 
conditions,  systematic variations in the apparent $\tau_{\rm dep}^{\rm mol}$ emerge both among and within galaxies
(see the width of the distribution in the top middle panel of Figure \ref{fig:histogram}). 
We show systematic variations of galaxy-average apparent $\tau_{\rm dep}^{\rm mol}$ as a function of
many host galaxy properties: stellar mass, rotation velocity, metallicity, dust-to-gas ratio, average gas surface
density, and morphology. These variations have the sense that low mass, low metallicity, late type galaxies exhibit 
shorter apparent $\tau_{\rm dep}^{\rm mol}$ than high mass galaxies. The trends persist, though weaker, 
even into the high mass region, $M_* > 10^{10}$~M$_\odot$, and agree well with those seen in the COLDGASS
sample \citep{SAINTONGE12}. We emphasize ``apparent'' because these variations appear to be a mixture of real changes in the 
rate at which gas forms stars and biases in physical parameter estimation. Adopting a CO-to-H$_2$ conversion factor that
depends on the dust-to-gas ratio can explain many of the strongest variations with host galaxy properties; note the
narrowing from the top middle panel of Figure \ref{fig:histogram} to the bottom middle panel. Our
ability to examine residual trends remains restricted by the limited precision with which we know $\alpha_{\rm CO}$,
but our best estimate is that correlations do remain between the real ($\alpha_{\rm CO}$-adjusted) $\tau_{\rm dep}^{\rm mol}$ 
and galaxy mass, average gas surface density, and perhaps several other quantities.

We also examine how $\tau_{\rm dep}^{\rm mol}$ varies as a function of local conditions. We find two strong relationships:
the apparent $\tau_{\rm dep}^{\rm mol}$ calculated for a fixed $\alpha_{\rm CO}$ varies systematically as a function of dust-to-gas ratio
and we observe systematically lower and widely varying $\tau_{\rm dep}^{\rm mol}$ in the inner kpc of our targets (see the top right
panel of Figure \ref{fig:histogram}. We interpret the first as indicating important variations in $\alpha_{\rm CO}$ and, as with the galaxy-integrated case, show
that application of a dust-to-gas ratio dependent $\alpha_{\rm CO}$ can explain much of the observed trend. The lower $\tau_{\rm dep}^{\rm mol}$
in galaxy centers appears real and robust to $\alpha_{\rm CO}$ considerations. Indeed, \citet{SANDSTROM12} find 
low $\alpha_{\rm CO}$ in the central parts of many of our targets, which implies even lower $\tau_{\rm dep}^{\rm mol}$. The resulting 
$\tau_{\rm dep}^{\rm mol}$ in the inner parts of galaxies varies widely (see the right panels in Figure \ref{fig:histogram}), providing strong 
evidence that environmental factors do drive $\tau_{\rm dep}^{\rm mol}$, and thus the star formation rate, in these regions.

Strong local trends beyond those linking $\tau_{\rm dep}^{\rm mol}$ to the dust-to-gas ratio and nuclear starbursts elude our present analysis, though the 
correlation of $\tau_{\rm dep}^{\rm mol}$ with integrated properties and weak dependence of scatter on scale strongly suggest their presence.
We suspect that the limited resolution and remaining imprecision in physical parameter estimation so far obscure these trends in HERACLES.
Comparison to existing {\em Herschel} data and spectroscopy of other molecular lines should improve our ability to understand the origins of the physical scatter
in $\tau_{\rm dep}^{\rm mol}$. Perhaps just as important, at kpc resolution significant averaging has already occurred, especially in regions of high 
$\Sigma_{\rm SFR}$. The star-forming ISM hosts many competing effects: for example, shear may both suppress collapse and lead to more frequent cloud 
collisions; high pressures may lead to both denser clouds and a substantial diffuse molecular ISM; spiral arms may both collect material and suppress collapse
via streaming motions. At our kpc resolution, such trends will be subtle as competing effects occur inside a single resolution element. In the longer term, higher 
resolution observations of a diverse sample of galaxies will be needed to diagnose the impact of dynamical effects on star formation.

These results fit into a broad picture of star formation in galaxies as follows. Within the disks of nearby galaxies, we find recent star formation correlated with molecular, 
rather than atomic or total, gas \citep{SCHRUBA11}. The ratio of CO emission to recent star formation appears roughly constant within massive, star-forming 
disk galaxies \citep[Section \ref{sec:scaling}][]{BIGIEL11} but when examined more closely, significant variations do emerge between this ratio and galaxy 
mass \citep[][Section \ref{sec:tdep_vary}]{SAINTONGE12} and dust-to-gas ratio or metallicity (Section \ref{sec:tdep_vary}). The interpretation of these trends 
depends critically on the behavior of the CO-to-H$_2$ conversion factor. We show that current best estimates may explain many of the observed 
trends in the CO-to-SFR ratio \citep[see also][]{BOLATTO11,SCHRUBA12}. However, we caution that significant work is still needed to bring conversion factor 
estimates to the precision needed to confidently interpret these trends. Our best estimate is that correlations in which the true molecular gas depletion 
time, $\tau_{\rm dep}^{\rm mol}$, increases with increasing galaxy mass or gas surface density do persist after accounting for $\alpha_{\rm CO}$ effects, but 
that these are comparatively weak. A sensible explanation for such trends is the emergence of a diffuse, unbound molecular medium at high gas surface densities
and high pressures, but direct evidence relating a local high molecular fraction to lower $\tau_{\rm dep}^{\rm mol}$ is weak in our present data set. We do find good
evidence for systematically low $\tau_{\rm dep}^{\rm mol}$ in galaxy centers, with a wide variation in the factor by which $\tau_{\rm dep}^{\rm mol}$ falls below the 
average for that disk. This supports the idea that in environments with high surface densities and short dynamical or orbital times, environmental factors may 
drive $\tau_{\rm dep}^{\rm mol}$ to a wide range of values at fixed average gas surface density. This is a more general formulation than the idea of distinct ``disk''
and ``starburst'' sequences, but qualitatively agrees with the picture from \citet{DADDI10} and \citet{GENZEL10}, with the nuclei of disk galaxies occupying a regime
intermediate between quiescent disks and merger-induced starbursts.

\acknowledgments  We thank the referee for a constructive report that lead to significant improvements in this paper. We also thank
Robert Feldmann, Scott Schnee, and David Whelan for feedback on drafts of this paper. We thank the {\em GALEX} NGS, SINGS, LVL, and
COLDGASS teams for making their outstanding 
datasets available. We thank staff of the IRAM 30m for their assistance carrying out the HERACLES
survey. We thank Deidre Hunter for sharing her H$\alpha$ image of NGC~4214. F.B., A.K.L., and F.W. gratefully acknowledge the Aspen
Center for Physics, where part of this work was carried out.  Support for A.K.L. for part of this project was provided by NASA through
Hubble Fellowship grant HST-HF-51258.01-A awarded by the Space Telescope Science Institute, which is operated by the Association of
Universities for Research in Astronomy, Inc., for NASA, under contract NAS 5-26555. K. S. is supported by a Marie Curie International Incoming Fellowship
J.C.M.M. acknowledges financial support from NASA JPL/{\em Spitzer} grant RSA 1374189 provided for the S4G project. A. B. wishes to acknowledge 
partial support from grants NSF  AST-0838178, NSF AST-0955836, as well as a Cottrell Scholar award from the Research Corporation for Science 
Advancement. We have made use of the Extragalactic Database (NED), which is operated by the Jet Propulsion Laboratory, California Institute of Technology,
under contract with the National Aeronautics and Space Administration. We also acknowledge use of the Lyon Extragalactic
Database (LEDA) and NASA's Astrophysics Data System (ADS). The National Radio Astronomy Observatory is a facility of the 
National Science Foundation operated under cooperative agreement by Associated Universities, Inc.

\newpage


\clearpage

\begin{appendix}

\section{Mean Line Ratio for HERACLES}

\begin{figure*}
\plottwo{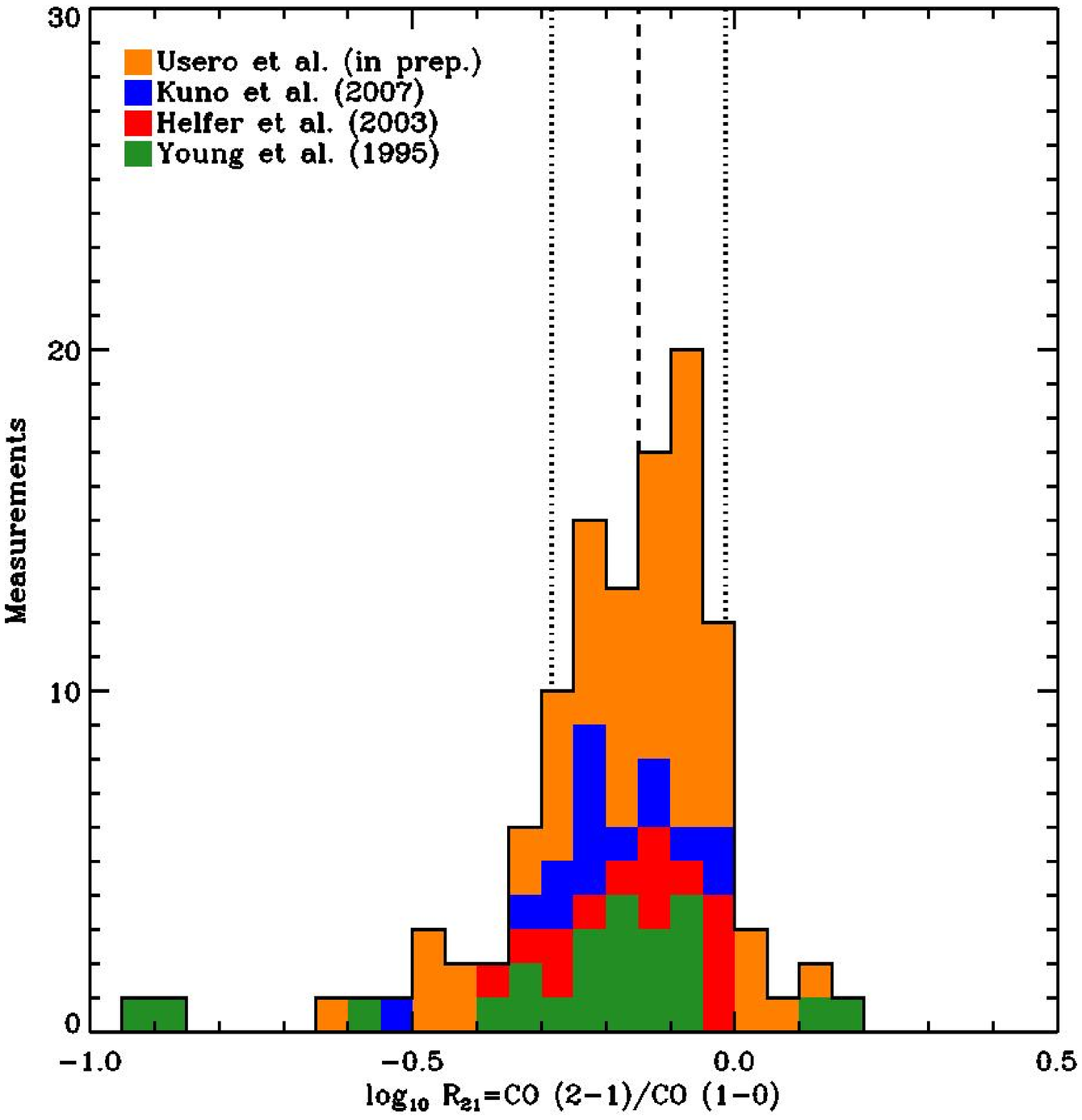}{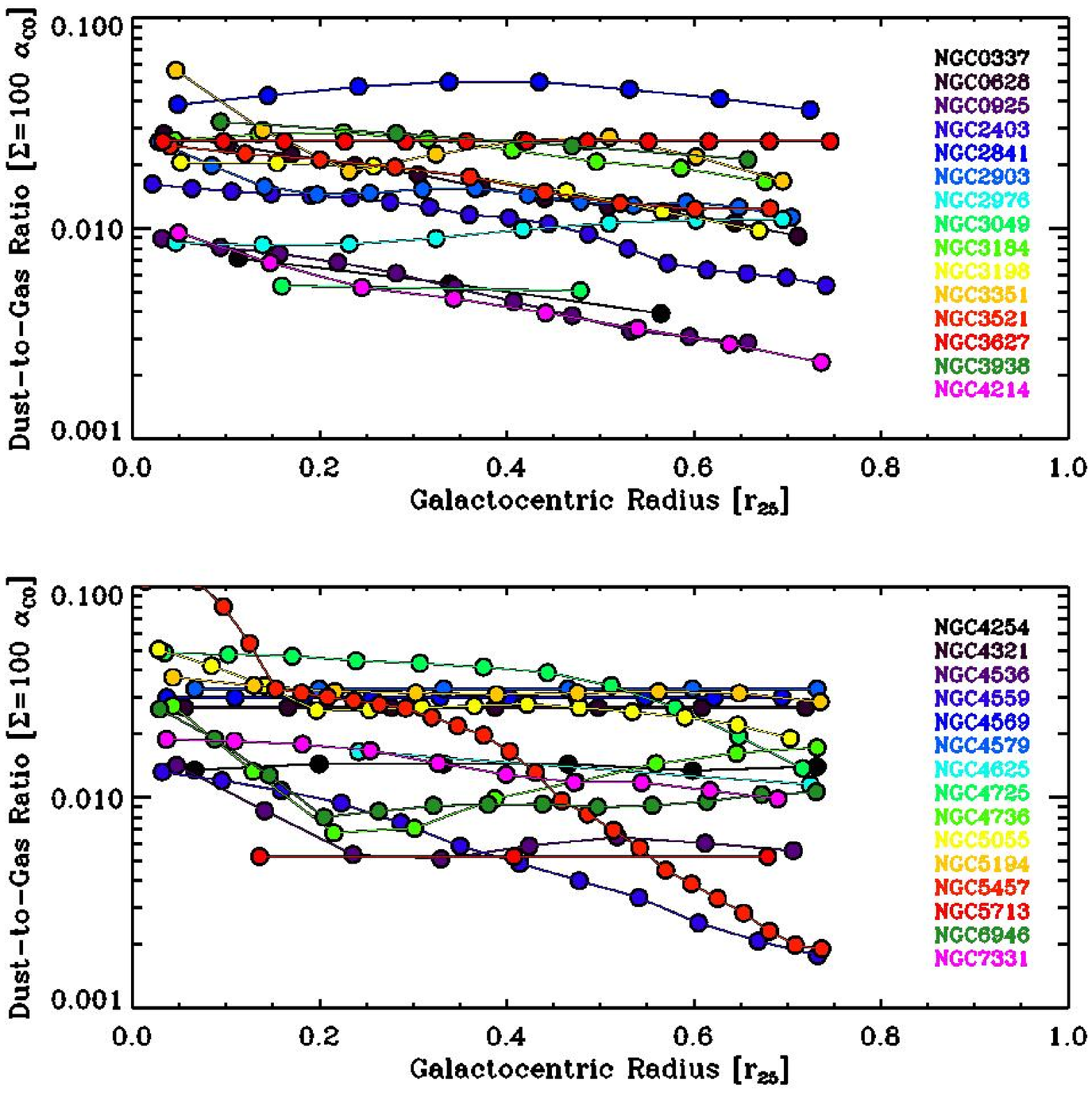}
\label{fig:linerat}
\caption{Line ratios and dust-to-gas ratios in HERACLES. ({\em left}) CO (2-1) / CO (1-0) line ratio comparing HERACLES integrated measurements to literature
CO (1-0) surveys: \citet{KUNO07}, \citet{HELFER03}, \citet{YOUNG95} and measurements for individual pointings obtained for follow-up spectroscopy 
by Usero et al. (in prep.). The ensemble of measurements has median $0.67$ with scatter $0.16$~dex. ({\em right}) Dust-to-gas ratio profiles derived comparing
IR SED modeling, HERACLES, and \hi\ data. We plot the dust-to-gas ratios derived fixed conversion factor in radial profile.}
\end{figure*}

HERACLES surveyed the CO(2-1) transition. Our adopted ``fixed'' $\alpha_{\rm CO} = 4.35$ has been derived considering mainly CO(1-0)
observations and most of the literature comparing recent star formation and molecular gas considers CO(1-0). Rosolowsky et al. (in
prep.) present a thorough analysis of the line ratio as a function of local conditions in HERACLES and explore the implications for
$\alpha_{\rm CO}$ and $\tau_{\rm dep}^{\rm mol}$. In this study we adopt a fiducial CO(2-1)/CO(1-0) ratio of 0.7. The left panel in
Figure \ref{fig:linerat} shows a histogram of CO(2-1)/CO(1-0) values measured for the HERACLES survey. We plot ratios derived from
comparing integrated HERACLES fluxes to those from the CO(1-0) surveys by \citet{YOUNG95}, \citet{HELFER03}, and \citet{KUNO07}. We also show
the results of pointed single-pixel spectroscopy carried out with the IRAM 30-m, part of a large spectroscopic database present by Usero et
al. (in preparation). These have been reduced using CLASS in a standard way and aperture-corrected to match beam areas using the
CO(2-1) distribution in HERACLES. The ensemble of measurements has median $0.67$, with the ratio taken in brightness temperature units,
and a scatter of $0.16$~dex, $\approx 40\%$. Some of this scatter will be due to calibration uncertainties in the CO(1-0) data. However, even comparing 
each data set to HERACLES separately, we find internal scatter in CO(2-1)/CO(1-0) of $20$--$40\%$. We discuss the 
most obvious environmental dependence of this ratio, the enhancement in galaxy centers in Section \ref{sec:centers} \citep[see also][]{LEROY09}. Based 
on Figure \ref{fig:linerat}, we adopt $0.7$ as a typical line ratio. In this paper, the primary application of this value is to apply a ``standard'' Milky Way CO(1-0) conversion
factor to the HERACLES CO(2-1) data.
 
\section{Dust-to-Gas Ratio and Conversion Factor Calculations}

The right panel in Figure \ref{fig:linerat} plots the dust-to-gas ratios, D/G, that we derive from a fixed $\alpha_{\rm CO}$ conversion factor and our {\em Spitzer} SED 
modeling.

\subsection{Correction Due to CO-Dark Gas}
 
We use the recent theoretical work by \citet{WOLFIRE10} to estimate the fraction of ``CO-dark'' molecular gas. This gas 
lies in in regions where carbon is mostly associated with \ion{C}{2} rather than CO, and so will not be readily traced by
maps of CO emission \citep[see][for a similar approach]{KRUMHOLZ11}. \citet{WOLFIRE10} present an expression
for the fraction of mass in this ``CO-dark'' phase,

\begin{equation}
\label{eq:xcowolf}
f_{\rm CO-dark} = 1 - \exp \left(\frac{- 4.0 \Delta A_{V,DG}}{\bar{A_V} (D/G^\prime)} \right)
\end{equation}

\noindent where $\Delta A_{V,DG}$ is the depth that the CO-dark phase extends into the cloud, measured in units of
visual extinction, and $\bar{A_V}$ is mean extinction through the whole cloud. $\bar{A_V}$ simply depends on the product of
cloud surface density --- or equivalently mean extinction through the cloud at solar metallicity, $\bar{A_V^0}$ --- and the 
dust-to-gas ratio, $\bar{A_V} (D/G^\prime) = \bar{A_V^0}~D/G^\prime$. \citet{WOLFIRE10} give an expression for $\Delta A_{V, DG}$ that 
depends weakly on the density and metallicity of the cloud 

\begin{equation}
\label{eq:deltaav}
\Delta A_{V,DG} = 0.53 - 0.045 \ln \left( \frac{G_0^\prime}{n_c} \right)-0.097~\ln Z^\prime~,
\end{equation}

\noindent Where $G_0^\prime$ is the radiation field relative to the Solar Neighborhood value, $n_c$ 
is the density of the cloud, and $Z^\prime$ is the metallicity (or dust-to-gas ratio) relative to the Galactic value. 
We neglect the second term, setting $G_0^\prime = 1$ and $n_c=1$~cm$^{-3}$ and use the $D/G^\prime$ in place of 
$Z^\prime$. $\Delta A_{V,DG}$ does not vary much with any of these quantities, so that  $f_{\rm CO-dark}$ mainly 
depends on the extinction through the cloud, $\bar{A_V} (D/G^\prime)$.

We scale the conversion factor by this expression, so that:

\begin{eqnarray}
\label{eq:xcowolf_app}
\alpha_{\rm CO} &\propto& \frac{1}{1 - f_{\rm CO-dark}} \\
\alpha_{\rm CO} &\propto& \left( \exp \left( \frac{- 4.0 \Delta A_{V,DG}}{\bar{A_V} (D/G)} \right) \right)^{-1} \\
\alpha_{\rm CO} &\propto& \exp \left(\frac{4.0 \Delta A_{V,DG}}{\bar{A_V} (D/G)} \right)
\end{eqnarray}
 
\noindent Adopting the conversion between dust and column density of \citet{BOHLIN78} and a fiducial average
molecular cloud surface density of $100$~M$_\odot$~pc$^{-2}$, so that $A_V \approx 4.9$, we calculate the correction for ``CO-dark'' gas via:

\begin{eqnarray}
\label{eq:codarkterm}
c_{\rm CO-dark} \left(D/G^\prime \right) &=& \alpha_{\rm CO} \left( D/G^\prime \right) / \alpha_{\rm CO} \left( D/G^\prime = 1 \right) \\
&=&  \alpha_{\rm CO} \left( D/G^\prime \right) / \alpha_{\rm CO} \left( D/G^\prime = 1 \right) \\
&=& \exp \left( \frac{4.0 \times 0.53}{ 4.9~\Sigma_{100}~D/G^\prime} \right) / \exp \left(\frac{4.0 \times 0.53}{ 4.9} \right) \\
&=& 0.65~\exp \left( \frac{0.4}{\Sigma_{100}~D/G^\prime} \right) 
\end{eqnarray}

\noindent where $\Sigma_{100}$ is the assumed universal surface density of molecular clouds normalized to 100~M$_\odot$~pc$^{-2}$ and $D/G^\prime$ is
the dust-to-gas ratio normalized to the Milky Way value. This is Equation \ref{eq:ccodark}.
 
 \subsection{Calculation of Dust-to-Gas Ratio and $\alpha_{\rm CO}$ From Profiles}
 
\section{Issues in Power Law Fitting}

\begin{figure*}
\plotone{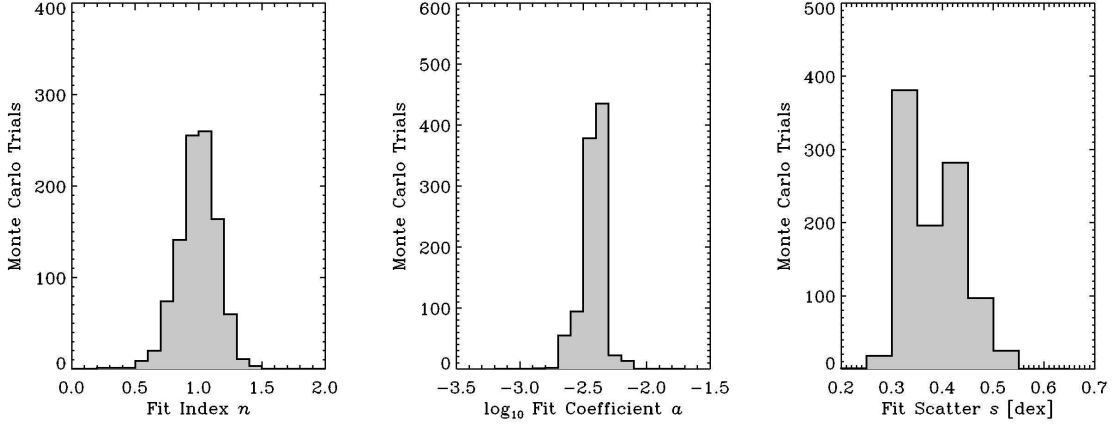}
\caption{Distribution of best-fit parameters for $1,000$ Monte Carlo iterations designed to test the 
impact of observation uncertainties, sample definition, and fitting technique on the best-fit power law. We 
plot the distributions of slope $n$, coefficient $a$, and intrinsic scatter $s$ across all iterations for our best 
$ \Sigma_{\rm SFR}$ estimate (H$\alpha$+24$\mu$m) and a fixed $\alpha_{\rm CO}$..}
\label{fig:fitunc}
\end{figure*}

\subsection{Monte Carlo Uncertainty Estimates}

\begin{deluxetable}{lccc}
\tablecaption{Monte Carlo Estimates of Uncertainty Power Law Fit Parameters} 
\tablehead{
\colhead{Source of Uncertainty} & \colhead{Scatter in $n$} & \colhead{Scatter in $\log_{10} a$} & 
\colhead{Scatter in $s$}
}
\startdata
Statistical Noise & 0.05 & 0.01 & 0.02 \\
Calibration Uncertainties & 0.12 & 0.06 & 0.04  \\
Robustness to Removal of Individual Data & 0.07 & 0.03 &  0.02 \\
Robustness to Removal of Galaxies & 0.10 & 0.06 & 0.05 \\
Choice of Fitting Parameters & 0.06 & 0.02 & 0.02 \\
\hline
Overall Uncertainty & 0.15 & 0.09 & 0.05
\enddata
\label{tab:unc_fits}
\tablecomments{Each entry gives the $1\sigma$ standard deviation in the parameter across 100 Monte Carlo 
iterations (1,000 for the overall uncertainty). We quote results for our best-estimate $\Sigma_{\rm SFR}$, H$\alpha$+24$\mu$m corrected for IR 
cirrus, and a fixed $\alpha_{\rm CO}$.}
\end{deluxetable}

If we treat our goodness-of-fit statistic (Equation \ref{eq:gof}) as $\chi^2$ then the implied statistical uncertainties on our fits are
very small \citep[the $\chi^2+1$ surface implies uncertainties order $1\%$, as in][]{BLANC09}. At some level, this is accurate: given our
fitting approach and data set, the best-fit power law is heavily constrained. However, this does not reflect our real best estimate of
the uncertainty in the underlying relationship between $\Sigma_{\rm SFR}$ and $\Sigma_{\rm mol}$. To estimate a more realistic
uncertainty, we carry out a series of Monte Carlo simulations. We begin with our measurements, estimates of $\Sigma_{\rm mol}$ and
$\Sigma_{\rm SFR}$ at each point. We then examine the effects of:

\begin{itemize}
\item {\em Statistical noise.} For each point, we add normally-distributed noise of the appropriate 
magnitude to each $\Sigma_{\rm mol}$ and log-normally distributed noise of magnitude $0.15$~dex 
to $\Sigma_{\rm SFR}$ (see Section \ref{sec:data}).
\item {\em Calibration uncertainties.} Calibration issues include uncertainties in the overall flux scale 
of the data or uncertainties in the conversion to physical parameters, e.g., due to variations in CO line ratios, 
dust properties, or stellar populations. In each case, these will tend to operate galaxy-by-galaxy. We take 
these to be log-normally distributed with a magnitude of $0.15$~dex \citep[][L12]{LEROY09}. We scale all of the data for each axis in 
each galaxy by a single, randomly generated factor.
\item {\em Robustness to removal of individual data.} We test the robustness of the results to the 
removal of individual data using a standard bootstrapping approach. We resample the data, allowing repeats, to 
produce a data set matched in size to the original.
\item {\em Robustness to removal of galaxies.} We also test the robustness of our fit to the removal of a whole galaxies using a 
bootstrapping approach. Instead of resampling the ensemble of measurements, we resample the list of galaxies, allowing repeats, to 
produce a sample of 30 (non-unique) galaxies.
\item {\em Choice of fitting parameters.} A subtle point in our fitting is how to handle data in cells with 
low (or zero) model probability. A single outlying datum can dramatically skew the results if this issue is not properly treated. 
By default, we follow \citet[][and G. Blanc priv. comm.]{BLANC09} and never allowing the weight for a single cell to be lower 
than expected for $N_{\rm model}^i = 1$ (i.e., we cap the denominator in Equation \ref{eq:gof} at this level). We test the impact of this 
choice and our adopted grid cell size (0.125 dex). We allow the bin size and the minimum weight per cell to vary by up to a factor of 
two either higher or lower, with equal probability across the range.
\end{itemize}

\noindent We test each of these sources alone and report the results for $\Sigma_{\rm SFR}$ traced by H$\alpha$ and 24$\mu$m and
$\Sigma_{\rm mol}$ calculated from a fixed $\alpha_{\rm CO}$ in Table \ref{tab:unc_fits}. Figure \ref{fig:fitunc} shows the distribution of
fitted parameters across all runs for our best $\Sigma_{\rm SFR}$ estimate. The strongest contributors to the overall uncertainty are
galaxy-to-galaxy calibration uncertainties or sample definition. Our large data set and good S/N makes statistical noise over individual
lines or sight or removal of individual data minor concerns. Choice of fitting parameters do impact the overall results, but do not dominate
the uncertainty.

In the end, we derive our overall uncertainty on the fits from the scatter in the best fit parameters across 1,000 Monte Carlo iterations
that include all of these effects. Because we can only add noise or decrease our sample size, each Monte Carlo iteration operates on a
data set of inferior quality to that used for our best estimates (Table \ref{tab:fit}). We consider these uncertainties realistic but conservative.

\subsection{Power-Law Index Biases in Bivariate Fits to Clipped Data}

\begin{figure*}
\plottwo{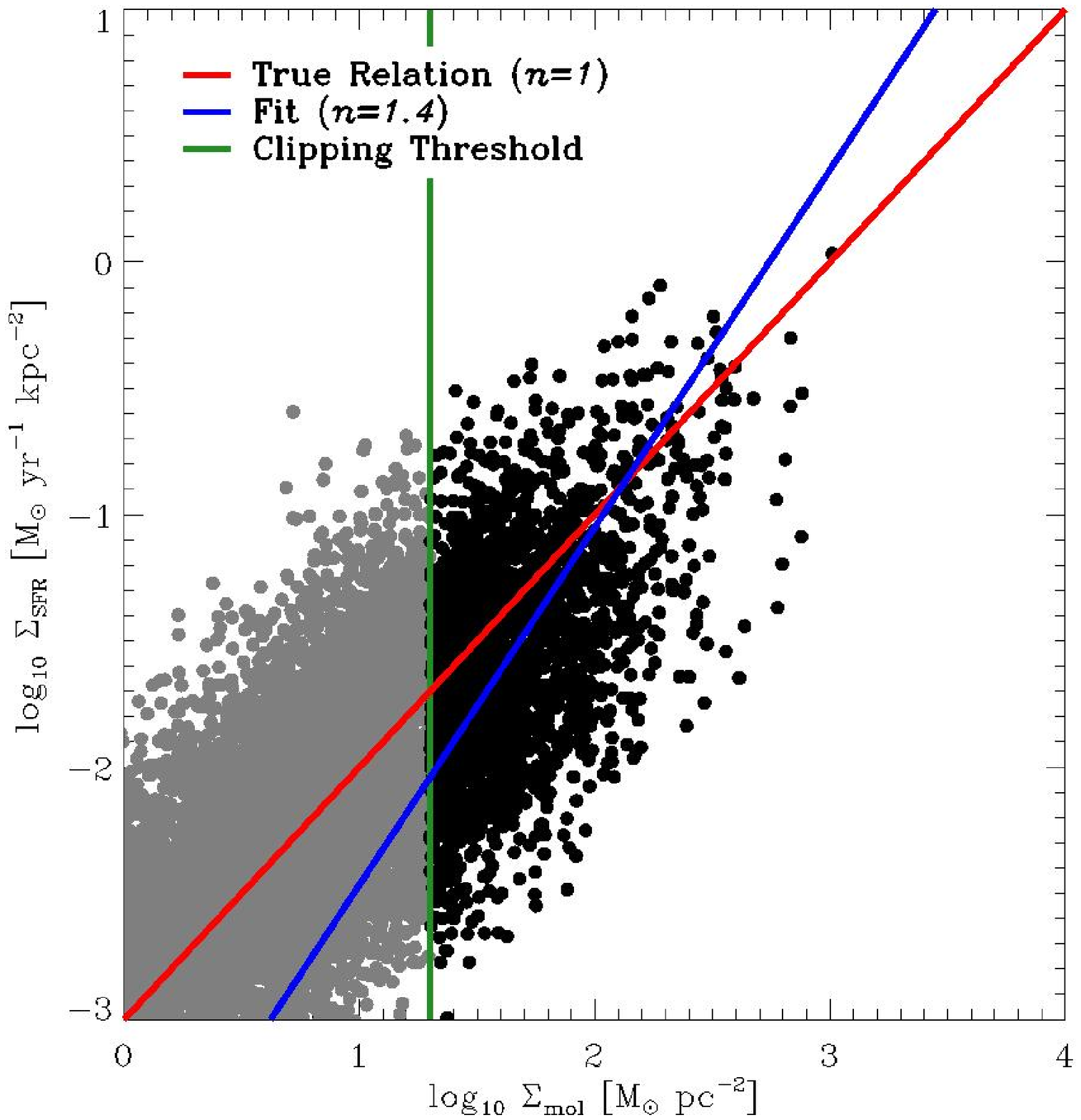}{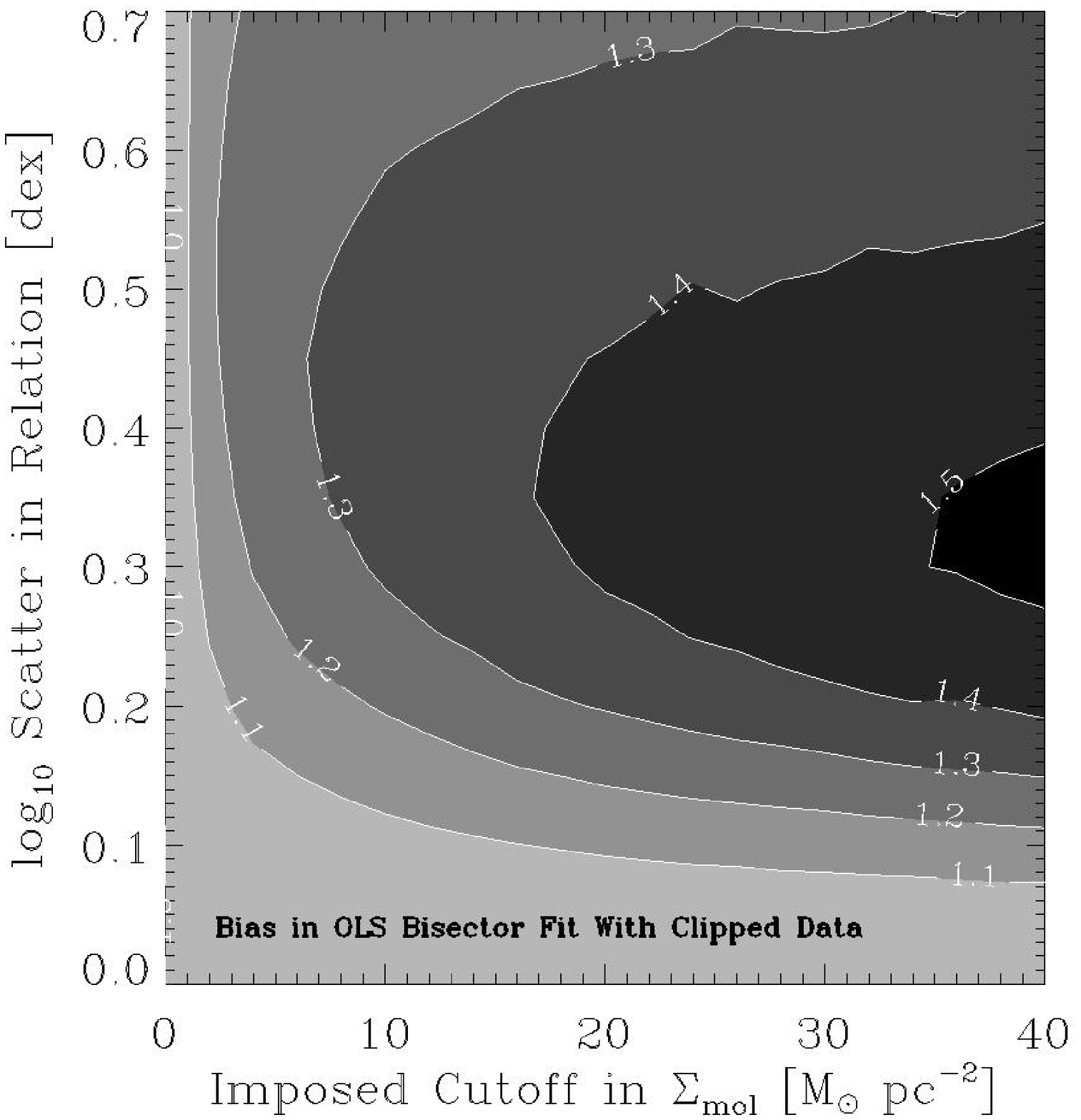}
\caption{({\em left}) Example of the biases introduced by clipping data 
on one axis (e.g., due to limited S/N) before carrying out a bivariate fit. Points 
show a simulated data set in which $\Sigma_{\rm SFR} \propto \Sigma_{\rm mol}$ 
by construction (red line) with $0.3$~dex scatter introduced to each axis. We clip 
data below $\Sigma_{\rm mol} = 20$~M$_\odot$~pc$^{-2}$ so that gray points are
discarded and black points are fit. We then fit the black points using the OLS bisector 
method and plot the result in blue. The result is a bias towards higher slope. ({\em right}) 
Derived power law index from using the OLS bisector to fit noisy, low dynamicrange data 
with clipping applied to one axis. The underlying model has an intrinsic $n=1.0$ slope with 
$\Sigma_{\rm mol}$ drawn from our data and log-normal scatter of the indicated magnitude 
applied to each axis. The grayscale and contours report the best-fit power law index after the 
data are clipped at the $x$-axis value in one axis and fit using the OLS bisector technique. Biases 
of several tenths up to as much as four tenths are possible within the range of values adopted by recent studies.}
\label{fig:fitbias}
\end{figure*}

Fitting a relationship between $\Sigma_{\rm SFR}$ and $\Sigma_{\rm mol}$ entails conducting 
a bivariate fit to data with limited sensitivity. The approach of \citet{BLANC09} incorporates upper
limits. However, it has been common practice to clip data at some signal-to-noise threshold in 
CO ($\Sigma_{\rm mol}$) and then to carry out a bivariate linear fit relating $\log_{10} \Sigma_{\rm SFR}$ to
$\log_{10} \Sigma_{\rm mol}$ above this threshold, e.g., using the OLS bisector \citep{ISOBE90} or 
FITEXY \citep{PRESS92} methods. This clip-and-fit approach can introduce a significant bias into the fit
power law index. We note this effect and its magnitude here.

The bias arises from the interaction of the clipping with fitting techniques that attempt to minimize the 
two-dimensional distance between a point and the fit line (in log-log space). Consider data
that are intrinsically defined by a power law, a line in log-log space, but exhibit significant scatter 
orthogonal to the line. Now consider data near the clipping threshold in $\Sigma_{\rm mol}$, the
quantity that typically represents the limiting observable. Data that scatter to high $\Sigma_{\rm SFR}$ 
and low $\Sigma_{\rm mol}$ will be discarded from the analysis due to the clipping. Data that scatter to
low $\Sigma_{\rm SFR}$ and high $\Sigma_{\rm mol}$ will be included in the analysis. Preferentially adding 
data at low $\Sigma_{\rm SFR}$ and high $\Sigma_{\rm mol}$ will tend to skew the fit towards steeper
slopes. The left panel in Figure \ref{fig:fitbias} illustrates the effect for simulated data.

We estimate the magnitude of the effect using a Monte Carlo simulation. We consider samples of 10,000 data 
points with $\Sigma_{\rm mol}$ randomly drawn from our data set. We assume an underlying linear relation, 
$\Sigma_{\rm SFR} \propto \Sigma_{\rm mol}$, so that the true $n=1$. We set $\tau_{\rm dep}^{\rm mol} = 1$
Gyr, but the choice is arbitrary. We realize 100 samples of 10,000 data points each. For each sample we introduce 
log-normal scatter of equal magnitude to each axis and then clip the data at a series of $\Sigma_{\rm mol}$ values 
from 0 to 40~M$_\odot$~pc$^{-2}$. We then carry out an OLS bisector fit to the data and record the best-fit
power law index. The right panel in Figure \ref{fig:fitbias} plots the average fit index across all Monte Carlo iterations 
as a function of the scatter introduced into the data ($y$-axis) and the threshold imposed ($x$-axis). We find that for 
commonly used thresholds ($\Sigma_{\rm mol} \sim 10$--$20$~M$_\odot$~pc$^{-2}$) and typical observed scatters 
($\sim 0.3$~dex) a significant bias can be introduced to the best fit index, often shifting it from its true value of $1.0$ 
(by construction) to $\sim 1.3$ or $1.4$.

Note that Figure \ref{fig:fitbias} illustrates the problem but that the exact magnitude of the bias will depend on the dynamic range,
noise, and underlying relationship in the data set studied. Also note that conversely, clipping based on $\Sigma_{\rm SFR}$ will tend to
bias the index in the opposite direction, to lower $n$. Clipping based on both $\Sigma_{\rm SFR}$ and $\Sigma_{\rm mol}$ may lead to
offsetting biases but does not represent a proper substitute for a rigorous treatment. Finally, we emphasize that although we do clip
data below $\Sigma_{\rm mol} = 5$~M$_\odot$~pc$^{-2}$ in our fitting (Section \ref{sec:index}) the bias describe here will not 
affect our results because we model the data distribution rather than carry out a bivariate fit.

\section{$\Sigma_{\rm SFR}-\Sigma_{\rm mol}$ Relations for Individual Galaxies}

\begin{figure*}
\plotone{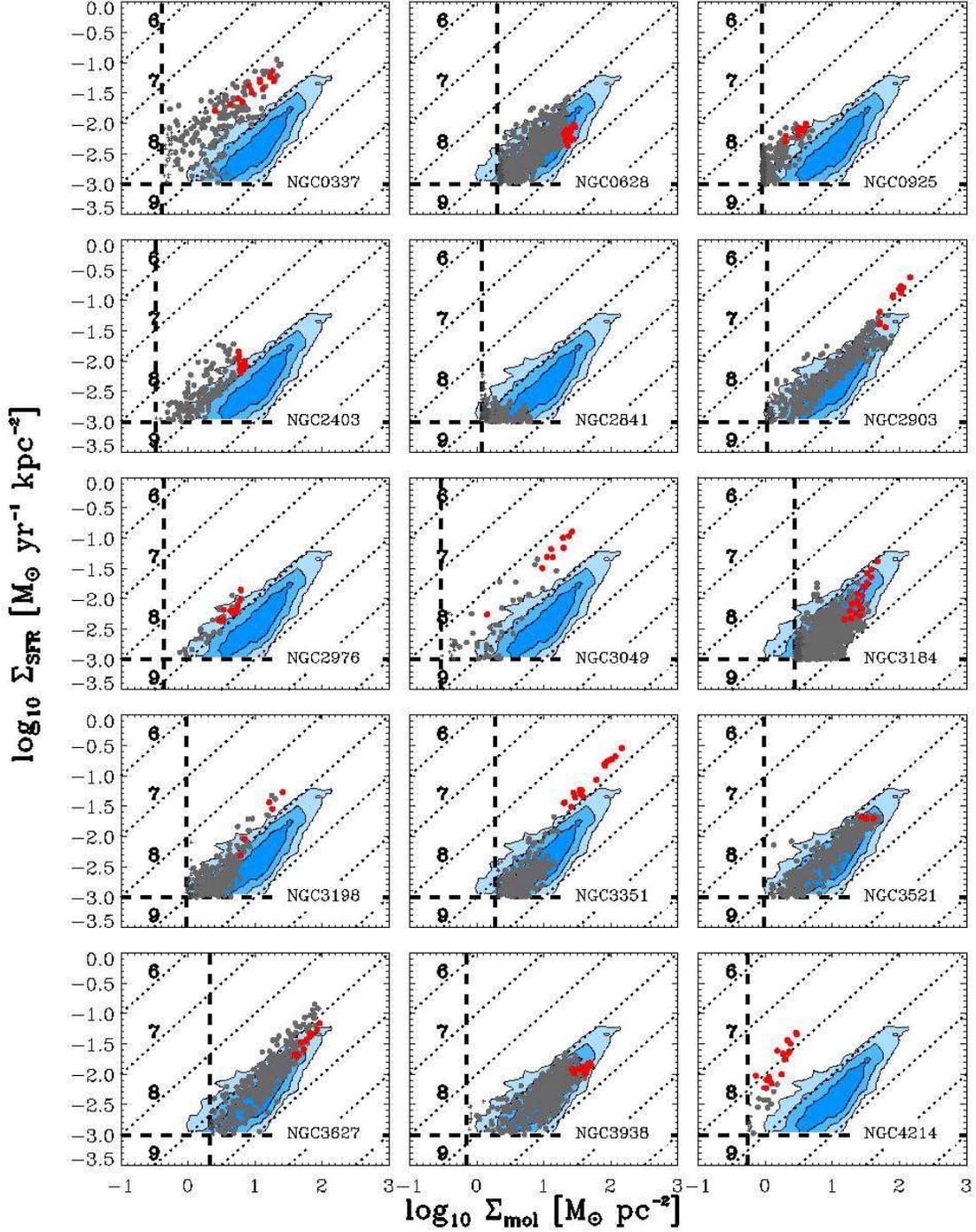}
\caption{$\Sigma_{\rm SFR}$, estimated from H$\alpha$+24$\mu$m, as a function of $\Sigma_{\rm mol}$, estimated from 
HERACLES CO(2-1) data and a fixed $\alpha_{\rm CO}$, for individual galaxies. Data from the designated galaxy appear
in gray, with data from the inner $0.1~r_{25}$ marked in red. In the background, we plot data density contours from the ensemble of
all measurements (Figure \ref{fig:combined}). Annotation and symbols are otherwise as Figure \ref{fig:combined}.}
\label{fig:individ_1}
\end{figure*}

\begin{figure*}
\plotone{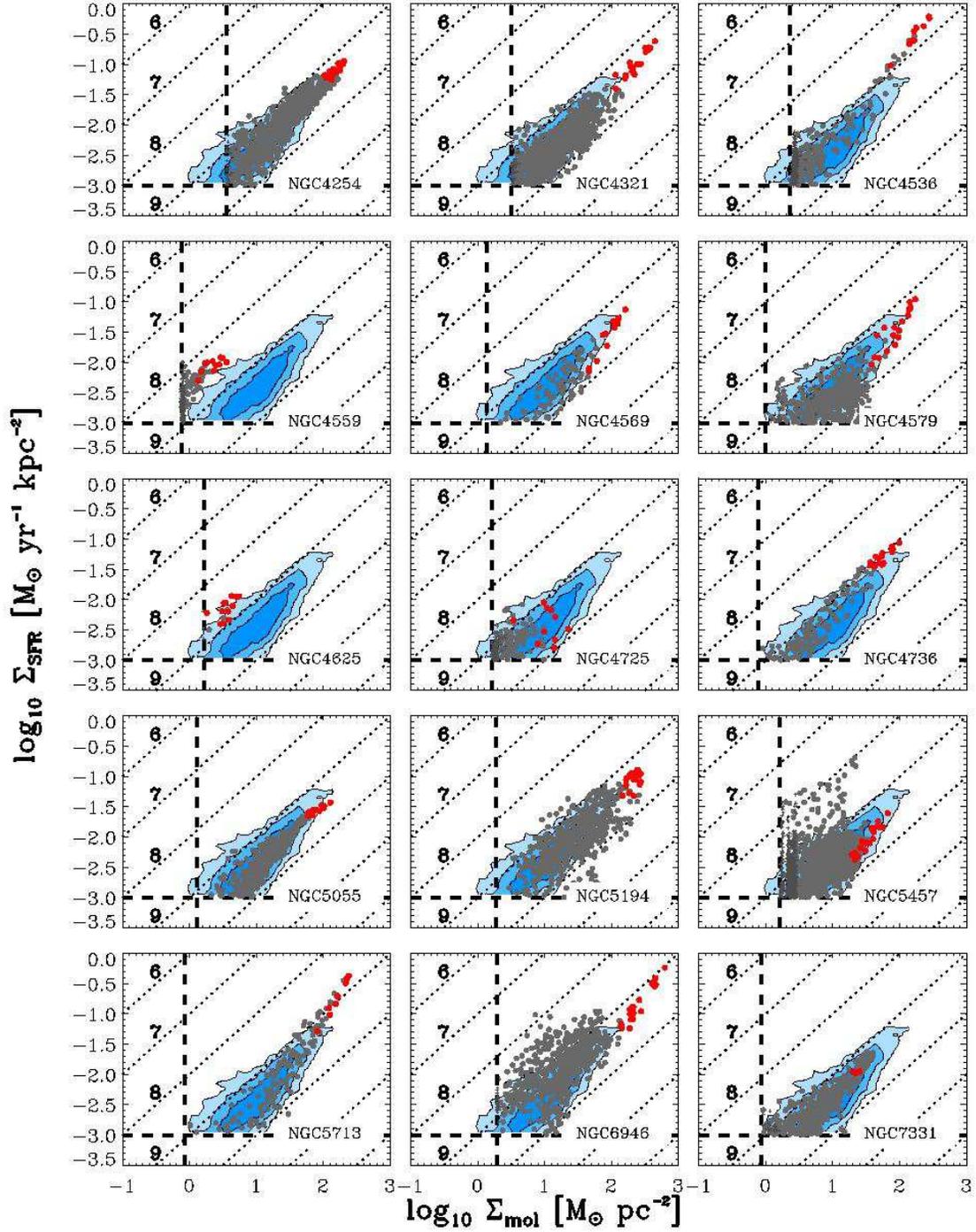}
\caption{As Figure \ref{fig:individ_1} for the remaining galaxies in our sample.}
\label{fig:individ_2}
\end{figure*}

Combining our measurements into a single data set obscures real differences among galaxies. Table \ref{tab:integrated}, Figure \ref{fig:fit},
and Section \ref{sec:global} show systematic variations in average $\tau_{\rm dep}^{\rm mol}$ and power-law index among galaxies. In 
Figures \ref{fig:individ_1} and \ref{fig:individ_2} we plot $\Sigma_{\rm SFR}$ as a function of $\Sigma_{\rm mol}$ for individual galaxies.
In the background, we plot contours for the combined distribution shown in Figure \ref{fig:combined}. We show results for our best SFR
tracer, H$\alpha$+24$\mu$m, and a fixed CO-to-H$_2$ conversion factor. Points from the central part of the galaxy, $r_{\rm gal} \leq 0.1~r_{25}$ appear
in red. Points from the rest of the galaxy appear in gray.

\end{appendix}

\end{document}